\documentclass[fleqn,usenatbib]{mnras}

\usepackage{newtxtext,newtxmath}

\usepackage[T1]{fontenc}
\usepackage{ae,aecompl,url}
\DeclareRobustCommand{\VAN}[3]{#2}
\let\VANthebibliography\thebibliography
\def\thebibliography{\DeclareRobustCommand{\VAN}[3]{##3}\VANthebibliography}

\usepackage{graphicx}	
\usepackage{amstext}

\usepackage{amssymb}	
\usepackage{geometry,lscape}
\usepackage{longtable}
\usepackage{parskip}

\newcommand{\NHI}{$N\rm _{H\,{\sevensize I}}$}

\newcommand{\lya}{Ly$\alpha$}

\newcommand{\mgii}{\mbox{Mg\,{\sc ii}}}

\newcommand{\oii}{[O\,{\sc ii}]}

\newcommand{\kms}{$\rm km~s^{-1}$}

\newcommand{\nqso}{28}
\newcommand{\nlls}{61}

\newcommand{\nem}{127}

\newcommand\HI{\ion{H}{I}}

\title[MAGG IV.]{MUSE Analysis of Gas around Galaxies (MAGG) - IV: The gaseous environment of $z\sim3-4$ Ly$\alpha$ emitting galaxies}

\author[Lofthouse et al.]{Emma K. Lofthouse$^{1,2}$\thanks{E-mail: emmakatherine.lofthouse@unimib.it}, 
  Michele Fumagalli$^{1,3}$\thanks{E-mail: michele.fumagalli@unimib.it}, 
  Matteo Fossati$^{1,2}$,
  Rajeshwari Dutta$^{1,2}$, 
  Marta Galbiati$^{1}$,\and
  Fabrizio Arrigoni Battaia$^{4}$, 
  Sebastiano Cantalupo$^{1}$, 
  Lise Christensen$^{5}$,
  Ryan J. Cooke$^{6}$,
  Alessia  Longobardi$^{1,2}$,\and
  Michael T. Murphy$^{7}$,
  J. Xavier. Prochaska$^{8,9}$
  \\
    $^{1}$Dipartimento di Fisica ``G. Occhialini'', Universit\`a degli Studi di Milano Bicocca, Piazza della Scienza 3, 20126 Milano, Italy \\
    $^{2}$INAF - Osservatorio Astronomico di Brera, via Brera 28, 20121 Milano, Italy \\
    $^{3}$INAF - Osservatorio Astronomico di Trieste, via G. B. Tiepolo 11, 34143 Trieste, Italy \\
    $^{4}$Max-Planck-Institut f\"ur Astrophysik, Karl-Schwarzschild-Str 1, D-85748 Garching bei M\"unchen, Germany \\
    $^{5}$Cosmic Dawn Center (DAWN), Niels Bohr Institute, University of Copenhagen,  Jagtvej 128,   2200 N Copenhagen,  Denmark\\
    $^{6}$Centre for Extragalactic Astronomy, Durham University, South Road, Durham, DH1 3LE, UK \\
    $^{7}$Centre for Astrophysics and Supercomputing, Swinburne University of Technology, Hawthorn, Victoria 3122, Australia \\
    $^{8}$Department of Astronomy and Astrophysics, University of California, Santa Cruz, CA 95064, USA \\
    $^{9}$Kavli Institute for the Physics and Mathematics of the Universe, 5-1-5 Kashiwanoha, Kashiwa 277-8583, Japan
}

\pubyear{\the\year}

\begin{document}
\label{firstpage}
\pagerange{\pageref{firstpage}--\pageref{lastpage}}
\maketitle

\begin{abstract}
We study the link between galaxies and \ion{H}{I}-selected absorption systems at $z\sim3-4$ in the MUSE Analysis of Gas around Galaxies (MAGG) survey, an ESO large programme consisting of integral field spectroscopic observations of 28 quasar fields hosting 61 strong absorbers with $\rm N_{\rm HI}\gtrsim 10^{16.5}~\rm cm^{-2}$. 
We identify \nem\ Ly$\alpha$ emitting galaxies (LAEs) around the absorbers, corresponding to a detection rate of 82$\pm$16 per cent. The luminosity function of these LAEs is $\approx5$ times higher in normalization than the field population and we detect a significant clustering of galaxies with respect to the gas, confirming that high column density absorbers and LAEs trace each other. 
Between 30 and 40 per cent of the absorbers are associated with multiple LAEs, which lie preferentially along filaments. Galaxies in groups also exhibit a three times higher covering factor of optically-thick gas compared to isolated systems. No significant correlations are identified between the emission properties of LAEs and the absorption properties of optically-thick gas clouds, except for a weak preference of brighter and multiple galaxies to reside near broad absorbers.
Based on the measured impact parameters and the covering factor, we conclude that the near totality of optically-thick gas in the Universe can be found in the outer circumgalactic medium (CGM) of LAEs or in the intergalactic medium (IGM) in proximity to these galaxies. Thus, LAEs act as tracers of larger scale structures within which both galaxies and optically-thick clouds are embedded. The patchy and inhomogeneous nature of the CGM and IGM explains the lack of correlations between absorption and emission properties. This implies that very large samples are needed to unveil the trends that encode the properties of the baryon cycle.\\
\end{abstract}




\section{Introduction}

Gas is a fundamental constituent in the life of galaxies. The very formation of galaxies hinges on the ability of gas to cool from the intergalactic medium (IGM) onto collapsing dark matter halos. Gas is the raw fuel that feeds and sustains the formation of stars, and it is a critical agent in the self-regulation mechanisms that operate inside and around galaxies, within the circumgalactic medium (CGM).
Therefore, a complete mapping of the different gas phases inside and around galaxies becomes essential to fully describe the processes that shape their evolution. 

A complete census of how gas is distributed and exchanged between galaxies across cosmic time is, however, far from trivial to obtain. The multiphase nature of gas requires concerted observational efforts across the entire electromagnetic spectrum, but some phases remain  particularly hard to detect especially at large cosmological distances due to the intrinsic diffuse nature of the gas and the weak electromagnetic signal associated. 

Recent advancements in instrumentation have however enabled a breakthrough in our ability to map gas as a function of cosmic time. The Atacama Large Millimeter/submillimeter Array (ALMA; \citealt{wootten2009}) has expanded our view of the cold atomic and molecular phase inside the interstellar medium (ISM) and the dense parts of the CGM, up to very high redshifts \citep[e.g.][]{franco2019,peroux2019,neeleman2017,neeleman2020}.
High-sensitivity integral field (IFU) spectrographs at 8~m class telescopes such as the Multi Unit Spectroscopic Explorer (MUSE, \citealt{bacon2010}) and the Keck Cosmic Web Imager (KCWI,  \citealt{morrissey2018}) are starting to yield a glimpse of the low-density and ionized hydrogen inside and around galaxies up to $z\approx 6$, including gas in filaments connecting multiple $z\approx 3$ star-forming galaxies \citep[e.g.][]{leclercq2017,wisotzki2018,kikuta2019,umehata2019, bielby2020,bacon2021, muzahid2021}.
Very recently, pathfinders of the Square Kilometre Array (SKA) such as MeerKAT \citep{jonas2009} are pushing the detectability of the more diffuse and widespread cold/warm neutral medium traced by atomic hydrogen, reaching for the first time cosmological distances. Stacking of 21cm signals have allowed detections out to $z\approx 1$ \citep[][]{chowdhury2020}. However, until the advent of the full SKA, these efforts remain limited to the current modest redshifts (z $\lesssim$ 1), and the $z\gtrsim 2$ Universe must be explored by other means, such as absorption spectroscopy. 

Indeed, thanks to large quasar surveys, and particularly the Sloan Digital Sky Survey \citep{york2000}, it has been possible to map in detail the distribution, kinematics, and chemical composition of fully neutral
and partially ionised \ion{H}{I} along the line of sight to high-redshift quasars. The detailed study of these intervening absorption line systems (ALSs), and in particular the damped Ly$\alpha$ absorbers (DLAs;~$N_{\rm HI} \ge10^{20.3}$) and the Lyman limit systems (LLSs; $10^{17.2} {\rm cm^{-2}} \le N_{\rm HI} < 10^{20.3} {\rm cm^{-2}}$) has been for several years a fruitful tool for the study of the high-redshift IGM, CGM, and ISM from $z\gtrsim 2$ all the way up to the end of the reionization epoch \citep[e.g.][]{peroux2003,wolfe2005, prochaska2009,rafelski2012,fumagalli2013,fumagalli2016}.

The most difficult aspect of this approach, however, has been to establish a direct connection between gas probed in absorption along the quasar pencil beam and the associated galaxies. Attempted searches in emission of galaxies near ALSs have been accumulating in the literature, but have traditionally resulted in a large number of non-detections. In the last decade, however, there has been an acceleration in successfully identifying associations between gas probed in absorption and galaxies seen in emission. At $z\approx 2-3$, surveys using multi-object spectroscopy, e.g. the Keck Baryonic Structure Survey \citep[KBSS;][]{rudie2012,trainor2012,rakic2012} and the VLT Lyman break galaxy (LBG) Redshift Survey \citep[VLRS;][]{bielby2011}, have shown that star-forming galaxies reside inside gas overdensities, and have revealed that metal-enriched gas exhibits complex kinematics modulated by a combination of outflow and inflows \citep[e.g.][]{steidel2010,rubin2010,crighton2011,rudie2012,turner2014,tummuangpak2014,bielby2017}. These surveys, albeit powerful, require a pre-selection of the targeted galaxy population and are thus likely to miss a large population of continuum-faint galaxies that emit strong emission lines \citep[e.g.][]{crighton2015}.

More targeted approaches based on long-slit spectroscopy have indeed revealed several emission line galaxies at close separation from ALSs, and in particular DLAs \citep[e.g.][]{moller2002,moller2004,fynbo2010,srianand2016}. 
Based on empirical scaling relations and model expectations \citep[][]{moller2004,ledoux2006,krogager2020}, most of these efforts have focused on metal-rich ALSs, leaving therefore mostly unexplored the population of galaxies in proximity of the more metal poor systems, which happen to be the more common ones \citep{fumagalli2016}. The advent of integral field spectrographs, starting with the pioneering efforts in the near infrared using e.g. SINFONI \citep{eisenhauer2003, bonnet2004} or OSIRIS \citep{larkin2006}, has transformed our ability to  search for galaxies in proximity to ALSs without the need for pre-selecting either the galaxies or the absorbers. Indeed, several successful examples are accumulating in the literature showing how sensitive searches can be conducted blindly around ALSs across a large range of column densities and metallicities, resulting in a high number of new detections \citep[e.g.][]{peroux2011,peroux2012, fumagalli2016b,fumagalli2017,mackenzie2019,bielby2020}. In parallel, searches for galaxies near ALSs with sub-millimeter interferometers have been equally successful \citep[e.g.][]{moller2018,klitsch2018,neeleman2017,kanekar2018,
neeleman2020}, underscoring the need of a multiwavelength approach to finally paint a complete view of the link between gas probed in absorption and galaxies detected in emission. 

In this paper, we exploit the power of MUSE and present a comprehensive study of the link between gas and galaxies at $z\gtrsim 3$ using the MUSE Analysis of Gas Around Galaxies (MAGG), an ESO Large Programme (PID 197.A-0384, PI Fumagalli) designed to blindly map the galaxy population in \nqso\ quasar fields with \nlls\ high \ion{H}{I} column density ALSs along the line of sight. With at least 5 hours of MUSE observations in each field complemented by high resolution and high signal-to-noise ratio ($S/N$) quasar spectroscopy, the MAGG survey is designed to yield the discovery of orders of magnitude more galaxies within a projected radius of $\approx 250~\rm kpc$ from strong ALSs compared to previous surveys. Moreover, differently from most of previous studies, MAGG targets high \ion{H}{I} column density ALSs ($\gtrsim 10^{16.5}~\rm cm^{-2}$) which are most likely to trace the denser parts of the CGM and IGM at $z\gtrsim 3$ \citep{faucherGiguere2011,fumagalli2011,vandeVoort2012} without selecting on absorption properties other than the hydrogen column density. This survey is therefore constructed to provide the most complete and unbiased snapshot of the link between gas and galaxies when the Universe was $\approx 4$ billion years old, just before the peak of the cosmic star formation history.   

The design of the MUSE survey, together with a detailed description of the observations and data reduction of both MUSE and ancillary high-resolution quasar spectroscopy, has already been presented in the first paper of this series \citep{lofthouse2020}. The MAGG data have also been used to study the link between lower redshift ($z\lesssim 1.8$) galaxies and metal absorption lines, and in particular \ion{Mg}{II} \citep{dutta2020} also as a function of galaxy overdensity \citep{dutta2021}, and to map the gas and galaxy environment of high-redshift quasars \citep{fossati2021}. We now turn to the core objective of the MAGG survey, i.e. the study of the gas-galaxy link at $z\gtrsim 3$. Throughout this work, we use the \citet{planck2016} cosmology, with $\Omega_m = 0.307 $ and $H_0= 67.7~\rm km~s^{-1}~Mpc^{-1}$. Unless otherwise stated, all distances are in proper units and we use the AB magnitude system.

\section{Data acquisition and reduction}\label{sec:data}

In this section we provide a brief overview of the acquisition and data reduction for the MUSE observations and for the high-resolution quasar spectroscopy. We also briefly summarise the steps taken to generate the galaxy catalogues. Additional details can be found in previous papers of this series, particularly \citet{lofthouse2020} and \citet{fossati2021}. In this work, we focus in more detail on the analysis of absorption line systems, including the measurements of the \ion{H}{I} and metal column densities and the procedures adopted to constrain the gas-phase metallicity of the ALSs.

\subsection{MUSE observations and data reduction}

The MAGG sample consists of  \nqso\ quasars, including five fields from the ESO archive such as data from \citet{borisova2016} as part of GTO observations (see \citealt{lofthouse2020}, their table 1 for full details including individual exposure times for each sightline and references to the ESO programmes). MUSE observations of the original MAGG fields have been carried out during ESO periods 97--103, integrating on-source for $\approx4$~h per field. Longer exposures are achieved for fields with available exposures in the ESO archives such as the GTO fields which have up to 10 hours observations. To improve the homogeneity of the sensitivity across the different MUSE spectrographs, each observing block included dithers and instrument rotations of 90 deg. Data have been acquired during clear nights at airmass $<$ 1.6, and the resulting image quality is on average 0.6-0.7 arcsec full width at half maximum (FWHM). 
For the data reduction we follow consolidated techniques, see e.g. \citet{lofthouse2020} and \citet{fossati2021}, that combine the ESO MUSE pipeline \citep{weilbacher2014} with post-processing tools, and in particular the {\sc CubExtractor} package, hereafter {\sc CubEx}  \citep[][and Cantalupo in prep.]{cantalupo2019}.
Bias subtraction, flat-field correction and the wavelength and flux calibrations are first completed using the ESO pipeline. Cubes are sky-subtracted and reconstructed on a regular 3D grid, with the astrometry registered on the Gaia DR2 reference frame \citep{gaia2018}.

Using {\sc CubEx}, we then correct residual imperfections in the sky subtraction and illumination corrections. In particular, we use the {\sc CubeFix} tool to balance the different response of spectrographs, stacks and slices in MUSE. The {\sc CubeSharp} tool is then used to suppress sky residuals arising from variations in the instrument line spread function. These tools are applied iteratively, to achieve optimal masking of the astrophysical sources in the field. In the end, we combine all the reduced exposures in final cubes using both  mean and median statistics, and we further produce two coadds containing only half of all the exposures, which we use to validate the detection of sources.
The wavelength solutions of the final cubes are all in air; however, when extracting spectra of continuum sources or of LAEs, we convert to vacuum wavelengths to be consistent with the quasar spectroscopy and hence the absorber redshifts (see next section).

While the pixel uncertainty is propagated during the various reduction steps, the uncertainty of the final cubes is found to not reproduce with sufficient accuracy the effective noise as measured on the data. We therefore bootstrap pixels in individual exposures to reconstruct an empirical noise estimate of each cube, as described in  \citet{fossati2019}. To minimise small fluctuations in the bootstrap procedure arising from the small number of individual exposures, we then correct the pipeline variance cube with a wavelength-dependent function computed from the bootstraps. 

\subsection{Archival quasar spectroscopy}

For the search and analysis of ALSs along the quasar sightlines, we collect archival high resolution and high $S/N$ spectroscopy from the Ultraviolet and Visual Echelle Spectrograph \citep[UVES;][]{uves2000}, High-Resolution Echelle Spectrometer \citep[HIRES;][]{hires1994} and the Magellan Inamori Kyocera Echelle instruments \citep[MIKE;][]{mike2003}. Where available, we also collect moderate resolution spectroscopy from ESI and X-Shooter, which is useful to extend our analysis in the near infrared (NIR) and to fill the gaps in wavelengths that are occasionally present in the high-resolution data. Details of the 62 individual spectra assembled for the \nqso\ quasars are provided in \citet{lofthouse2020}.

All data have been reduced with optimised and instrument-specific pipelines, as detailed in \citet{lofthouse2020}. Although differences in the adopted algorithms and procedures are present, the reduction for all the instruments follow a similar set of procedures. First, basic calibrations (bias subtraction, flat-fielding, and where applicable dark subtraction) are applied to the 2D frames, and a vacuum wavelength solution is computed. After sky subtraction, the objects are extracted on an order-by-order basis and data are combined, also co-adding multiple exposures if present. If applicable, spectra are then flux-calibrated and continuum normalized. Observations from the same instruments but using different setups are further combined into a single spectrum by resampling the data
onto a common wavelength solution. We do not attempt to combine data across different instruments, which we analyse separately as discussed in the following sections.

\section{Data analysis and catalogue preparation}\label{sec:dataanalysis}

\begin{table}
\centering
\caption{Summary of the absorption line properties identified in the quasar spectra, including the N(\ion{H}{I})-weighted redshift, the total \ion{H}{I} column density, the inferred metallicity, the interquartile range between 16\% and 84\% of the posterior metallicity distributions and the number of LAEs (see text for details). Note that the quoted error on the HI column density is the value input into our photoionisation model while the error on the metallicity is the interquartile range on the model output rather than the traditional 1$\sigma$ error as these distributions are typically not Gaussian.}\label{tab:als}
\begin{tabular}{cccccc}
\hline
Quasar & $z_{\rm HI}$ &  $\log N_{\rm HI}$& $\rm [M/H]$ & IQR  & $\rm N_{LAE}$ \\
       &              & (cm$^{-2}$)          &   \\
\hline
J010619$+$004823 & 4.23536 & $17.2\pm 0.1$  &  -3.18  & $^{+ 0.20} _{-0.18}$ & 1 \\
                 & 4.06780 & $17.1\pm 0.1$  &  <-2.78 & & 1 \\
                 & 3.72887 & $18.05\pm 0.60$ & >-0.70 & & 1 \\
                 & 3.32100 & $19.2\pm 0.1$  &  -2.61 & $^{+ 0.47}_{-0.67}$ & 3 \\
J012403$+$004432 & 3.76618 & $16.8\pm0.1$   &  -2.51 & $^{+ 0.15}_{-0.16}$ & 2 \\ 
                 & 3.07777 & $20.20\pm0.05$ &  -2.09 & $^{+ 0.05}_{-0.05}$ & 2 \\ 
J013340$+$040059 & 4.11223 & $18.2\pm 0.3$  &  -2.97 & $^{+ 0.20}_{-0.16}$ & 5  \\
                 & 3.99526 & $19.8\pm0.1$   &  -2.56 & $^{+ 0.04}_{-0.03}$ & 1 \\
                 & 3.77240 & $20.4\pm0.1$   &  -0.85 & $^{+ 0.09}_{-0.07}$ & 1 \\
                 & 3.69164 & $20.5\pm0.1$   &  -2.56 & $^{+ 0.03}_{-0.02}$ & 2 \\
J013724$-$422417 & 3.66541 & $19.15\pm 0.15$&  -2.88 & $^{+ 0.06}_{-0.06}$ & 2 \\
                 & 3.10030 & $19.75\pm 0.05$&  -1.69 & $^{+ 0.04}_{-0.04}$ & 1 \\
J015741$-$010629 & 3.38595 & $19.2\pm 0.1$  &  -1.56 & $^{+ 0.05}_{-0.05}$ & 8  \\
J020944$+$051713 & 3.98786 & $17.6 \pm 0.2$ &  -2.00 & $^{+ 0.15}_{-0.13}$ & 1 \\
                 & 3.86337 & $20.4\pm0.1$ &    -2.65 & $^{+ 0.06}_{-0.05}$ & 0 \\
                 & 3.70700 & $19.2\pm0.1$   &  -2.96 & $^{+ 0.13} _{-0.19}$ & 1 \\
                 & 3.66592 & $20.45\pm0.05$ &  -2.35 & $^{+ 0.03} _{-0.03}$ & 1 \\
J024401$-$013403 & 4.03100 & $17.40\pm0.05$  &  -3.69 & $^{+ 0.15} _{-0.14}$ & 4 \\ 
                 & 3.96620 & $18.3\pm0.6$   &   -2.70 & $^{+ 0.02} _{-0.02}$ & 3  \\
J033413$-$161205 & 4.23449 & $17.55\pm0.10$ &   -3.03 & $^{+ 0.11} _{-0.10}$ & 0  \\
    		     & 4.14700 & $17.2\pm 0.1$&     -2.71 & $^{+ 0.19} _{-0.21}$ & 3 \\
                 & 3.55655 & $20.7\pm0.1$ &     -1.09 & $^{+ 0.01} _{-0.01}$ & 1 \\
J033900$-$013317 & 3.11540 & $19.60\pm 0.05$&   -1.57 & $^{+ 0.03} _{-0.03}$ & 1 \\
                 & 3.06222 & $20.4 \pm 0.15$ &   -1.05 & $^{+ 0.05} _{-0.04}$ & 1 \\
J094932$+$033531 & 3.92880 & $17.0\pm 0.1$  &    -0.77 & $^{+ 0.11} _{-0.11}$ & 2 \\
                 & 3.31137 & $19.80\pm 0.05$&    >-0.55 & &  1\\
J095852$+$120245 & 3.22320 & $17.3\pm 0.1$  &    -2.97 & $^{+ 0.14} _{-0.13}$ & 2 \\
                 & 3.09622 & $17.2\pm 0.1$  &    <-2.75 & &  1\\
J102009$+$104002 & 3.05410 & $17.10\pm 0.05$&    -2.94 & $^{+ 0.37} _{-0.22}$ & 0 \\ 
J111008$+$024458 & 3.47555 & $17.55\pm 0.05$&    -2.85 & $^{+ 0.15} _{-0.15}$ & 2 \\ 
                 & 3.44172 & $17.4\pm0.1$   &    <-2.75 & & 0 \\
J111113$-$080402 & 3.81137 & $18.2\pm0.3$   &    -3.19 & $^{+ 0.14} _{-0.12}$ & 5 \\ 
                 & 3.60766 & $20.30\pm0.05$ &    -2.11 & $^{+ 0.04} _{-0.04}$ & 5 \\
                 & 3.48185 & $19.95\pm0.05$ &    -2.24 & $^{+ 0.04} _{-0.04}$ & 1 
\end{tabular}
\end{table}

\begin{table}
\centering
\contcaption{
}
\begin{tabular}{cccccc}
\hline
Quasar & $z_{\rm HI}$ &  $\log N_{\rm HI}$& $\rm [M/H]$ & IQR & $\rm N_{LAE}$ \\
       &              & (cm$^{-2}$)          &   \\
\hline
J120917$+$113830 & 3.02224 & $19.2\pm0.1$    &    -0.36 & $^{+ 0.11} _{-0.07}$ &  0\\
J123055$-$113909 & 3.31990 & $16.5\pm0.1$    &    -2.69 & $^{+ 0.59} _{-0.45}$ &  4\\
J124957$-$015928 & 3.52694 & $17.71\pm0.05$  &    -3.61 & $^{+ 0.47} _{-0.54}$ &  5\\
J133254$+$005250 & 3.42111 & $19.1\pm0.1$    &    -2.39 & $^{+ 0.02} _{-0.02}$ &  1\\
                 & 3.08390 & $19.25\pm0.15$   &   -2.49 & $^{+ 0.09} _{-0.06}$ & 1 \\
J142438$+$225600 & 3.38275 & $16.9\pm0.1$    &    -2.86 & $^{+ 0.64} _{-0.65}$ &  1\\
J162116$-$004250 & 3.10494 & $19.7\pm0.1$  &      -1.27 & $^{+ 0.07} _{-0.06}$ &  6\\
J193957$-$100241 & 3.57226 & $18.00\pm 0.05$ &    -2.59 & $^{+ 0.04} _{-0.04}$ &  0\\
J200324$-$325144 & 3.55212 & $18.00\pm 0.05$ &    -1.72 & $^{+ 0.04} _{-0.04}$ &  4\\
                 & 3.18780 & $19.85\pm0.05$  &    -1.59 & $^{+ 0.05} _{-0.05}$ & 3 \\ 
                 & 3.17243 & $19.80\pm0.05$  &    -3.33 & $^{+ 0.03} _{-0.02}$ & 1 \\
J205344$-$354655 & 3.17249 & $18.2\pm0.6$    &    -1.50 & $^{+ 0.14} _{-0.20}$ &  4\\
                 & 3.09420 & $19.05\pm 0.10$ &    -1.07 & $^{+ 0.07} _{-0.06}$ &  4\\
                 & 2.98882 & $20.1\pm0.1$    &    -1.86 & $^{+ 0.10} _{-0.10}$ &  0\\
J221527$-$161133 & 3.89620 & $17.40\pm0.05$  &    -3.73 & $^{+ 0.19} _{-0.17}$ & 2\\ 
                 & 3.70152 & $19.40\pm0.05$  &    -1.89 & $^{+ 0.03} _{-0.03}$ & 0 \\ 
                 & 3.66185 & $20.1\pm0.1$    &    -2.18 & $^{+ 0.07} _{-0.06}$ & 1 \\
                 & 3.46525 & $19.0\pm0.2$   &     -3.60 & $^{+ 0.15} _{-0.15}$ & 3 \\
                 & 3.30557 & $19.0\pm0.1$    &    <-3.42 & & 3 \\
J230301$-$093930 & 3.31186 & $17.80\pm0.15$  &    -3.04 & $^{+ 0.10} _{-0.09}$ & 1 \\
J231543$+$145606 & 3.27319 & $20.30\pm0.05$  &    -2.48 & $^{+ 0.03} _{-0.03}$ & 5 \\ 
                 & 3.13528 & $19.95\pm0.05$  &    -2.61 & $^{+ 0.05} _{-0.05}$ & 3 \\
J233446$-$090812 & 3.22578 & $17.7\pm 0.1$   &    -2.81 & $^{+ 0.10} _{-0.08}$ &  1\\
                 & 3.05723 & $20.40\pm0.05$  &    -1.66 & $^{+ 0.04} _{-0.03}$ & 7 \\ 
J234913$-$371259 & 4.08985 & $18.80\pm0.05$  &    -3.48 & $^{+ 0.11} _{-0.16}$ & 1 \\
                 & 3.69231 & $20.15\pm0.10$  &    -2.55 & $^{+ 0.05} _{-0.05}$ & 2 \\
                 & 3.58203 & $19.1\pm0.1$    &    -4.17 & $^{+ 0.08} _{-0.07}$ & 0
\end{tabular}
\end{table}

\begin{figure}
    \centering
    \includegraphics[width=0.5\textwidth]{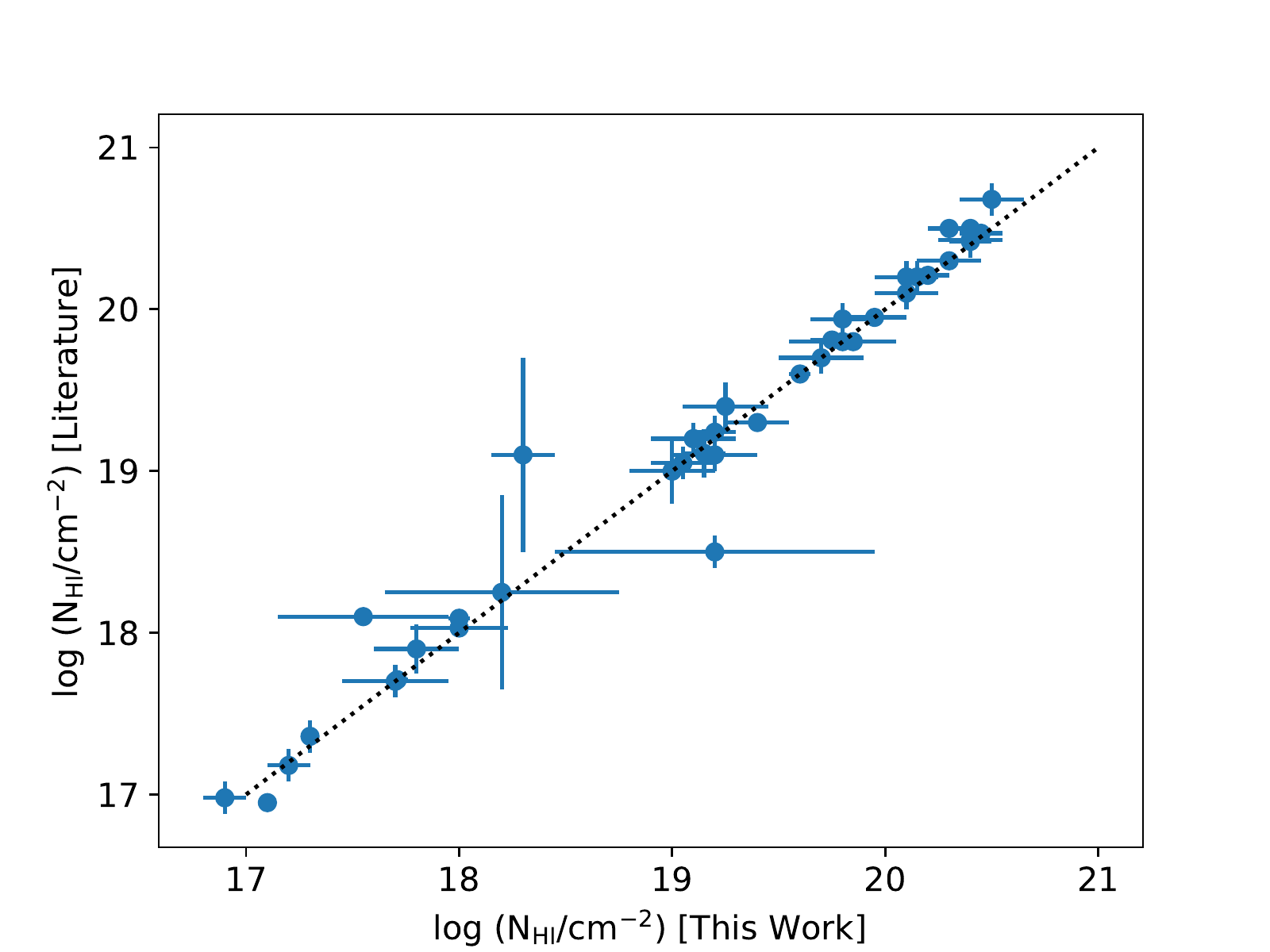}
    \caption{Comparison of the \ion{H}{I} column densities measured in this work with the column densities reported in the literature (see text for details). A tight correlation is found over a wide range of column density.}
    \label{fig:hicfrlit}
\end{figure}

\begin{figure*}
    \centering
    \includegraphics[width=\textwidth, trim =0cm 0.5cm 0cm 0cm,clip]{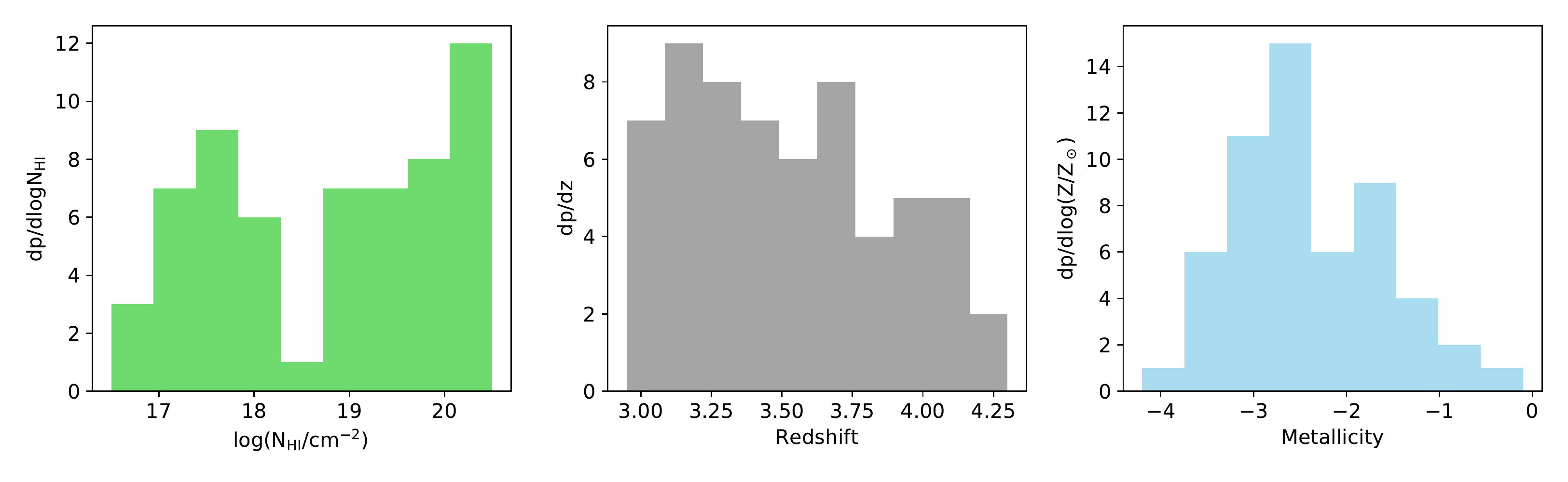}
    \caption{Summary of the statistical properties of the \nlls\ ALSs identified along the \nqso\ quasar sightlines. From left to right, we show the \ion{H}{I} column density, the redshift and the metallicity distributions.}
    \label{fig:alssummary}
\end{figure*}

\subsection{System selection and HI determination}\label{sec:himeasure}

The core objective of our analysis is to investigate the connection between gas and galaxies at $z\approx 3-4$, where MUSE is most effective at uncovering Ly$\alpha$ emission lines (due to wavelength coverage and the absence of sky lines) and where we can search for high-column density absorption line systems ($N_{\rm HI} \gtrsim 10^{16.5}~\rm cm^{-2}$) which are tracers of the denser IGM and of the CGM \citep{faucherGiguere2011,fumagalli2011,vandeVoort2012}.
 
These absorption line systems are selected purely based on their \ion{H}{I} column density, to ensure we assemble a representative sample that is not biased e.g., towards metal rich systems \citep[see e.g.,][]{berg2021}.
 By scanning each spectrum, we identify the highest column density systems ($\log (N_{\rm HI}/\rm cm^{-2}) \gtrsim 19)$ thanks to their distinctive Voigt profile in the Ly$\alpha$ and, when present, Ly$\beta$ transitions. Lower column density systems ($16.5 \lesssim \log (N_{\rm HI}/\rm cm^{-2}) \lesssim 19$) are instead recognised by their Lyman series and the flux decrement at the Lyman limit. Because of this selection, our sample of absorbers does not follow a statistical distribution of column densities (i.e., the $f(N_{\rm HI},z)$), as we can more easily identify systems with $\log (N_{\rm HI}/\rm cm^{-2}) \gtrsim 19$ at all wavelengths, while we are limited to the identification of  $16.5 \lesssim \log (N_{\rm HI}/\rm cm^{-2}) \lesssim 19$ systems for which we have complete coverage of the Lyman limit (in other words, we have a longer search path for systems with damping wings). Moreover, MAGG sightlines are selected to preferentially contain LLSs at $z\gtrsim 3$, which naturally bias the incidence of LLSs above what expected in random sightlines.  

In this work, we expand the initial analysis with the {\sc PYIGM} package\footnote{\url{https://github.com/pyigm/pyigm}} presented in \citet{lofthouse2020}, moving to a custom-built tool that allows the simultaneous analysis of spectra from different instruments. Similarly to what is possible within the {\sc PYIGM} package, this tool allows us to construct full models of the Lyman series plus Lyman limit of every absorber, by varying their column density, Doppler parameter and redshift. Each model is then convolved with the appropriate kernel to account for the resolution of each instrument. Within this tool, each model can then be compared to both the high-resolution and the moderate-resolution spectra that are available for each quasar \citep[see table 2 in][]{lofthouse2020}.
By using the full set of data available, we revisit the search for high-column density absorbers of \citet{lofthouse2020}, and expand the catalogue originally presented to include 9 new absorbers (generally systems for which a refined fit of the Ly$\alpha$ line revealed a column density above the selection threshold), for a total of \nlls\ systems in \nqso\ sightlines.

For all of the systems, in accordance to well-consolidated practice in the literature \citep[e.g.][]{fumagalli2013,fumagalli2014,prochaska2015,berg2021}, we visually fit the continuum level and then systematically vary the redshift and column density to obtain the best model that describes the full Lyman series and the transmission at the Lyman limit. Within this model, we consider as the main component the one that dominates the optical depth at the Lyman limit and within the Lyman series. 
In most cases, a single component is sufficient to account for the absorption in high-order Lyman series lines. We then add a second component when clearly required by the data (that is, when high order lines separate in distinct components with column density of $\gtrsim 1/10$ that of the main component), following the same methodology. For these systems, we quote the total column density over all components and the resulting column-density weighted redshift. 
For the high HI column density systems, we rely primarily on the presence of damping wings in the Ly$\alpha$ absorption profile. After fitting the continuum level around the absorption profile, we adjust the model column density and Doppler parameter to find the best fit to the observed profile.
The redshift and column density are, in most cases, well constrained by the unsaturated high-order lines in the Lyman series, the presence of damped wings in the Ly$\alpha$ lines, and/or the residual transmission at the Lyman limit. For the cases where high-order lines are saturated and there is no evidence of damped wings in the Ly$\alpha$ profile, we quote an \NHI\ range where the lower bound is set by the lack of transmission at the Lyman limit and the upper bound is set by the onset of the damped wings. The value we adopt is the mid-point of this interval and the error expresses the half-width of the interval \citep[see also][]{prochaska2015}. 

A summary of all the systems identified with their column density is presented in Table~\ref{tab:als}. Fig.~\ref{fig:hicfrlit} shows a direct comparison of our \ion{H}{I} determination with the one reported in the literature for the same systems \citep[][]{zafar2013, neeleman2013, riemer2015, prochaska2015, berg2021,lehner2021}. Despite some scatter and a couple of outliers (which arise from the fact that some literature work relied on data at lower-resolution, or in a narrower wavelength range), we find excellent agreement. We note that some of the systems with $\log (N_{\rm HI}/\rm cm^{-2})\lesssim 19$ have large errors. This is a result of the Lyman series lines being saturated limiting the ability to determine an accurate column density. In this case, the error bars represent the range of values allowed by 
the observations. The left and middle panels of Fig.~\ref{fig:alssummary} summarise the \ion{H}{I} column density and the redshift distributions of our ALS sample respectively. 

\begin{table*}
\caption{Column density of metal lines associated with the absorption system at $z=3.69164$ along the J013340$+$040059 sightline. Tables showing the column densities of metals for all the other systems are included in the online material. The columns are as follows: (1) name of the ion; (2) rest frame wavelength of the line in \AA; (3) observed equivalent width of line in \AA; (4) instrument used to obtain the spectrum from which we calculate the values shown in this table; and (4) the adopted column density of the ion in log(N/cm$^{-2}$) from the analysis of all transitions as described in the text. }
\label{tab:metalfits}
\begin{tabular}{cccccc}
\hline \hline Ion & Wave (\AA) & EW (\AA) & Instrument & log($N/~\rm cm^{-2}$) \\
\hline &  &  &  &  & \vspace{-0.3cm} \\
CIV & 1548 & 0.413$\pm$0.008 & UVES & 13.56$\pm$0.07 \\
 & 1550 & 0.255$\pm$0.005 & UVES &  \\
AlII & 1670 & 0.437$\pm$0.007 & UVES & 12.47$\pm$0.02 \\
AlIII & 1854 & - & X-shooter& <12.1 \\
MgII & 2796 & 2.2$\pm$0.02 & X-shooter &  >13.26 \\
 & 2803 & 1.83$\pm$0.02 & X-shooter & >13.26 \\
FeII & 1608 & 0.171$\pm$0.003 & UVES & 13.3$\pm$0.1 \\
 & 2344 & 0.47$\pm$0.02 & X-shooter&  \\
 & 2374 & 0.19$\pm$0.01 & X-shooter&  \\
 & 2382 & 1.06$\pm$0.02 & X-shooter&  \\
 & 2600 & 1.0$\pm$0.02 & X-shooter& \\
SiII & 1526 & 1.62$\pm$0.01 & UVES & 14.3$\pm$0.3 \\
 & 1808 & 0.0541$\pm$0.0007 & UVES &  \\
SiIV & 1393 & 0.289$\pm$0.007 & X-shooter  & 12.88$\pm$0.05 \\
 & 1402 & 0.153$\pm$0.003 & X-shooter &  \\
\hline &  &  &  &  &  \\
\end{tabular}
\end{table*}

\begin{figure*}
\includegraphics[width=0.95\textwidth, trim=1cm 1cm 1cm 1cm, clip]{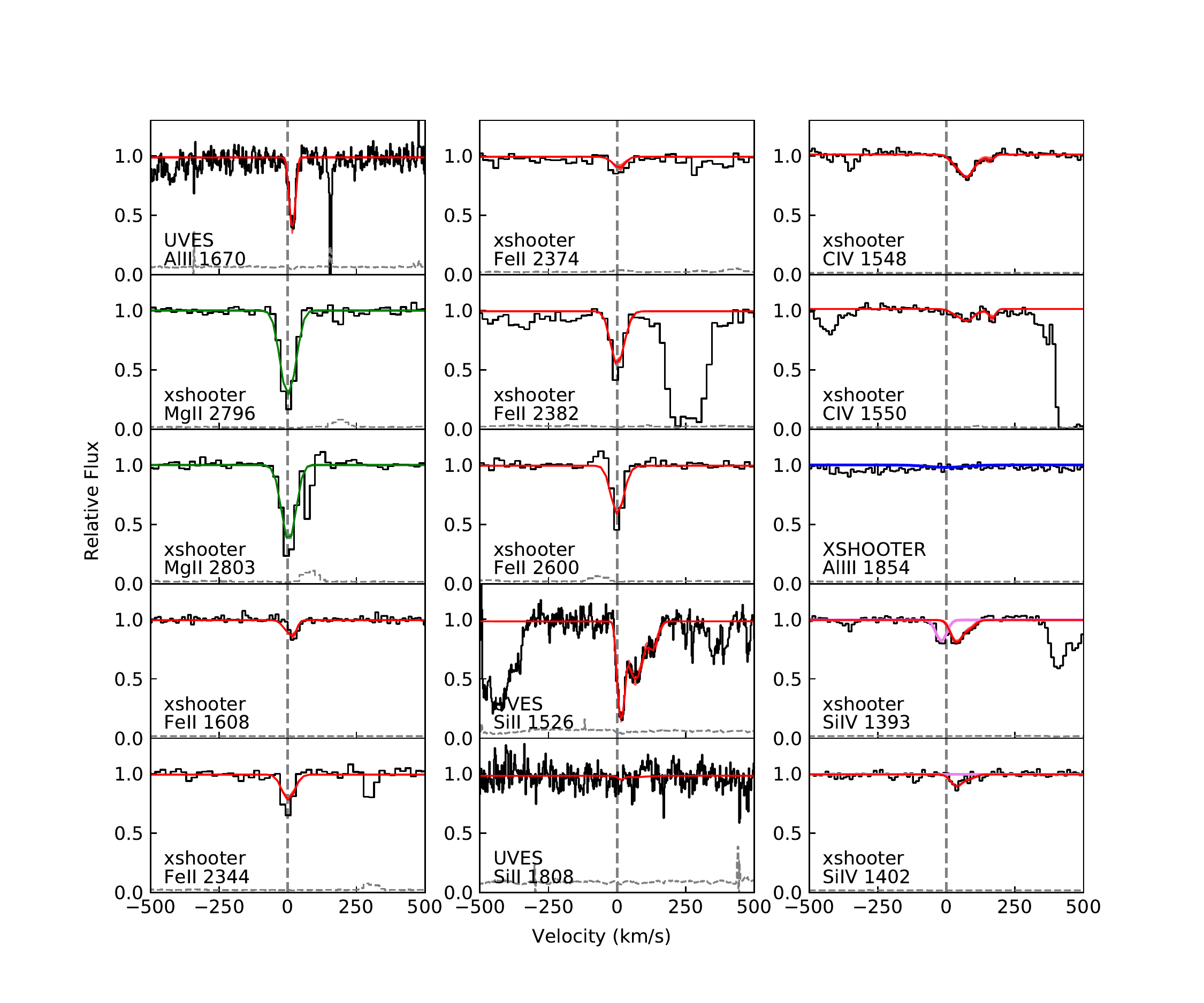} 
\caption{Example of fits to metal lines for a single system (J013340$+$040059, $z=3.69164$). Models for all the remaining systems are provided in the online material. Fits from {\sc MC-ALF} are shown in red for unsaturated line, in green for lines that are saturated and in blue for upper limits. Filler components, which arise from absorption due to unrelated systems that overlap with the metal associated with the absorption system, are shown in pink. The error spectra of the data are shown in each panel as the horizontal grey dashed line. The vertical dashed grey line marks the systemic velocity of the HI absorption.} 
\label{fig:metalfits}
\end{figure*}

\subsection{Models of individual ions and metallicity estimates}\label{sec:metalmeasure}

To identify the presence of metal absorption lines associated with each ALS we use the {\sc Linetools} package\footnote{https://github.com/linetools/linetools}. 
Given a quasar spectrum and the redshift of the absorber, the routine {\sc lt\_xabssys} simultaneously displays $\pm$1000~\kms~ windows around the expected position of potential metal lines (e.g., \ion{Si}{II}, \ion{Mg}{II}, \ion{Fe}{II}, \ion{Al}{II}, \ion{Al}{III}, \ion{C}{IV}, \ion{Si}{IV}). From this, we select those which show detected absorption and determine the wavelength range that needs to be fit in the following analysis. We exclude lines which fall within the \lya\ forest as the heavy blending with the hydrogen lines makes it difficult to identify the underlying metal components. 
  
To fit the identified metal lines we utilise the Monte Carlo Absorption Line fitter code ({\sc MC-ALF}, Longobardi et al., in prep; see also \citealt{fossati2019}). 
For all the transitions in common for a given ion, in each absorber, {\sc MC-ALF} uses a Bayesian statistical method to determine the posterior probability distribution of the models that describe the data. 
Specifically, the code uses the wavelength range specified using {\sc lt\_xabssys} as a prior on the redshift interval to consider. Within a given range for the number of components (based on visual inspection) that can be used to describe the absorption profile, {\sc MC-ALF} returns the posterior of each model computed for a  progressively increasing number of components together with the corresponding Akaike Information Criterion (AIC). We then select fits with an AIC within 5 of the lowest value and adopt the best model from this pool following visual inspection. Where the models cannot provide a good description of the data due to blending with lines not associated with the absorber under consideration, {\sc MC-ALF} allows us to add in filler Voigt components to account for these additional sources of optical depth. These lines (shown in pink in Fig.~\ref{fig:metalfits}) contribute to the overall fit (shown in red in Fig.~\ref{fig:metalfits}) to the observed spectrum but are not included in the calculation of the metal column density as they arise from unassociated systems at different redshifts to the absorption system we are studying. 
The code further allows the continuum to vary by $\pm 2$ per cent, to account for small offsets in the applied normalization. We also use a flat prior on the Doppler parameter between 1 and 35 \kms~ and a flat prior on the column density in the range $12 \le \log (N/\rm cm^{-2}) \le 16$. 
While the redshift and Doppler parameter of components are not constrained to be the same for different low/high ions, we tend to find that the different ions in the same system but with similar ionisation states have  similar profiles with comparable number of components. 

For each component of the adopted model, {\sc MC-ALF} returns a column density. In cases where multiple components are required, the total column density is calculated as the sum of all these components. For ions where all of the available transitions are saturated as determined from visual inspection of the line profiles, it is not possible to determine an accurate measurement of the column density with this approach and we instead use the apparent optical depth method (AODM; \citealt{savage1991}) to measure lower limits. Specifically, we adopt the most stringent lower limit on the column density by performing the AODM analysis of the weakest transition that is not blended. In cases where there is no detectable line within $\pm$1000~\kms~ of the absorber redshift, we instead determine a $2\sigma$ upper limit on the column density using the AODM. For this calculation we use a velocity window of $\pm$100~\kms~  centred on the redshift of the absorber as this is typical of the $\Delta v_{90}$ (i.e. the velocity range which encompasses 90 per cent of the optical depth of the line) for our systems. 
As an example, the column densities for the different metals along with our fits to the spectra are shown in Table~\ref{tab:metalfits} and Fig.~\ref{fig:metalfits}, respectively, for the system at $z=3.69164$ in the J013340$+$040059 sightline. Tables and profile models for all the remaining systems are provided in the online material.

\begin{figure}
    \centering
    \includegraphics[width=0.5\textwidth]{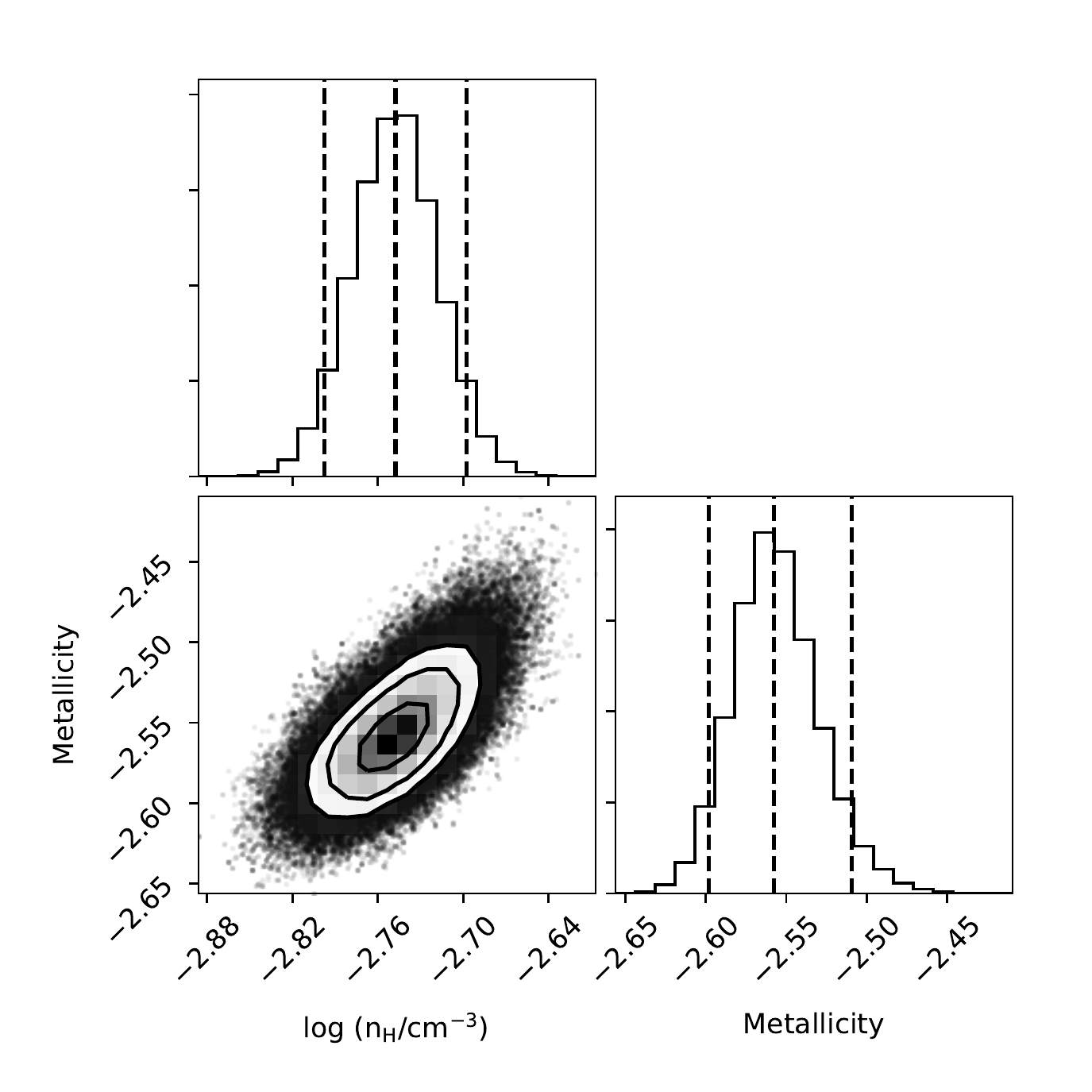}
    \caption{Example of the results from the photoionisation modelling to determine the metallicity of the absorption system. Here we show the posterior probabilities for the metallicity and the density for the system at $z=3.69164$ in J013340$+$04005. Results for all the absorption systems in our sample are included in the online only material.}
    \label{fig:corner}
\end{figure}

\subsubsection{Photoionisation modelling}\label{subsubsec:photmodel}

The gas in the lower column density absorbers in our sample (LLSs and sub-DLAs) is not fully neutral. As a result, the measurements acquired in the previous section cannot be converted trivially into a value for metallicity as in DLAs. In these cases, we must apply ionization corrections when converting the observed column densities to metallicities. 
To do this, we follow the technique described in \citet{fumagalli2016} (see also \citealt{crighton2015}). We include a brief overview here and refer the reader to \citet{fumagalli2016} for a more detailed description. 

Using {\sc cloudy} \citep[][c17.01]{ferland2013}, we construct a grid of models using the assumptions of the minimal model in \citet{fumagalli2016}. Briefly, this consists of a 1D single-phase gas slab with a constant density (n$_{\rm H}$), which is illuminated on one side by a redshift-dependent UV background radiation \citep{haardt2012}.
The metallicity and density are given flat priors and allowed to vary within the ranges $-4.0 <  \log~Z/Z_\odot < -1.0$ and $-4.5 <  \log~(n_{\rm H}/~\rm cm^{-3}) < 0.0$, respectively. The redshift and HI column density are defined by Gaussian priors based on the observed values and errors except for where the LLSs have saturated Lyman series lines and it is not possible to accurately determine a value of the column density (see the systems in Table ~\ref{tab:als} and Fig.~\ref{fig:hicfrlit} with larger errors). For these cases we use a top-hat function for the prior where the limits are defined by the minimum and maximum values permissible by the observations. 

We compare the observed column densities and limits for each absorber to the grid of models using {\sc emcee} \citep{foreman-mackey2013}. This provides posterior probability distribution functions (PDFs) for the metallicity and density of each of our \nlls\ absorbers. An example for the system shown in the previous section (at $z=3.69164$ along the J013340$+$040059 sightline) is provided in Fig.~\ref{fig:corner} with all the figures for the remaining systems provided as online-only material. The distribution of metallicity for the entire sample is further shown in the right panel of Fig.~\ref{fig:alssummary} and listed in Table~\ref{tab:als}. 
We note that the errors shown in this table for the \ion{H}{I} column density are the values that are input to our model. On the other hand, the uncertainty on the metallicity are an output of the model. For these we use the 16 per cent to 84 per cent interquartile range (IQR) as the posterior metallicity distributions are typically not Gaussian.
Therefore, while in some cases it seems that the metallicity error is smaller than for the \ion{H}{I} column density from which it is derived, we confirm that this IQR is larger than the IQR of the corresponding output \ion{H}{I} column density.

As 16 of the quasars in the MAGG sample are also in the HD-LLS sample studied by \citet{prochaska2015}, we can test for consistency both in the determination of the ion column densities and in the inferred metallicity. Regarding the ion column densities, we find good agreement with the values reported in \citet{prochaska2015}. Of the 22 absorption systems that we are able to compare, the MAGG metallicities of 19 agree within $<2\sigma$ with the published values. The three other systems have some of the smallest errors in our sample resulting from strong constraints on both high and low ionisation absorption lines. We note that both studies resort to the same technique for inferring the metallicity, but our work relies on a new set of {\sc cloudy} models. 

Before proceeding with our analysis, we re-iterate some caveats of this modelling which have been already discussed in the literature \citep{fumagalli2016b, berg2021}. First, and foremost, this modelling treats the gas slab as a single phase, which is clearly not an ideal approximation given the multiphase nature of the CGM. While more advanced modelling has been introduced in the literature \citep[e.g.][]{buie2020,haislmaier2021}, this approach has yet to scale to the wide range of redshift, metallicity, and density of interest to this work. We therefore resort to this simpler approach, keeping in mind that we are modelling in most cases a cold/cool metal phase (traced by low ions), which we attribute to all the neutral hydrogen measured in absorption. The inclusion of \ion{Si}{IV} and \ion{C}{IV} deserves a special note, as these ions straddle the separation between a cool and photoionized phase and a warmer, possibly collisionally ionized, one. As shown in appendix 1 of \citet{fumagalli2016} (see their figure A3), the inclusion of an additional \ion{C}{IV} contribution produces an appreciable difference to the model results. At the current precision, however, this difference is comparable to $\approx 2\sigma$ of the recovered posteriors. We therefore choose to include these higher-ionization ions, mindful that an (unknown) contribution from a second phase is likely present.     

As a second approximation, we do not apply a full model for dust depletion. While clear evidence of a small but non-negligible contribution of dust has emerged in the literature of absorption line systems, especially for DLAs \citep[e.g.][]{deCia2016,murphy2016}, we currently have no detailed model to break the degeneracy between the intrinsic abundance pattern and the effect of dust depletion for LLSs. As shown in figure 9 of \citet{fumagalli2016}, models with and without dust depletion yield similar results for LLSs, with more scatter for $\log Z/Z_\odot > -1.5$ (as expected). Despite the good approximation, strictly speaking, the inferred metallicity is related only to the gas phase \citep[see e.g.][]{berg2021}. 

Finally, we note that our operational definition of LLSs assumes that the detected metal abundance across the entire velocity range can be associated with the entire (or most dominant, see above) \ion{H}{I} component. This approach smooths the fluctuations in the metallicity across components that are known to be present in these absorbers \citep[e.g.][]{prochter2010,fox2013}, but is meant to capture the mean metallicity of the entire absorber and it is the only feasible measurement when lacking coverage of high-order Lyman series lines \citep{kacprzak2019}. 

\begin{figure}
    \centering
    \includegraphics[width=0.5\textwidth]{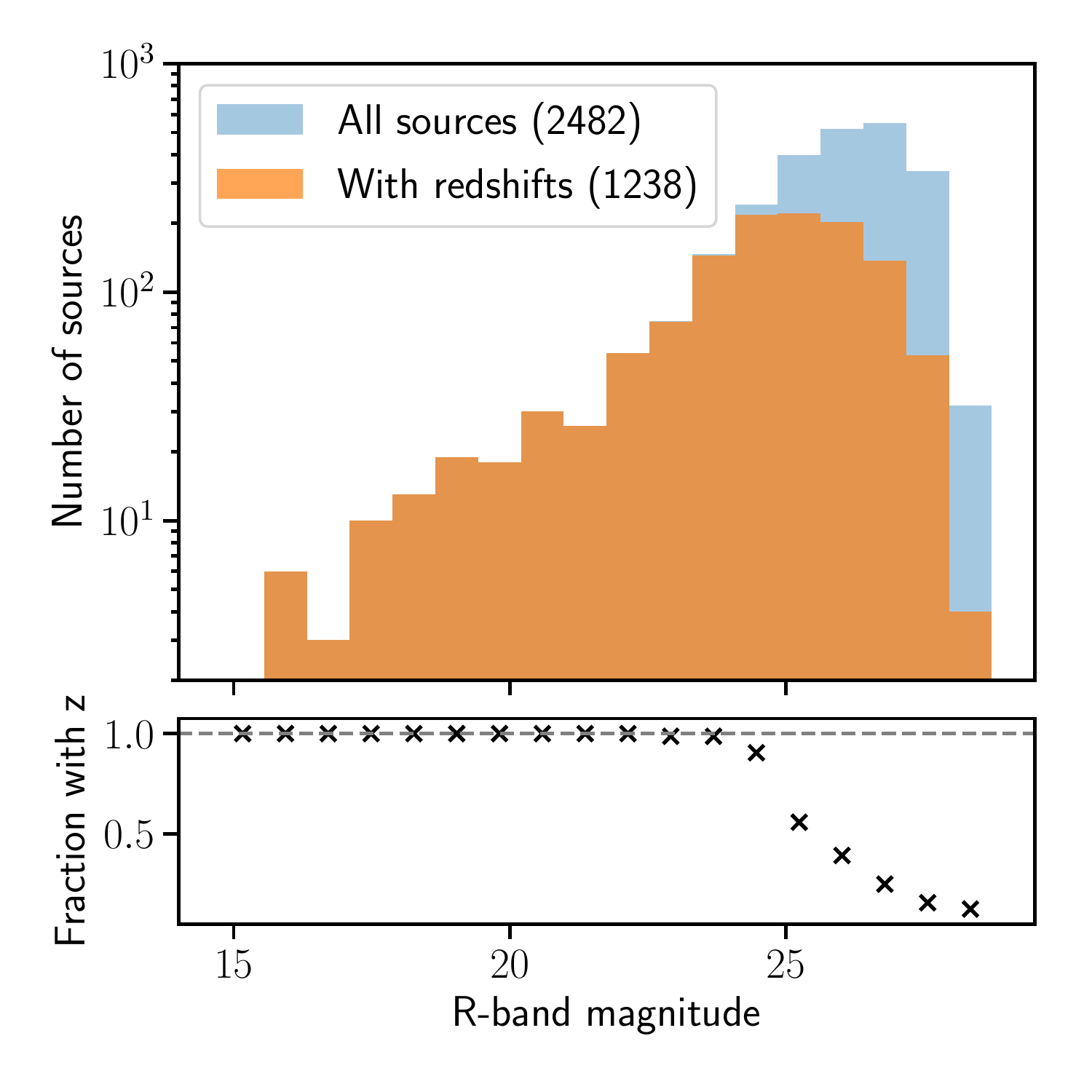}
    \caption{
    Number of detected sources in all 28 MAGG fields as a function of R-band magnitude. We are able to obtain reliable redshifts for 1238 sources for a spectroscopic survey that is  $\approx 90$ per cent complete down to $\approx 24.85$~mag.}
    \label{fig:cont_mag}
\end{figure}

\begin{figure}
    \centering
    \includegraphics[width=0.50\textwidth]{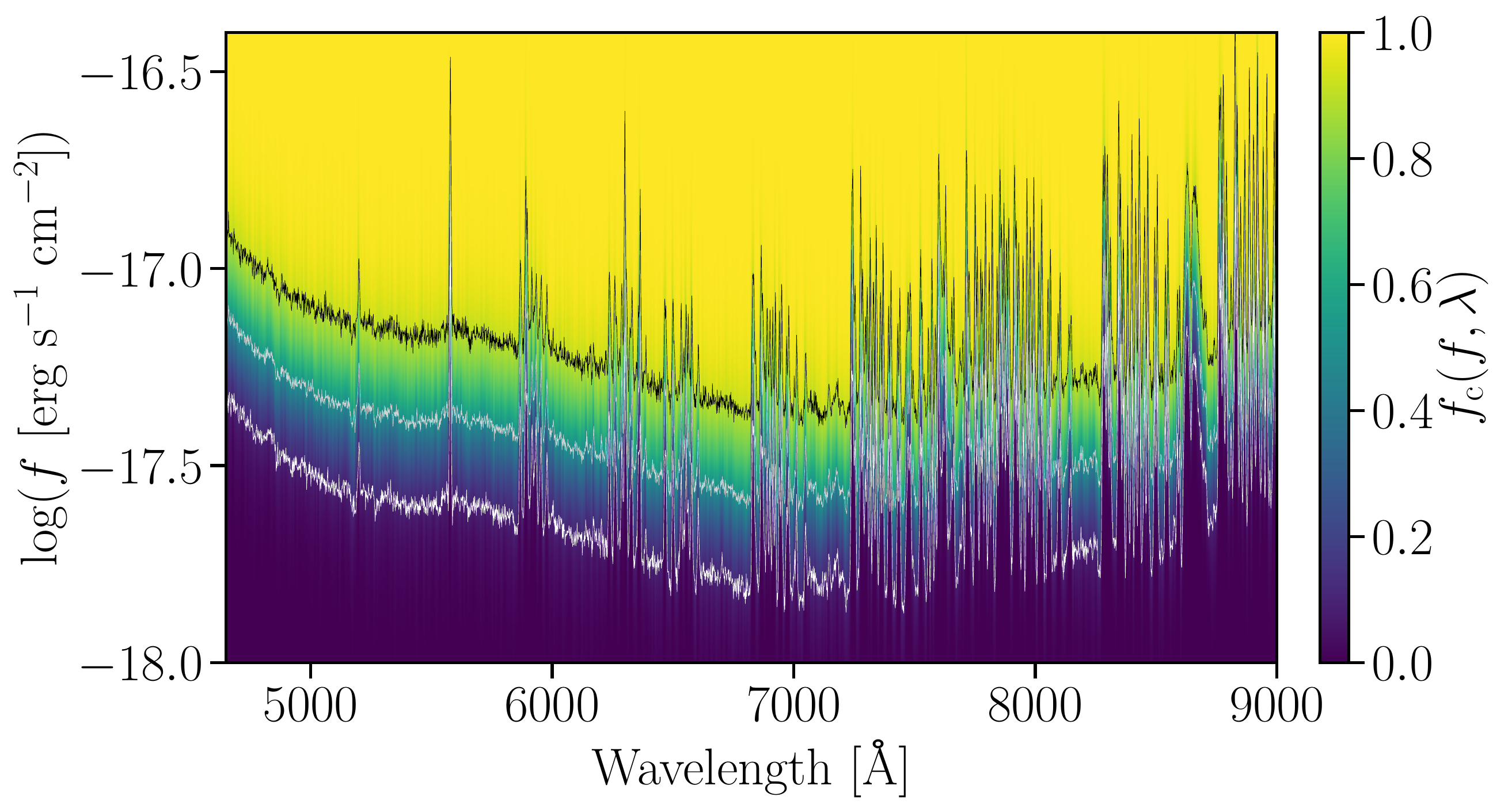}
    \caption{Recovery fraction of real LAEs as a function of wavelengths and flux in the MAGG sample. The black, grey, and white lines show, respectively, the 90, 50, and 10 per cent completeness limit.}
    \label{fig:flxcomp}
\end{figure}

\subsection{Detection of galaxies in MUSE data}\label{subsec:detectgals}

For each field, we search for any galaxies associated with one of the absorbers previously identified in the quasar spectra. We use two methods to detect galaxies. Firstly, we search for galaxies detected via their continuum emission in white light images constructed from the MUSE datacube. Secondly, we search for line emitters which are too faint to be identified in the continuum images. Specifically in this work, we look for \lya~ emitters (LAEs) at the redshift of the absorbers. See \citet{dutta2020} for a study using similar methods to detect \oii~ emitters in the MAGG data at lower redshifts, namely $0.8 < z < 1.5$. We briefly describe the two methods used below and refer to \citet{lofthouse2020} for a more detailed description of our methodology. 

For the continuum source search, we use SE{\sc xtractor} \citep{bertin1996} on the white light images created from the MUSE datacubes to identify all sources within the field of view (FOV) using the same parameters as in \citet{lofthouse2020}. Across all 28 fields, this results in a catalogue of 2482 sources.
1D spectra of the galaxies are extracted from the MUSE datacubes using the 2D segmentation map from SE{\sc xtractor}. We use these spectra and the {\sc MARZ} tool \citep{hinton2016}, specifically the customized branch by M. Fossati\footnote{matteofox.github.io/Marz}, to measure redshifts for each of these sources. 

Fig.~\ref{fig:cont_mag} shows the number of detected sources as a function of $R$-band magnitude. 
Of the 2482 sources identified by SE{\sc xtractor} we are able to determine reliable redshifts for 1238 sources with $\approx 90$ per cent completeness down to $\approx 24.85$~mag.  These sources were assigned a confidence flag in {\sc MARZ} of either 6 indicating a star or 3 and 4 indicating they showed multiple features which enabled a reliable estimate of the galaxy redshift according to the classification scheme followed in \citet{bielby2019} and \citet{lofthouse2020}.

In addition to identifying continuum-detected sources, we search for LAEs associated with each absorber.
In this work, we broadly define LAEs as high-redshift galaxies ($z\gtrsim 3$) displaying a significant Ly$\alpha$ emission, without imposing cuts in equivalent widths as traditionally found in the literature based on narrow band imaging. This additional search allows us to detect fainter galaxies with bright emission lines, which would otherwise be missed in the continuum search.
To detect the LAEs, we first run the {\sc CubEx} 3D detection algorithm on the full MUSE datacube, restricting the selection of sources to those with integrated $S/N>5$. Here the $S/N$ is based on the effective noise, as described in detail in section 3.2.5 of \citet{lofthouse2020}. We also refer the reader to this paper for the full description of the parameters used to reduce the detection of nonphysical sources, e.g. cosmic rays. 

This produces a catalogue of potential sources across the full wavelength range of MUSE. For each absorber, we restrict the catalogue to a velocity window of $\pm1000$\kms~ and for each source we visually inspect its spectrum, 3D segmentation map and images (constructed from mean, median and independent half cubes). This visual identification was performed by at least three authors (EKL,MFo,RD) and sources are only included where there is consensus that it is a real source. During the identification, we also classify sources into the following confidence groups (which expands the original classification presented by \citealt{lofthouse2020}):
\begin{itemize}
\item[--] 1: Highest confidence LAEs, showing e.g. bright and asymmetric lines and/or other absorption lines, with integrated $S/N > 7$;
\item[--] 1.5: Considered likely to be Ly$\alpha$ but less certain than confidence 1 sources, with integrated $S/N > 7$;
\item[--] 2: Likely to be Ly$\alpha$ but with $5 < S/N < 7$;
\item[--] 2.5: Tentative identification with $5 <  S/N < 7$, but generally more noisy as visible on the images or affected by, e.g., sky residuals;
\item[--] 3: Poor quality sources, e.g. overlapping with the edge of MUSE cube and for which reliable photometry is not possible. 
\end{itemize}
We exclude sources identified as confidence 3 in the following analysis.

In addition to the flux above a given $S/N$ per voxel, directly provided by {\sc CubEx}, we also estimate the total flux using a curve of growth method following \citet{fossati2021} and, originally, \citet{marino2018}. In brief, we create a narrow-band image by collapsing the datacube within 15\AA~ of the LAE redshift. We then compute the flux in circular apertures, centred on the {\sc CubEx} coordinates of the LAEs, in steps of 0.2~arcsec with any nearby sources masked. The radius of the circular aperture is increased until the flux is no longer increasing by at least 2.5 per cent of the value at the previous radius. The flux at this final radius is taken to be the total Ly$\alpha$ flux. 
The median radius used is 7 pixels (1.4~arcsec) with a range between 3 pixels and up to 17 pixels for the largest sources.
All LAEs and fluxes are visually inspected to ensure they are correctly centred and any unassociated sources of flux are masked. The Ly$\alpha$ flux for each source has been corrected for Milky Way extinction using extinction values for each of the MAGG fields derived  from the dust map by \citet{schlegel1998} and the extinction curve from \citet{fitzpatrick1999}.

To establish the source redshifts we proceed according to the line shape. Specifically, LAEs fall into three categories. Firstly, the emitters which do not have any significant asymmetry in their profile. In this case we fit the line with a Gaussian profile and take the peak as the redshift of the LAE. In the second case, we have emitters with asymmetric line profiles, which we visually 
inspect to obtain the wavelength of the peak and then use this to determine the galaxy's redshift. Finally, a small number of LAEs (13) showed double-peaked lines with peak separations of around a few 100\kms. For these cases we take the wavelength of the red peak as the redshift of the galaxy, for consistency with what is measured for the single peak lines which are assumed to be the most redshifted component. 
See \citet{fossati2021} for further details. 

In some cases, emitters that are spatially close can be blended and are identified by {\sc CubEx} as a single source. This is a particular problem in the regions that fall within the nebula around the quasar, as discussed in more detail in \citet{fossati2021}. Briefly, we separate these sources by generating a composite image of the Ly$\alpha$ emission on which we run {\sc SExtractor}. A {\tt DEBLEND\_CONT} of 0.05 is used to balance the requirement to efficiently separate sources without over-shredding the signal into a high number of sources which are not real. To further prevent over-separation of sources, we require each one to have an integrated $S/N>5$ and to be composed of at least 9 pixels. For each source that satisfies this, we generate a 3D segmentation map from which we extract the spectrum. 

As a last step, we measure the completeness of our LAE catalogue using the procedure described by \citet{fossati2021}, which consists of injecting LAEs from the MUSE Ultra Deep Survey \citep[MUDF;][]{fossati2019} which is $>5\times$ deeper than the MAGG observations, into the cubes. The recovery fraction as a function of flux and wavelengths for the MAGG cubes is shown in Fig.~\ref{fig:flxcomp}. Our LAE survey is 90 per cent complete for line fluxes $\approx 10^{-17.2}~\rm erg~s^{-1}~cm^{-2}$, and $\approx 50$ per cent complete for fluxes $\approx 10^{-17.4}~\rm erg~s^{-1}~cm^{-2}$. The completeness for mock point sources is instead a factor $\approx 3$ deeper, as shown by \citet{fossati2021} and, originally, by \citet{herenz2019}. 
We note that only three sources are identified around absorbers in the continuum search, but these sources are subsequently also identified in the emitter search. In addition,  four continuum sources 
that were detected by SE{\sc xtractor} but whose redshifts were not reliably determined in {\sc MARZ}
are subsequently confirmed at $z>3$ as a result of the emitter search. Therefore, in the following analysis we will generally refer to LAEs as the galaxy population that is the subject of our study.
The fact that our search yields a large and complete sample of LAEs confirms that MUSE is highly effective for the detection of emission line galaxies. This means, however, that our sample is incomplete with respect to dusty galaxies, passive galaxies, and/or active galaxies (such as some LBGs) showing only Ly$\alpha$ in absorption.

\section{The environment of optically-thick gas clouds}\label{sec:env}

With the catalogues of galaxies and absorbers in hand, in this section we first explore the general properties of the environment of the optically-thick \ion{H}{I} systems, before turning to a detailed investigation of the correlations between physical parameters of the absorbers and galaxies. 

\begin{figure}
    \centering
    \includegraphics[width=0.5\textwidth]{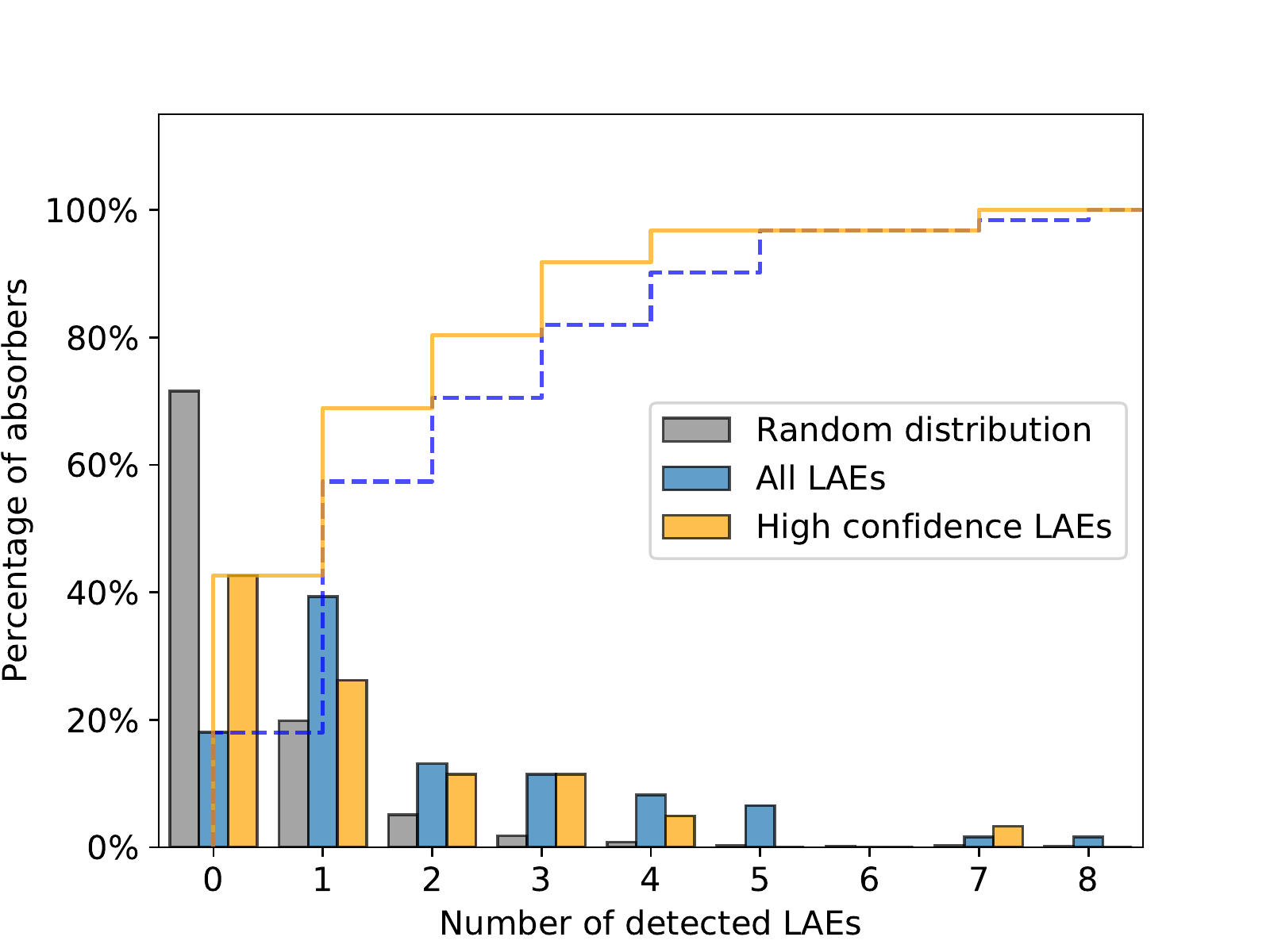}\label{emitterHist}
    \caption{Number of LAEs detected within the MUSE FOV and at a velocity separation of $<1000$~\kms\ around each absorber. In blue we show all detected LAEs while in orange we include only those that were classified as high confidence (confidence 1 and 1.5) which have $\rm S/N > 7$.
    The solid orange and dashed blue lines show the cumulative number of absorbers for the all the galaxies and the high confidence sample respectively. For comparison, we also show the number of $\rm S/N > 7$ LAEs expected from a random distribution in grey. }
    \label{fig:nEmitter_hist}
\end{figure}

\begin{figure*}
    \centering
    \includegraphics[width=\textwidth]{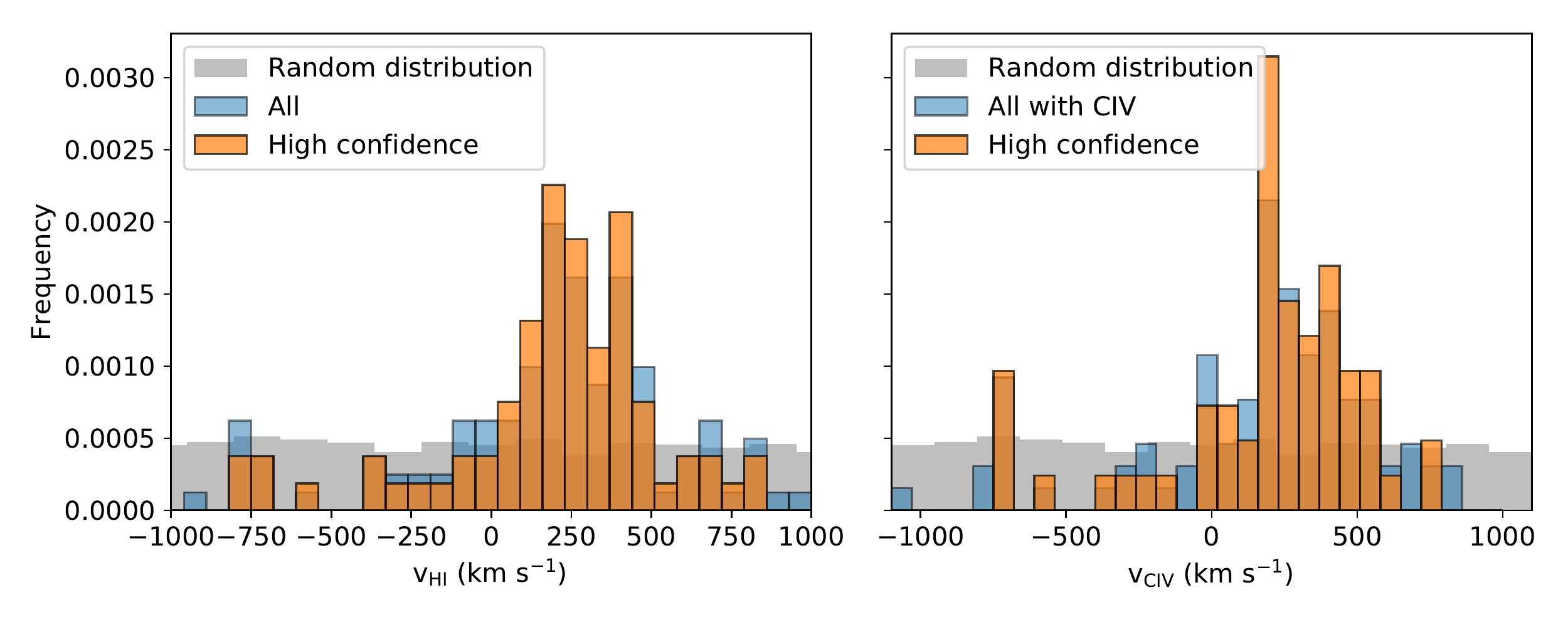}\caption{Velocity separation between the absorbers and the LAEs detected within the MUSE FOV. In blue we show all detected LAEs while in orange we include only those that were classified as high confidence (confidence 1 and 1.5). For comparison, we also show the results obtained from a random distribution of galaxies (in grey). The left figure uses the \HI\ to determine the redshift of the absorber while the right figure shows the velocity separation when using the redshift determined from the \ion{C}{IV} absorption lines. For all histograms, there is a peak in the velocity distribution at $\sim$250~\kms.}
    \label{fig:vel_sep}
\end{figure*}

\subsection{Detection rate of galaxies around absorbers}\label{subsec:DR}

The emission line selection identified \nem~LAEs across all confidence groups within $\pm1000$~\kms\ of an absorber. The catalogue of these galaxies is included in the next paper in the MAGG series (Galbiati et al., in prep). Of these 55 (44 per cent) are flagged as confidence 1 (i.e. the highest confidence group), as shown in Fig.~\ref{fig:nEmitter_hist}. A further 21 systems are assigned confidence 1.5, 36 systems are confidence 2, 3 are confidence 2.5 and only 12 in the lowest confidence group (confidence 3 which are not used in the analysis performed in this study). 
The majority of the absorption systems have only one or two detected galaxies with a decreasing frequency up to the highest number of detected LAEs of 8. 
There are only 11 absorbers (out of the full sample of 61) with no detected galaxies within the MUSE field of view and within the specified velocity range. 
This corresponds to a detection rate of 82$\pm$16 per cent when considering all LAEs and 57$\pm$12 per cent  when considering only the highest confidence (confidence 1 and 1.5) LAEs. 
To check whether the presence of sky lines has impacted the non-detection of galaxies, we investigate where these 11 absorbers with no associated LAEs fall at wavelengths of low transmission as seen in Fig.~\ref{fig:flxcomp}. Only 3 of these absorbers are found at redshifts close to significant sky lines. The remaining 8 absorbers are all in regions with good sky transmission, and so we would have detected any galaxies above the typical flux limits quoted if they were present. Therefore we conclude that the lack of detections is an intrinsic properties and not a selection effect. 

For comparison, we also show the results expected from a random distribution of LAEs. To calculate this we randomly choose a field and a redshift between 3 and 4.5. Given this position, we search for any LAEs within the MUSE FOV and $\pm$1000~\kms at $S/N >7$. This process is repeated 1000 times and the resulting histogram is shown in grey. We find no associated LAEs in $72\pm4$ per cent of the cases, a single galaxy in 18$\pm$1 per cent and a small number of multiple associations (less than 10 per cent). This is much lower than the percentage of multiple associations found around the absorbers (42 per cent).

While it has previously been common to consider a one-to-one association between absorbers and galaxies, the large FOV of MUSE, combined with its sensitivity, enables multiple galaxy detections, as seen here, of which many are plausibly associated to the absorbers. 
Indeed, such multiple detections are seen in other MUSE fields both at high redshift and low redshift \citep[e.g.][]{fumagalli2016,mackenzie2019,dutta2020,peroux2019}. Multiple detections have also been seen around high-redshift DLAs using the Hubble Space Telescope \citep{moller1998} and in lower redshift studies focusing on \mgii\ absorption systems \citep[e.g.][]{bergeron1992,whiting2006,kacprzak2010}.

\begin{figure}
    \centering
    \includegraphics[width=0.5\textwidth]{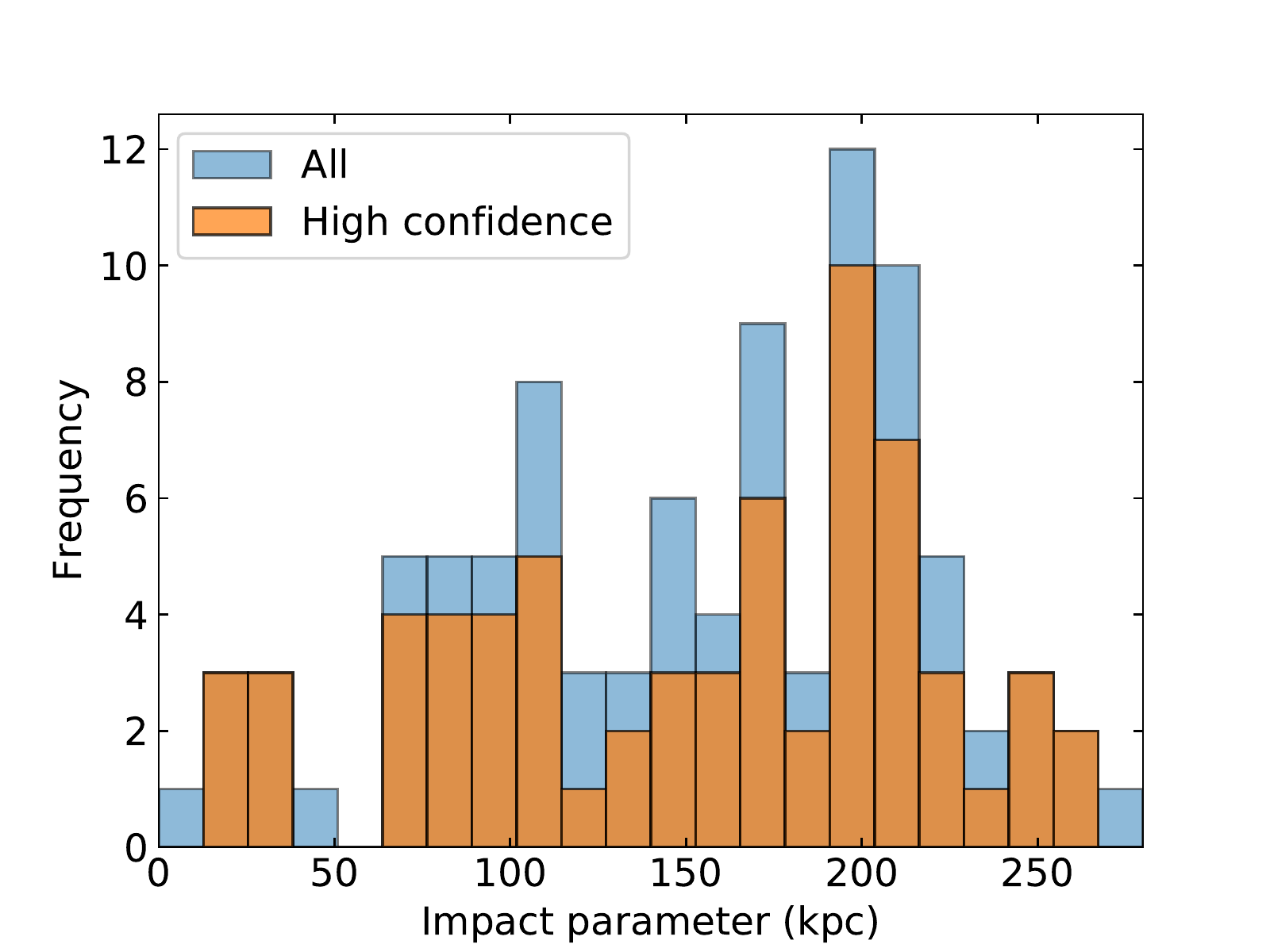}
    \caption{The distribution of impact parameters for all the identified LAEs within 1000~\kms of an absorber. In blue we show all detected LAEs while in orange we include only those that were classified as high confidence (confidence 1 and 1.5). }
    \label{fig:IPhist}
\end{figure}

\begin{figure*}
    \centering
    \begin{tabular}{cc}
    \includegraphics[width=0.47\textwidth, trim= 0cm 0cm 0.5cm 0cm,clip]{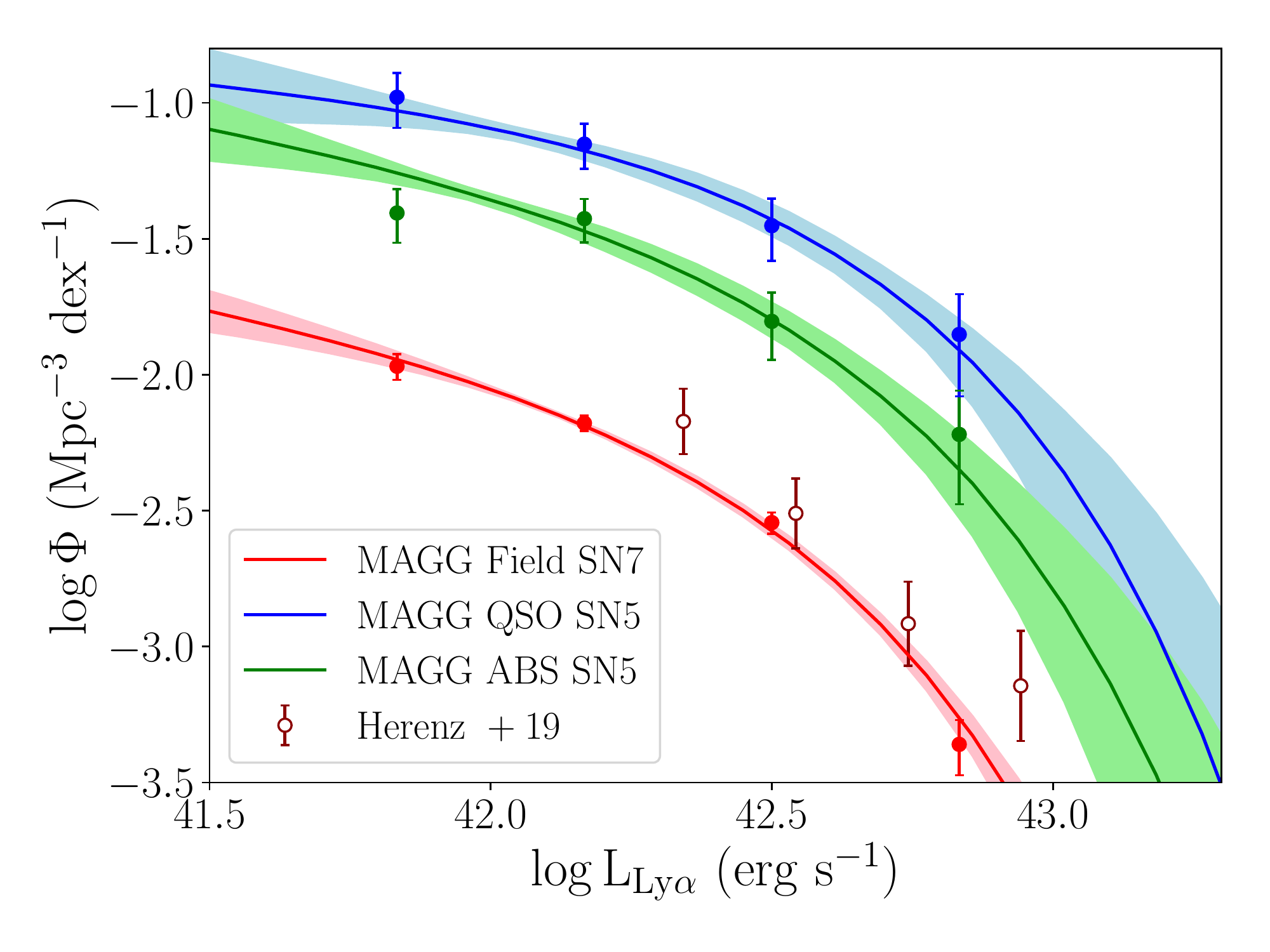} &
    \includegraphics[width=0.47\textwidth, trim= 0cm 0cm 0.5cm 0cm,clip]{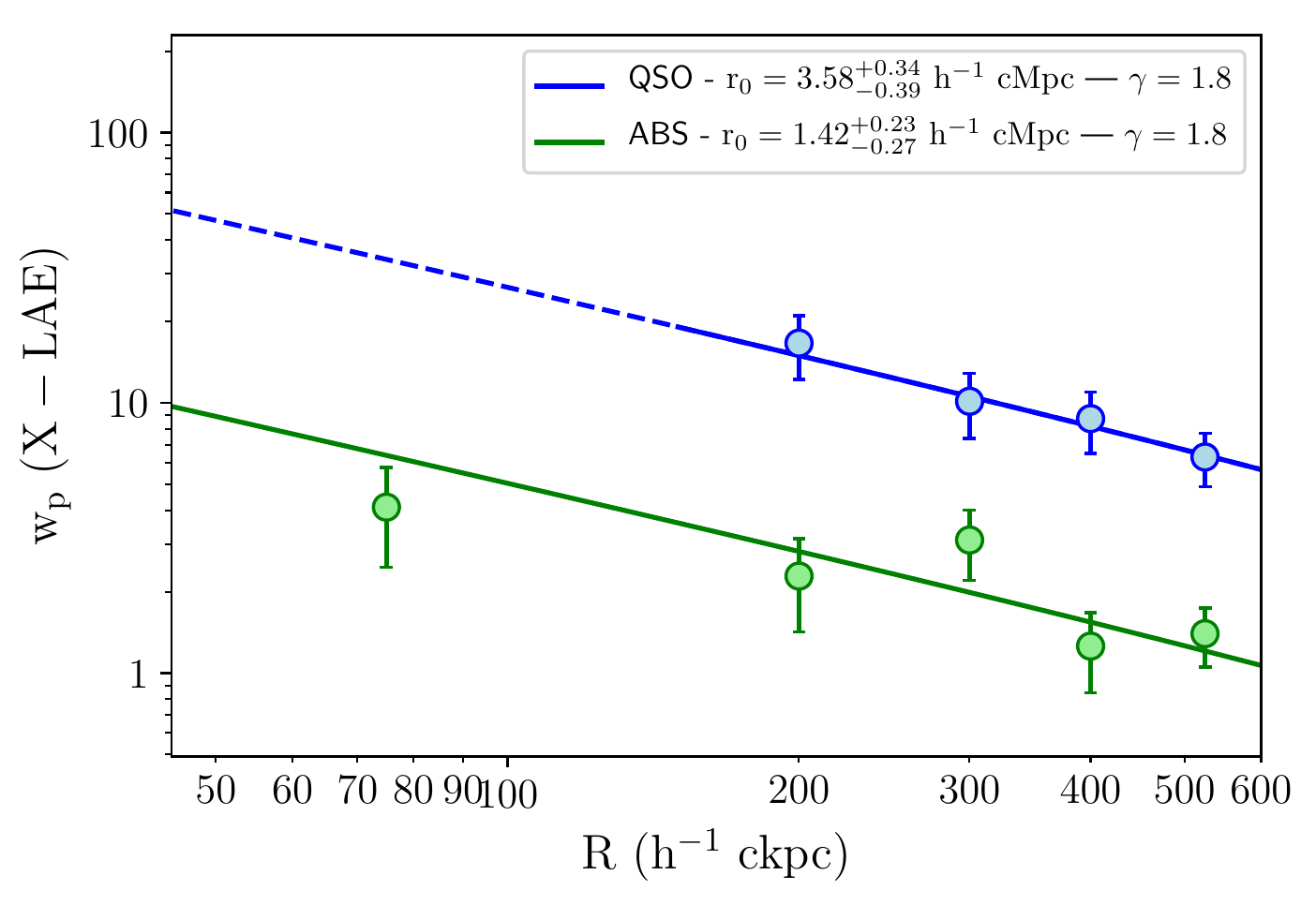}    
    \end{tabular}
    \caption{\textit{Left:} Differential LAE luminosity functions for galaxies in different environments in the MAGG sample. In red we show field galaxies which are selected at $S/N>7$ across the entire MUSE cubes. In green we show the LF for galaxies within $\pm 500$~\kms\ of optically-thick \ion{H}{I} absorbers and in blue are galaxies within $\pm 500$~\kms\ of the quasars. Markers shown with error bars indicate the 1/V$_{\rm max}$ binned values while the solid lines are Schechter fits with a 1$\sigma$ error shown by the shaded region.  Results from \citet{herenz2019} using their real source selection function are shown by the open red points. 
    \textit{Right:} Angular cross-correlation function between absorbers and LAEs (green) and quasars and LAEs (blue). Binned values are shown with 1$\sigma$ Poisson error bars and the solid line shows the data fit to Equation~\ref{eqn:Xcorr}.}
    \label{fig:LF_xcor}
\end{figure*}

\subsection{Establishing the association between LAEs and ALSs}

To investigate the environment of these absorbers on a more statistical ground, we compute a series of metrics to establish how closely related are LAEs to the ALSs in these fields.

\subsubsection{Clustering in velocity}\label{subsec:velclustering}

To identify LAEs with high completeness, we have adopted a generous velocity window of $\pm 1000~\rm km~s^{-1}$ to establish a link with ALS, which corresponds to a line-of-sight distance of $\sim3~$Mpc at $z\sim3$.
We now evaluate more in detail how LAEs cluster in velocity space with respect to ALSs, to rule out the possibility of a substantial spurious association.

Fig.~\ref{fig:vel_sep}~(left) shows the velocity separation between the LAEs and the absorbers where the redshift is determined from the HI absorption. We find a clear peak in the number of detected LAEs at velocity separations around 250~\kms\ in both the full sample of LAEs (blue) and when restricted to only the highest confidence galaxies (orange). 
The bulk of the emitters are distributed within $\approx 500$~\kms\ of this peak, with very few LAEs scattering at larger velocity separation.

The small offset relative to the velocity of the absorbers is an expected and known property due to scattering of the Ly$\alpha$ photons \citep[e.g.][]{steidel2010,verhamme2018}. Similar offsets have been seen in previous works. For example,  \citet{muzahid2020} find an average velocity difference of $169\pm10$~\kms~ between their sample of 96 LAEs at $z\sim3.3$ and nearby absorption systems while \citet{shibuya2014} find that the Ly$\alpha$ line in LAEs is shifted by $\sim$ 230~\kms relative to the nebular emission lines in the same galaxies. Similarly, when using the \ion{C}{IV} absorption lines (where available) to determine the redshift of the absorber, we again find a peak around 250~\kms~ (Fig.~\ref{fig:vel_sep}, right). We find no significant offset between the velocity of the \HI\ and the metals as traced by \ion{C}{IV}.

This clustering around 250~\kms~ strongly indicates that, despite the large velocity window and possible contamination from projection effects, there is a clear physical association in velocity space between the absorbers and the galaxies detected in emission. We clarify at this point that, in the context of this work, we define an ``association'' as a link between absorbers and galaxies that are not arising at random, as clearly shown in Fig.~\ref{fig:vel_sep}. 
The grey distribution in this figures shows the results for a random arrangement of LAEs, unrelated to absorbers. This was calculated by randomly choosing a field and a redshift between 3 and 4.5 as with the random sample used in Fig.~\ref{fig:nEmitter_hist}. Given this position, we search for any LAEs within the MUSE FOV at $S/N >7$. This process is repeated 1000 times to obtain the distribution shown.
This random sample shows a broadly uniform distribution across the full velocity range with no peak as observed around the absorbers. By comparing the random and observed distributions within $\pm 500~\rm km~s^{-1}$ of the peak, we infer an overdensity of $\approx 6$ near LLSs. 

In previous works, the focus has often been on identifying the ``host galaxy'' to the ALSs, more strictly defined as the galaxy within which the absorbing gas resides. In our analysis, for the reasons provided below, we do not attempt to make this strict identification and hence the definition of an association should be regarded in broader terms. This is especially true given that the projected distances between galaxies and ALSs (the impact parameter, see Fig.~\ref{fig:IPhist}) is generally large, ranging from $\approx 18$~kpc to $\approx 275~$kpc, with a median value of 165~kpc.

\subsubsection{Number density and luminosity function}

Having established that LAEs are clustered in velocity with respect to the ALSs, we now consider in detail how the detected number density compares with expectation from a field population.
Based on the Ly$\alpha$ luminosity function at $z\sim3-4$, we would expect to find an average of between 0.3 and 1 source\footnote{For a discussion on the differences among the various luminosity functions obtained with different techniques, see \citet{herenz2019}.} at a luminosity of $\ge 4.5\times 10^{41}~\rm erg~s^{-1}$ within the $\approx 120~\rm Mpc^3$ comoving volume searched around each absorber \citep{grove2009,cassata2011,drake2017,herenz2019}. This luminosity is chosen as it is the value at which we are $\approx 50$ per cent complete for extended sources based on the tests shown in figure 9 of \citet{lofthouse2020}.
As an example, for an absorber at $z=3.5$, a $\pm1000$~\kms search window equates to a line-of-sight distance of $\sim 6.6 ~\rm Mpc$. This is assuming expansion purely due to the Hubble flow and $H_0= 67.7~\rm km~s^{-1}~Mpc^{-1}$ \citep{planck2016}, which gives $2000~{\rm km~s^{-1}}~/~(67.7~\rm km~s^{-1}~\rm Mpc^{-1}\times(1+3.5)) = 6.56 Mpc$, or $\sim 30~\rm cMpc$ (comoving Mpc). The MUSE FOV at this redshift is $0.45 \times 0.45~\rm Mpc$ giving a search area of $[0.45 \times (1+3.5)]^2 \approx 4~\rm cMpc$. Combining this search area with the line-of-sight distance gives a total search volume of $30 \rm Mpc^2 \times 4 \rm Mpc \sim 120 \rm ~cMpc^3$.

Therefore, for the \nlls~absorbers studied, the random distribution of LAEs at this redshift would lead to a detection of $\approx 18 - 60$ LAEs within the full MAGG sample. We note that apart from two absorbers in J200324$-$325144 which have a separation of $\Delta z \approx 0.015$, there is no overlap in the search areas around any of the other absorbers. The identification of \nem~LAEs in the MAGG sample is $\approx 2-7$ times greater than the number expected from a random distribution. This is in line with the inference from the clustering in velocity above, and it offers convincing evidence of an excess of galaxies, indicating that, on average, these LAEs are physically associated to optically-thick ALSs.

For a more formal comparison, we use the luminosity function (LF) to compare the environment around the absorbers to that of the field. The LF is defined as the number of galaxies at a given luminosity and redshift per unit volume. Given that the LF of LAEs is known to show no significant evolution over the redshifts studied \citep[][]{herenz2019}, we assume any variation due to redshift is negligible and bin all our LAEs together. 
To construct the LF for the galaxies we use the 1/V$_{\rm max}$ method as in \citet{schmidt1968} and \citet{felten1976} with a luminosity dependent selection function \citep{fan2001,herenz2019}.
This can be defined as
\begin{equation}
    \phi(\langle \tilde L_{\rm Ly\alpha} \rangle) = \frac{1}{\Delta \tilde L_{\rm Ly\alpha}}\sum_i \frac{1}{V_{\rm max,i}}
\end{equation}
where $\langle \tilde L_{\rm Ly\alpha} \rangle$ is the average Ly$\alpha$ luminosity in a given bin, 
$\Delta \tilde L_{\rm Ly\alpha}$ is the width of that bin and
$V_{\rm max,i}$ is the volume of the survey which has been adjusted for the selection function \citep{johnston2011}. See \citet{fossati2021} for more details on the methodology used here. 

Motivated by the clustering in velocity space previously shown, we restrict the LAEs to only those within $\pm$500 \kms~ of the peak of the Ly$\alpha$ emission as in Fig.~\ref{fig:vel_sep} (i.e. within $-245$ and 745 \kms~ of the absorber redshift), given that galaxies beyond this are increasingly more likely to be chance rather than physical associations.
We fit the LF with a parametric Schechter function \citep{schechter1976} defined as
\begin{equation}
    \Phi(L) = {\rm ln}(10)\Phi^* 10^{(\tilde L-\tilde L^*)(1+\alpha)}{\rm exp}(-10^{(\tilde L-\tilde L^*)})
\end{equation}
We show the LF of galaxies around the absorbers in Fig.~\ref{fig:LF_xcor} (left). The points with errors are the 1/V$_{\rm max}$ binned values and the solid line shows the Schechter function fit. 
Also shown is the LF for galaxies detected around the quasars (blue) as in \citet{fossati2021} using a search window of 500~\kms\ corrected by 245\kms due to the scattering of Ly$\alpha$ and with the same $S/N$ cut of 5. The quasar redshift here is taken to be the peak of the Ly$\alpha$ emission.
Given that the redshifts for both the QSO and the associated LAEs are both derived from Ly$\alpha$, radiative transfer effects are expected to have similar impacts on the measured redshifts and therefore compensate for each other. Indeed, \citet{fossati2021} discuss this effect and show that the average velocity offset between the LAEs and QSO is close to zero. 

To compare with the field population, we include the LF by \citet{herenz2019} and we also compute the LF for the $\approx 1,100$ LAEs found in the MAGG cubes up to a $S/N>7$ (MAGG V; Galbiati et al., in prep), which are identified with the same procedure described here. The \citet{herenz2019} LF is derived from observations of 237 LAEs over $2.9<z<6$ from the MUSE-Wide survey \citep{urrutia2019}.

Compared to the two field LFs, we see an increase in the number of LAEs near absorbers, with a normalisation that is about a factor $\approx 5$ higher than the field. This is in line with the excess noted above with simple counting arguments.  
Conversely, the shapes of the LF in the field and near absorbers are the same, suggesting that LAEs near optically-thick ALSs have comparable luminosity to the field. 
By comparing with the quasar sample in \citet{fossati2021}, we find again a similar shape but with a higher normalisation than the absorber sample, demonstrating that the excess of LAEs near ALSs is bracketed by values in the field from below and values near quasars from above.

\begin{figure*}
    \centering
    \includegraphics[width=0.33\textwidth]{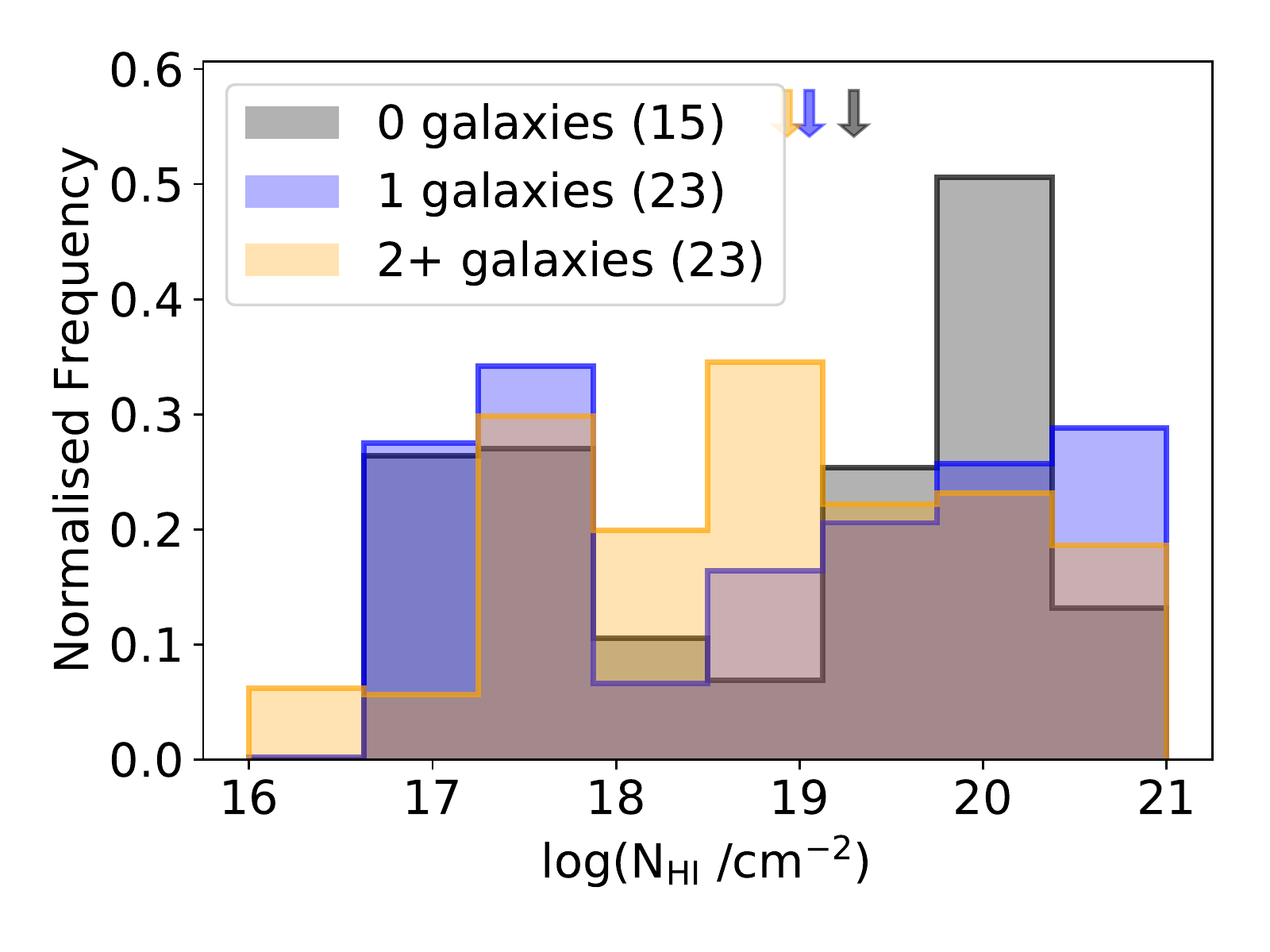}
    \includegraphics[width=0.33\textwidth]{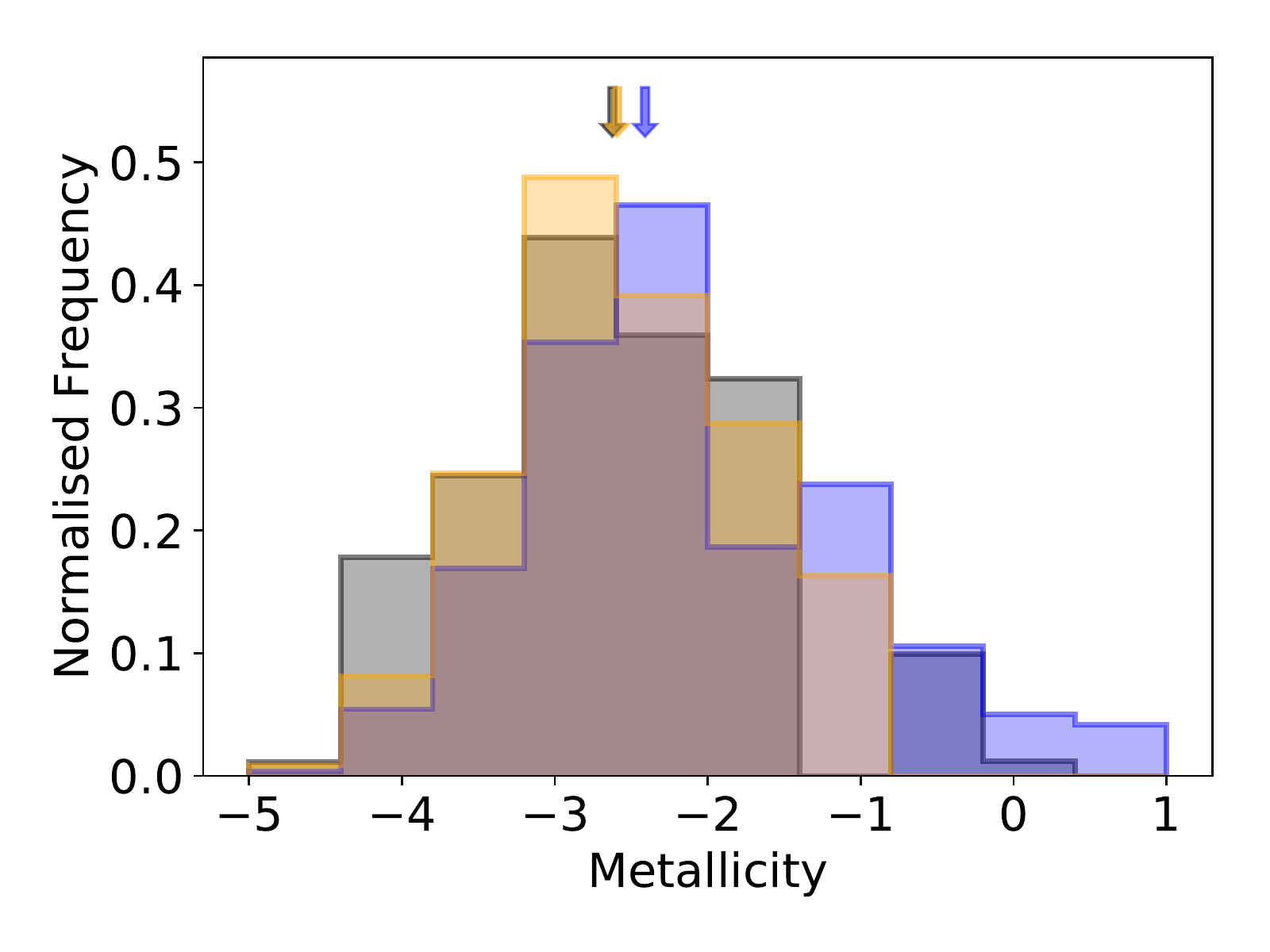}
    \includegraphics[width=0.33\textwidth]{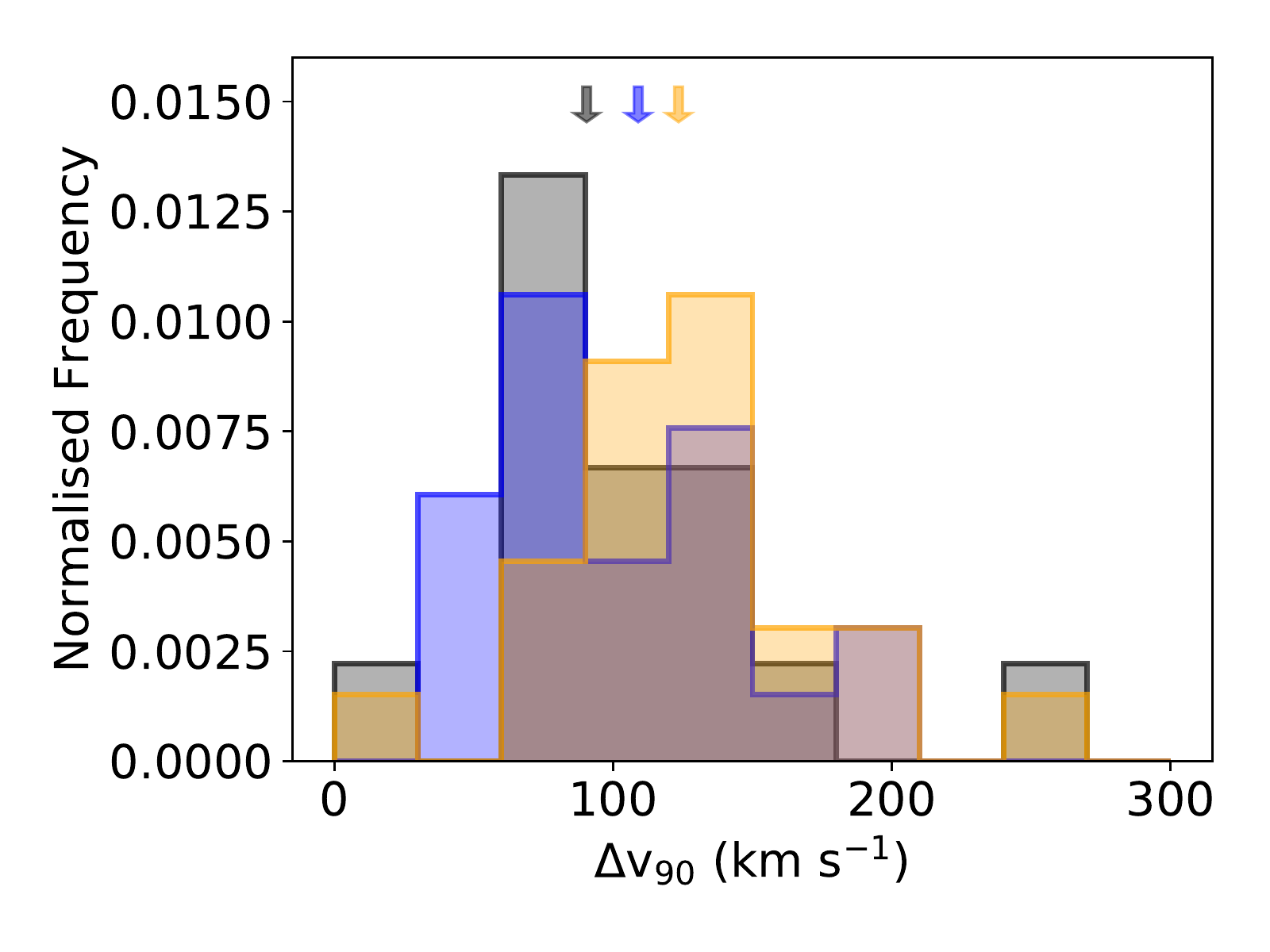}

    \caption{\textit{Left:} Distributions of $N_{\rm HI}$ column density for absorbers where we detect: no galaxies within the MUSE FOV and within $\pm 500$~\kms\ of the absorber redshift (grey); detect a single galaxy association (blue); and detect multiple galaxy associations (orange). The number of absorbers in each sample are indicated by the number in brackets in the legend. \textit{Middle:} Metallicity distributions for the same three groups. 
    \textit{Right:} $\Delta v_{90}$ distributions for each group.
    The median values of each distribution are indicated by the correspondingly coloured arrows at the top of each panel. For this analysis we use the full LAE sample (confidence $< 3$). }
    \label{fig:DR_NHI_met}
\end{figure*}

\subsubsection{Projected cross-correlation function}

An additional way to assess statistically an overdensity of LAEs around the absorption systems is to compute the cross-correlation function between the absorbers and LAEs. This can be thought of as the increased probability of detecting a galaxy in a given volume at a specified distance relative to the probability of detecting a similar galaxy at a random position. 
The quasar-LAE angular cross-correlation has been calculated for the MAGG sample by \citet{fossati2021}. We follow the same methodology here (see also \citealt{trainor2012}) for the absorbers and compare our results. 
The reduced angular cross-correlation function within a given velocity window (in our case $\pm$500~\kms\ as motivated by the clustering seen in Fig.~\ref{fig:vel_sep} and adjusted for the Ly$\alpha$ offset of $\approx 250~$\kms) is defined as
\begin{equation}
    w_p^{\rm AL}(R) = (R_0^{\rm AL} / R)^\gamma \times {}_{2}F_1 (1/2,\gamma/2;3/2;-z_w^2/R^2)\:,
    \label{eqn:Xcorr}
\end{equation}
where $R$ is the impact parameter between the absorbers and the LAEs, ${}_{2}F_1$ is the Gaussian hypergeometric function which corrects the power law for the effects of truncations introduced by the use of a finite redshift window, and 
$z_w = (500~{\rm km~s^{-1}})(1+z)H(z)^{-1}$ is the half-width of the redshift window in units of cMpc. 

The results of the cross-correlation are shown in the right panel of Fig.~\ref{fig:LF_xcor}. The green markers show the values binned in circular annuli centred on the absorbers up to 600 $\rm h^{-1}$~ckpc where this upper limit is defined by the MUSE FOV. The solid line shows the fit to Equation~\ref{eqn:Xcorr}. Also shown are the results for the quasar-LAE cross-correlation (blue). 
The slope of both fits, $\gamma$, is fixed to 1.8 following the convention of other similar studies \citep[see e.g.][]{diener2017,garciavergara2019,fossati2021} as a result of the small range of distances which can be probed within a single MUSE pointing. The resulting $R_0$ is insensitive to the precise choice of $\gamma$ as discussed in \citet{diener2017}.
We report a clustering length $R_0$ = 1.42$^{+0.23}_{-0.27}$ h$^{-1}$~cMpc for the absorber-LAE angular cross-correlation. This weak but positive signal indicates the presence of clustering on the small scales probed within the MUSE FOV, consistent with the results seen in the velocity distributions and the LAE luminosity functions. 
In comparison to the quasars, which have a $R_0 = 3.58^{+0.34}_{-0.39}$ h$^{-1}$ cMpc, the clustering around the absorbers is weaker than around the quasars, and it is more similar to the value inferred for the autocorrelation of LAEs ($R_0 = 2.09^{+0.46}_{-0.47}$ h$^{-1}$ cMpc; Galbiati et al. in prep.). In particular, we note that in the case of the ALSs, LAEs are distributed over a larger range of radii, diluting more the amplitude of the cross-correlation function on small scales compared to the larger contrast with respect to the field that emerges when studying the LF that marginalises over the angular separation. 

Overall, from the evident clustering in velocity, the excess in the normalisation of the LF and a positive clustering from the projected cross-correlation, we conclude that there is strong statistical evidence that the environment around optically-thick absorption systems is richer in galaxies than the field, similarly (but less prominently) to what is found near the quasars themselves.
The environment and clustering around hydrogen absorption systems has previously been investigated using LBGs. For example, using VLRS, \citet{bielby2011} and \citet{crighton2011} find an increase in the neutral hydrogen gas within few Mpc of galaxies. This is consistent with other LBG surveys, such as KBSS where a definite clustering between hydrogen and metal absorbers is found with respect to continuum-detected galaxies  \citep[e.g.][]{adelberger2003,adelberger2005b,rudie2012}.
Our study extends this analysis to the lower masses probed by LAEs, using spectroscopic redshifts from MUSE that are more precise than those from narrow-band surveys and employing large enough samples to statistically study the cross-correlation between absorbers and such spectroscopically-selected galaxies.

\subsection{Dependence of absorption properties on galaxy environment}

Having established a correlation between LAEs and optically-thick \ion{H}{I} selected ALSs, we now investigate whether there is a link between the richness of the galaxy environment (i.e. the number of LAEs identified next to ALSs) and the properties of the absorbers themselves, namely the \ion{H}{I}  column density or the metallicity. 

Fig.~\ref{fig:DR_NHI_met} (left) shows the distribution of the HI column densities for absorbers split into three categories: absorbers where we do not detect any LAEs within the MUSE FOV and within $\pm 500$~\kms\ of the peak of the velocity distribution seen in Fig.~\ref{fig:vel_sep} (grey), absorbers where we detected a single galaxy within this volume  (blue) and absorbers where we detect two or more galaxies (orange). While they are not shown in the figures due to low numbers, we specifically check the values for systems with a very high number of detected LAEs (5 or more) and find that they do not stand out as having particularly high or low values and they broadly follow the distributions of the `two or more galaxies' sample shown. 
For this analysis, we consider LAEs with confidence class $<3$. 
The median value of each distribution is indicated by the corresponding arrows at the top of the figure.
All of the categories show broadly the same distribution in \ion{H}{I} column density, spanning almost the entire range covered by our sample and with no clear peak in their distributions. We see no evidence of a global correlation between $N_{\rm HI}$ and the wider galaxy environment. Indeed, both the highest ($\log (N_{\rm HI}/\rm cm^{-2}) \approx 21.1)$ and lowest  ($\log (N_{\rm HI}/\rm cm^{-2}) \approx 16.5)$  column density systems studied in MAGG show multiple galaxy associations.

To statistically test whether there is any difference in the $N_{\rm HI}$ distributions, we use the Kolmogorov-Smirnov (KS) test. The $p$-value, indicating the probability of the null hypothesis that the distributions are drawn from the same overall distributions, is $>0.7$ for all the comparisons between the three distributions. In particular, comparing the distribution of $N_{\rm HI}$ for absorbers where we detect no galaxies and where we detect 2 or more galaxies results in a $p$-value of 0.84. Even when restricting the sample to only the highest confidence LAEs, the KS test gives a $p$-value of >0.5 between all distributions.
From this, we conclude that statistically there is no dependence of the $N_{\rm HI}$ on the number of detected galaxies and it is therefore not possible to predict the number of LAEs, and hence the environment in which an absorber lies, purely from the \ion{H}{I} column density of the system. 

We investigate next whether the metallicity of the absorber can provide any hints to the environment. It may be expected that higher metallicity systems would have more associated galaxies (or be closer to a galaxy, which is explored in Section~\ref{HIZcorrelations}) as the cumulative effects of stellar feedback or the removal of enriched gas by environmental processes would provide a source of metals for the absorbing gas. 
Fig.~\ref{fig:DR_NHI_met} shows the distribution of metallicities for each of the same three categories: no galaxies, single galaxy, and multiple galaxy associations. The photoionisation modelling used to determine the metallicity of the absorbers from the observed column densities, as described in Section~\ref{subsubsec:photmodel} provides a PDF of the metallicity of each absorber. We combine these PDFs for all of the absorbers in each of the three categories to produce the histograms shown in the figure. As with the \ion{H}{I} column density, we see no significant differences in the distributions for each of the three categories with each distribution spanning a wide range of metallicities. 
As with the $N_{\rm HI}$ distributions, we show the median values with a vertical arrow at the top of the figure. 

All three of the distributions show the same shape and have very similar median metallicities. The group of absorbers with no associated galaxies have a median metallicity of $\log (Z/Z_\odot) \approx -2.62$, absorbers with a single detected galaxy have a median metallicity of $\log (Z/Z_\odot) \approx -2.41$ and the absorbers with 2 or more detected galaxies have a median metallicity of $\log (Z/Z_\odot) \approx -2.59$. 
This result still holds when we repeat the analysis using only the highest confidence LAEs in our sample (confidence $<2$). 
We repeat the KS test for our metallicity distributions, again finding $p$-values that indicate that the distributions are consistent with each other, with probabilities over 0.5 for all the comparisons between each of the three distributions. 
The similarity in the distributions means that it would be impossible to predict the galaxy environment based purely on the metallicity of the absorber alone. 

A final absorber property we investigate is the kinematics, in this case via the use of the $\Delta v_{90}$ parameter. This parameter is the velocity range which encompasses 90 per cent of the optical depth of an unsaturated line \citep{prochaska1998,ledoux2006}. In the right hand panel of Fig.~\ref{fig:DR_NHI_met} we show the distributions of the $\Delta v_{90}$ parameter for the same three samples as in the previous plots (``no galaxies'', ``one galaxy'' and ``two or more galaxies'').
These distributions are calculated using the average $\Delta v_{90}$ value for all lines in a given absorber. However the analysis has been repeated using only \ion{Si}{II} to investigate the kinematics of the low ionisation gas and separately with \ion{C}{IV} for the higher ionisation gas. These distributions show the same results as the average $\Delta v_{90}$. 

As with the  \ion{H}{I} column density and metallicity, the distributions for each of the three samples look broadly the same with the distributions of each one covering almost the full $\Delta v_{90}$ range. 
The median values for the sample with no detections and a single detection are $\approx90$~\kms and $\approx100$~\kms respectively while the sample for two or more detections shows a slightly higher value of $\approx125$~\kms. 
Indeed the KS test between the `no galaxies' and `one galaxy' gives a $p$-value of 0.74 indicating a high likelihood that they are drawn from the same distribution. On the other hand, the KS test between the `no galaxies' and `two or more galaxies' samples results in a $p$-value of 0.07. This difference between the samples is more prominent when considering only the high confidence LAEs which gives a $p$-value of 0.002. 

We conclude from this analysis that the absorption systems studied here are found within a wide range of environments and there does not appear to be any clear correlation between the global properties of absorbers such as their \ion{H}{I} column density or metallicity and the richness of its galaxy environment. While there does appear to be a tendency for absorbers within rich environments with multiple galaxy detections to have higher $\Delta v_{90}$, the distribution spans the full range of values and hence it is not possible to determine the richness of an absorber's environment purely from the properties of the absorber itself.

\begin{figure}
    \centering
    \includegraphics[width=0.5\textwidth]{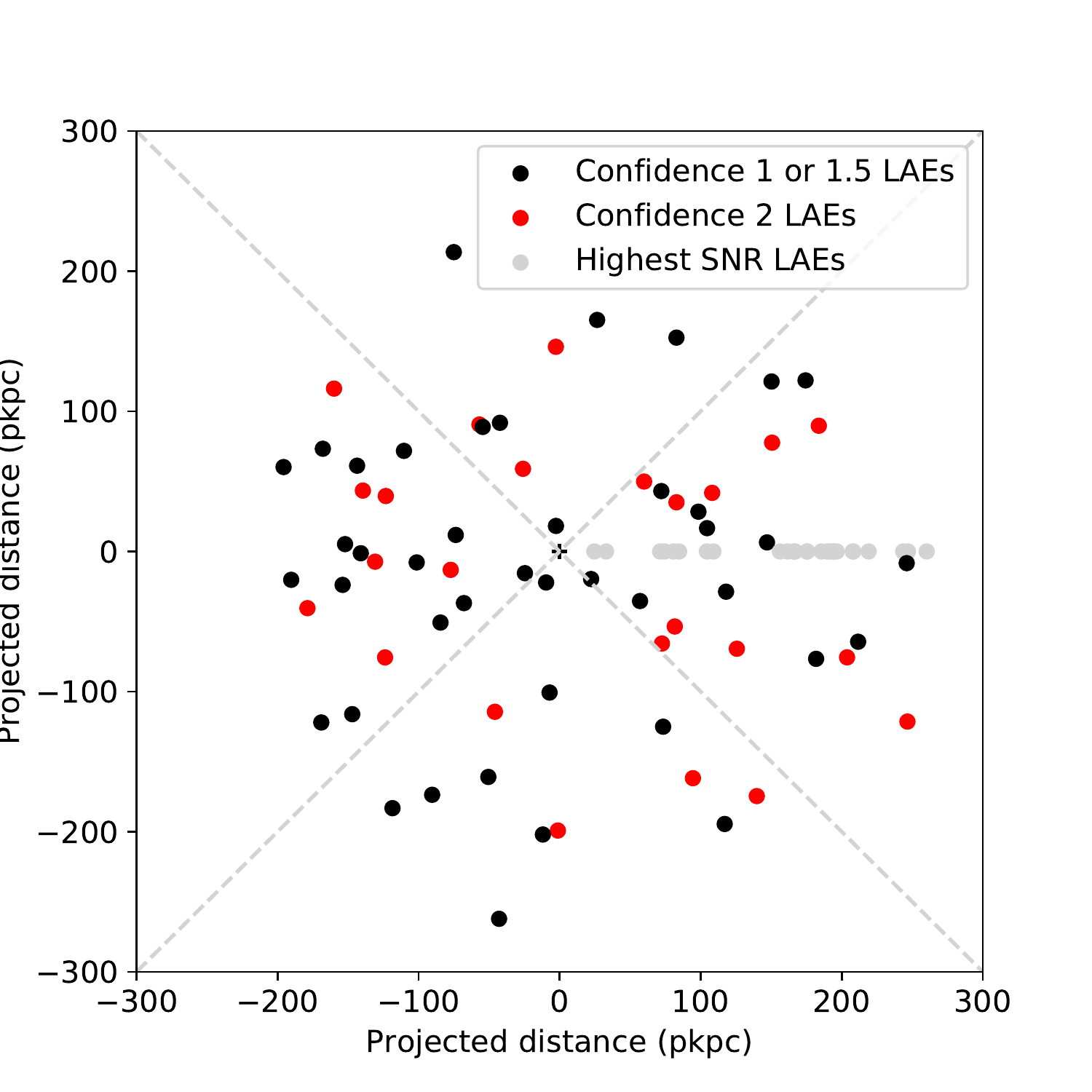}
    \caption{Spatial distribution of LAEs relative to the position of the absorption system (at the centre). For each field we rotate the positions of the LAEs such that the highest $S/N$ LAE is along the positive x-axis (grey points). All other LAEs are identified as either high confidence (black; confidence 1 or 1.5) or lower confidence (red; confidence 2 or 2.5). Systems with only a single LAE are not included in this plot.}
    \label{fig:align}
\end{figure}

\begin{figure}
    \centering
    \includegraphics[width=0.5\textwidth]{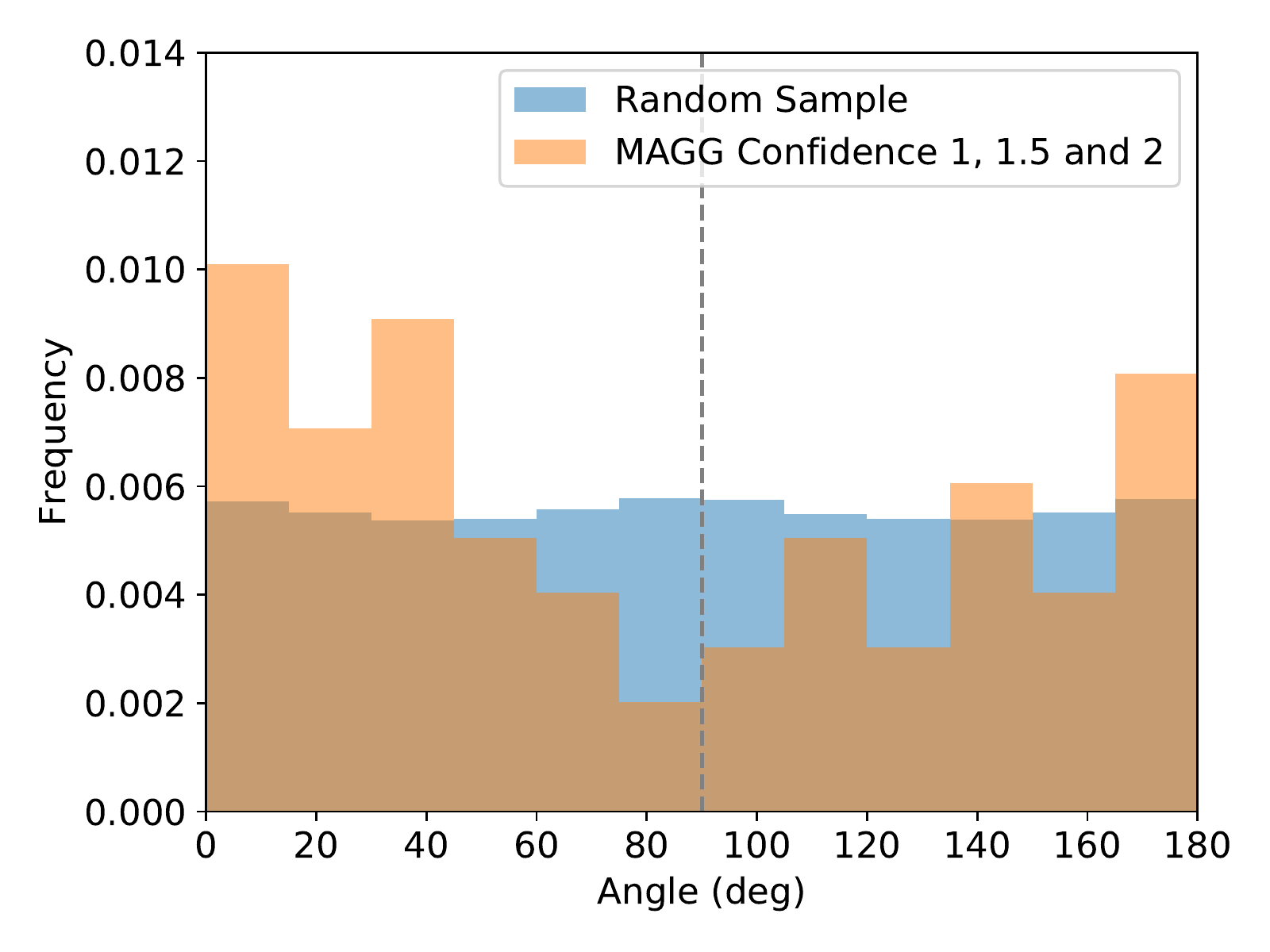}
    \caption{Number of galaxies at a given angle from the axis between the absorber and the closest LAE (orange; shown along the positive x-axis in Fig.~\ref{fig:align}). For comparison, we also show (in blue) the distribution of angles expected for a random distribution of galaxies.}
    \label{fig:filmetric}
\end{figure}

\subsection{Spatial distribution and alignment of LAEs}\label{subsection:align}

One of the possible scenarios for the formation of LLSs is that they originate from gas within the cosmic filaments that connect galaxies. Various studies have attempted to probe such structures combining absorption spectroscopy and galaxies in emission as in MAGG \citep{moller2001,fumagalli2016b}, and in recent years we have now begun to observe these filaments directly through the detection of their very faint emission \citep[e.g.][]{umehata2019, bacon2021, daddi2021}. Indeed, for the first sightline studied in MAGG and presented in \citet{lofthouse2020}, we conclude that this is the most likely case for the extremely metal-poor LLS observed at $z\approx3.53$ (see also \citealt{fumagalli2016b}, \citealt{moller2001}, \citealt{mackenzie2019} for other examples).
If this is the case for the majority of the absorption systems, we would expect to see a preferential alignment between the positions of the LLSs and the galaxies detected with MUSE. 

To test for this effect, Fig.~\ref{fig:align} shows the positions of all the detected LAEs in groups of two or more LAEs relative to the position of the absorption systems, which are at the centre of the figure. This analysis is therefore limited to $92/127$ (72 per cent) LAEs out of the full sample. The positions have been rotated around this point such that the highest $S/N$ LAE in each group is positioned along the positive x-axis (grey points). If the LLSs are lying within a filamentary structure we would expect an overdensity of galaxies along the horizontal axis (i.e. the axis between the absorber and the first aligned galaxy) relative to the number along the vertical axis. 

Dividing the area into four sections, defined by cones with a 45 degree opening angle to the x-axis (see grey dashed lines in Fig.~\ref{fig:align}), we find 79 galaxies (80 per cent) along the horizontal axes (i.e. left/right cones) compared to only 20 on the perpendicular axis (top/bottom cones). When removing the highest $S/N$ LAEs from this calculation, as they are by definition along the x-axis, we still have 50 (71 per cent) galaxies in the left/right quadrants. 
We also test this result by restricting the calculation to only our highest confidence sources (confidence 1). Again, we still observe an excess along the horizontal axis with 29 galaxies in the left/right cones compared to only 11 in the top/bottom. Based on Poisson statistics the probability of obtaining this result with a random distribution of galaxies is $\approx$1 per cent. 

To investigate this result further, we explore the distribution of angles between the LAEs and the axis defined by absorbers and the \textit{closest} LAEs. 
Fig.~\ref{fig:filmetric} shows the number of galaxies at a given angle from this axis (orange).  We also show the result that would be expected from a random distribution of galaxies (blue) which is broadly flat across all the possible angles. This random distribution was calculated by drawing coordinates at random in the range -250,+250 kpc in each direction. We then computed the angle distribution as normal and repeat the process 10,000 times. 

For the LAEs near LLSs, we see an excess of galaxies near 0 degrees, i.e. along the same axis as the closest LAE, and at 180 degrees, i.e. in the opposite direction with respect to the quasar relative to the closest LAE but along the same axis. In contrast, there are significantly fewer galaxies around 90 degrees than would be expected from a random distribution. A KS test between the observed LAEs results and the random distribution gives a probability of 0.08 that the samples are drawn from the same overall distribution. 
From this, we conclude that there is evidence for a preferential alignment of LAEs around absorption systems at $z\sim3-4$. 
In contrast, \citet{fossati2021} performed a similar analysis for LAEs around quasars at $z\sim3-4.5$ in MAGG and found no strong evidence for any spatial alignment of LAEs in these environments.

\begin{figure*}
    \centering
    \includegraphics[width=1\textwidth, trim = 2.5cm 2.5cm 3cm 3cm, clip]{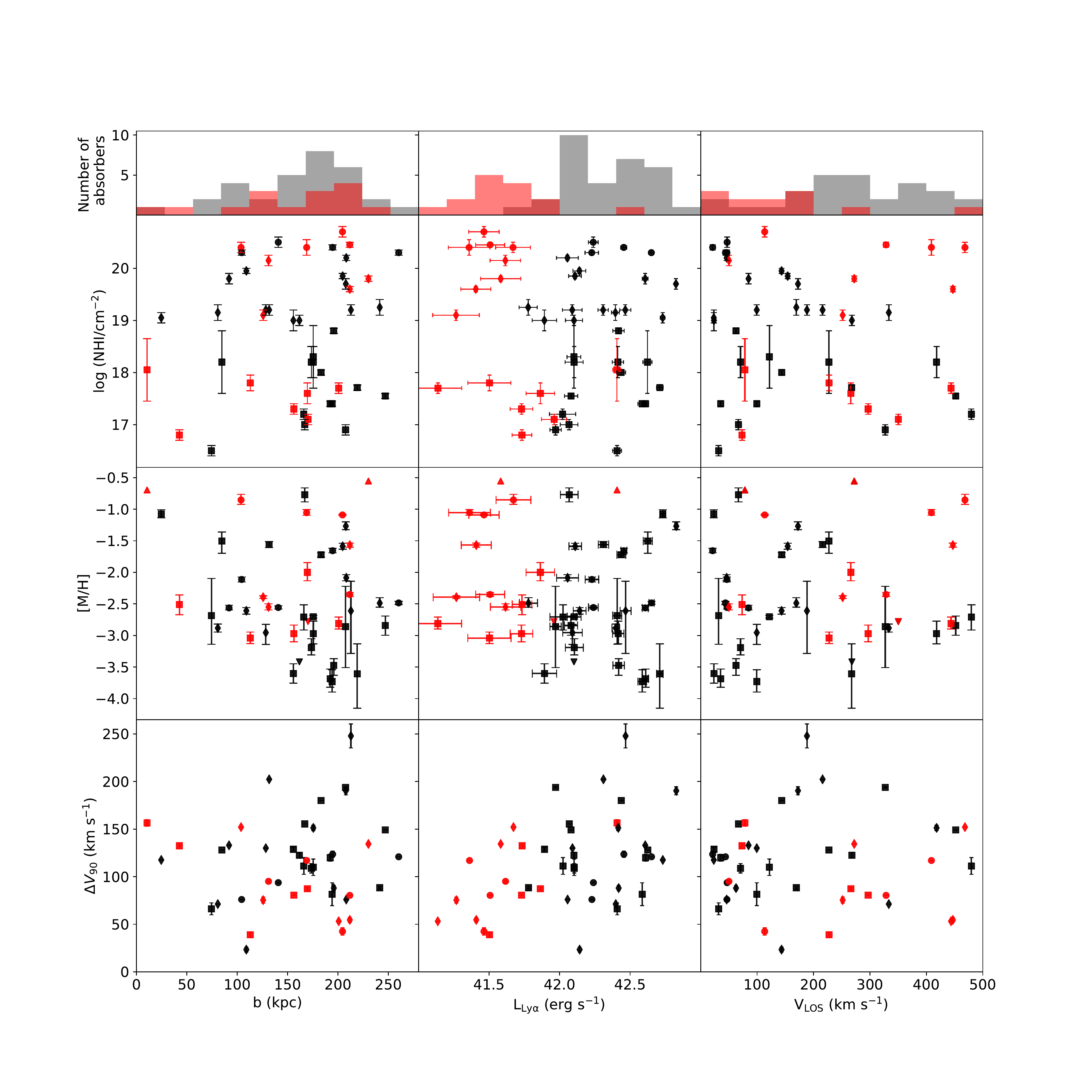}
    \caption{Correlations between the absorber properties and the properties of the brightest LAE. The top row shows the relationship between the absorber column density and impact parameter (left), Ly$\alpha$ luminosity (middle) and line of sight velocity separation (right). The middle row shows the absorber metallicity against the same three galaxy properties while the bottom row shows the $\Delta$v$_{90}$. In each plot we show the high confidence LAEs (i.e., confidence 1 and 1.5) in black and the lower confidence LAEs in red. Additionally we use different marker shapes to denote the absorber column densities with circles for DLAs, diamonds for sub-DLAs, and squares for LLSs. Upper and lower limits are shown by triangular markers. At the top of the figure we also show histograms of the number of absorbers within bins of the galaxy properties (impact parameter, luminosity and velocity separation). These histograms are coloured with grey denoting high confidence sources and red denoting the lower confidence sources.}
    \label{fig:cor_bright}
\end{figure*}

\begin{figure*}
    \centering
    \includegraphics[width=1\textwidth, trim = 2.5cm 2.5cm 3cm 3cm, clip]{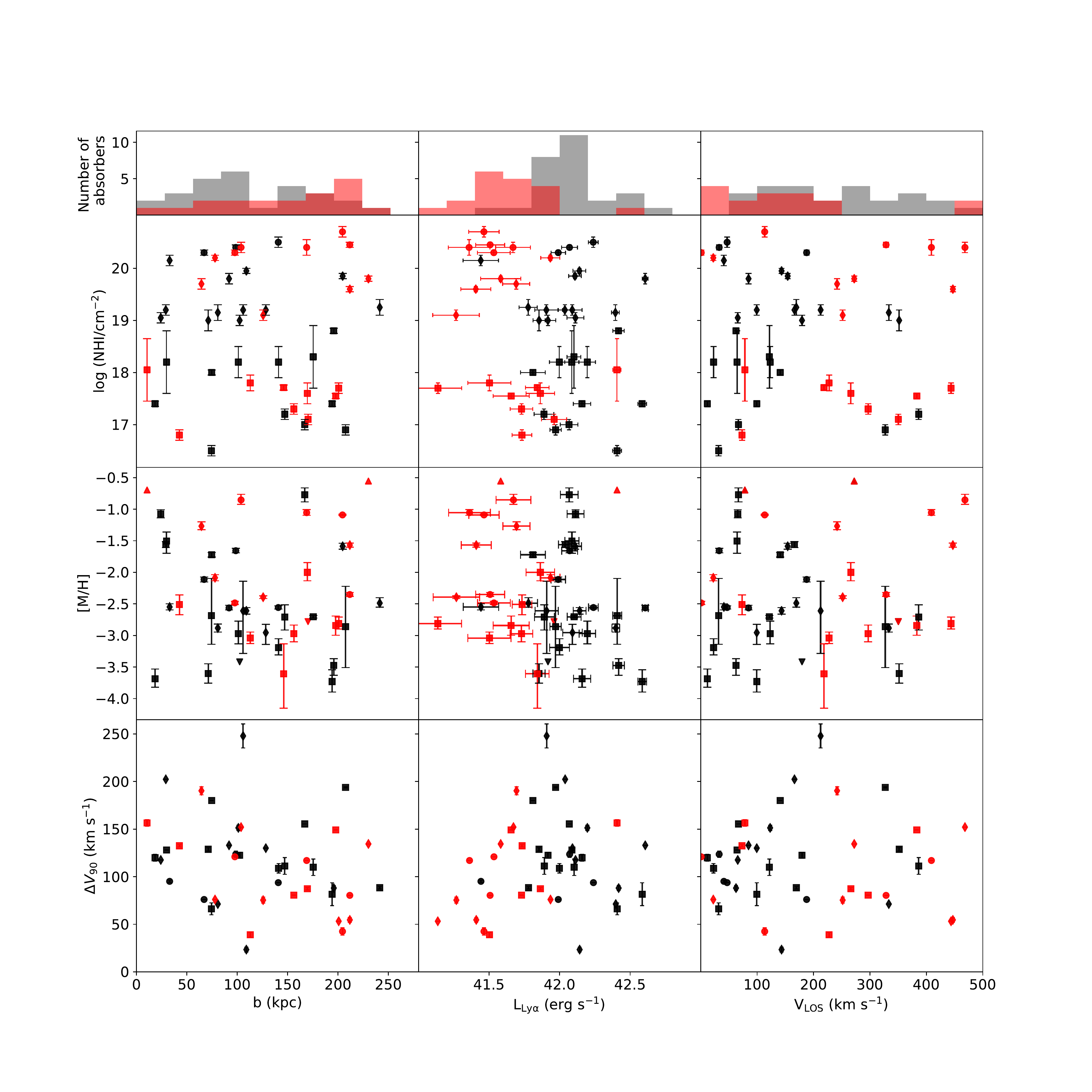}
    \caption{As Fig.~\ref{fig:cor_bright} but selecting the closest LAE in impact parameter rather than the brightest.} 
    \label{fig:cor_close}
\end{figure*}

\section{Correlations between absorbers and LAE\texorpdfstring{\MakeLowercase{s}}{s}}\label{HIZcorrelations}

In the previous sections we have explored globally the relation between the properties of the optically-thick gas and galaxies.
With the large sample of systems in MAGG, it is also possible to study the relationship between individual absorbers and galaxies in a statistical sense and investigate the presence of any correlations between absorption and emission properties. 

As shown in Section~\ref{sec:env}, using MUSE observations we often detect multiple galaxies around a single absorber.
This presents the issue of deciding which of the multiple galaxies should be used to compare properties with the absorber. The two widely-used options are to use the 'brightest' galaxy or the 'closest' galaxy. By choosing the brightest galaxy we are interested in investigating the relevance of what is assumed to be the highest star forming and most massive galaxy which is likely to be the central galaxy of the halo where gas is detected. 
Caveats to this approach are the use of Ly$\alpha$ which does not trace the entire star-forming population, it is an imperfect indicator of star formation, and the fact that there is non-negligible scatter in the underlying galaxy main sequence.
Alternatively, by choosing the closest galaxy to the quasar line of sight, we focus on the galaxy that is not necessarily the central one (hence, we could select satellites) but the one that might have the largest influence given its proximity to the absorbing gas. As shown above, all the galaxies considered are highly clustered with the absorbers along the line of sight, which mitigates the impact of projection effects.  In this case, however, we do not know the true location along the line of sight, although we can test for correlations with respect to the line of sight velocity.  
In the following, we show the results of using both of these options and discuss the impact of this choice on our conclusions. 

We now proceed to examine, in turn, how the absorption properties (\HI\ column density, metallicity, and velocity width) relate to properties defined by the galaxies in emission (relative distance between absorber and galaxy or impact paramater, velocity separation along the line of sight between absorber and galaxy, and Ly$\alpha$ luminosity). 
By definition, this experiment is only limited to the subset of $49/61$ absorbers that are associated with at least one galaxy detected via Ly$\alpha$ emission. 
(as noted in Section~\ref{subsec:velclustering}, associations here are considered within the entire FOV and within $\pm 500$~\kms of the peak of the LAE velocity distribution relative to the absorbing gas). 
Results are shown in Fig.~\ref{fig:cor_bright}-\ref{fig:cor_close} for the case of brightest and closest galaxies, respectively. Highest confidence sources are in black and the lower confidence sources are in red. Different symbols denote intervals of column densities for the absorbers (circles for DLAs, diamonds for sub-DLAs, and squares for LLSs).
While we cannot investigate the correlations specifically for the handful of LAEs that are also continuum-detected sources due the limited statistics, we do check whether they show any tendencies to lie in certain regions of these plots. Beyond being at the higher end of the luminosity range, as expected, they appear to show no preferences in any of the properties studied here and fall within the scatter of the overall population.

\subsection{Correlations with impact parameter}\label{subsec:IP}

Considering the relationship between the \HI\ column density of the absorber, N$_{\rm HI}$, and the impact parameter, in the parameter space probed in MAGG which ranges from lower \ion{H}{I} column density systems at $10^{16.8}~\rm cm^{-2}$ up to DLAs, we do not see any evidence of a correlation with absorbers. At all column densities, absorbers are found near galaxies and are scattered across impact parameters of $\approx 50-250$ kpc. This applies to both the results when we select the brightest LAEs in the FOV where multiple galaxies are detected (Fig.~\ref{fig:cor_bright}; top left) and where we select the closest galaxy to the absorber (Fig.~\ref{fig:cor_close}; top left). To quantify this, we conduct a Kendall's Tau test on the two data sets \citep[e.g.][]{brown1974,isobe1986}. This provides a $p$-value indicating the probability of obtaining the observed data given that the two variables are uncorrelated. For the brightest LAE and closest LAE sample, we obtain $p$-values of 0.46 and 0.98 respectively indicating that it is highly likely that the \NHI\ is uncorrelated with impact parameter.

We examine next the comparison between the metallicity of the absorber and the impact parameter between the absorber and the brightest and closest galaxies.
When selecting based on the brightest detected galaxy we see a potential absence of sources at very low metallicities and small impact parameters (middle left panel of Fig.~\ref{fig:cor_bright}).
This hints at an admittedly weak trend where metallicity increases towards smaller separations, although with a large scatter. This trend, however, vanishes once we consider the galaxies at the closest separation  (middle left panel of Fig.~\ref{fig:cor_close}).
Finally, considering the correlation between the velocity width of the absorbers ($\Delta \rm v_{90}$) and the impact parameter, we observe no clear correlation between the two properties both when considering the brightest LAE (bottom left panel of Fig.~\ref{fig:cor_bright}) and the closest LAE (bottom left panel of Fig.~\ref{fig:cor_close}). We find both small and large velocity widths across the full range of impact parameters up to $\approx 250$~kpc.

\subsection{Correlations with Ly\texorpdfstring{$\alpha$}{Lya} luminosity}\label{subsec:lya}

Next, we move on to investigating the correlations between the absorber properties and the Ly$\alpha$ luminosity of the selected LAE. 

The \ion{H}{I} column density of the absorbers is shown as a function of LAE luminosity in the top middle panels of Fig.~\ref{fig:cor_bright} and Fig.~\ref{fig:cor_close}. As for the impact parameter results, we see no clear correlation between the \ion{H}{I} column density and the galaxy luminosity, both for the brightest and for the closest LAEs. This result remains when we restrict the sample to only the highest confidence LAEs (confidence 1 and 1.5; black) which are typically found at the largest luminosities.

The centre plots of each figure show the relationship between the absorber metallicity and galaxy Ly$\alpha$ luminosity. Similarly to the results for the metallicity versus impact parameter, when selecting the brightest LAE we see a hint of a negative correlation between the two parameters with an absence of systems at low metallicities and low luminosities. However a Kendall's $\tau$ test on this data produces a $p-$value of 0.45 which is not sufficiently low to reject the null hypothesis and confirm the presence of a relation. When using the closest LAE instead, we see similar results but a Kendall's $\tau$  test on this sample now produces a lower probability (0.15) of there being no correlation between the two parameters. 

When comparing the Ly$\alpha$ luminosity with the average $\Delta v_{90}$ of the absorber's metal lines, as shown in the bottom middle panels of Fig~\ref{fig:cor_bright}, a positive correlation emerges such that we do not see fainter galaxies around absorbers with high $\Delta v_{90}$ values. This result remains the same when we use both the brightest and the closest LAE. This may be the result of the brightest LAEs tracing the most massive haloes where the velocity dispersion is highest and hence the gas exhibits higher velocity, or a reflection of higher velocities in the CGM of galaxies with higher star formation, as expected for instance due to the presence of outflows.

\subsection{Correlations with line of sight velocity}\label{subsec:vel}

Finally, we focus on the correlation between absorption properties and the line of sight velocity difference between gas and galaxies, where the velocity separation has been systematically adjusted by 245~\kms\ to account for the shift in the Ly$\alpha$ due to scattering (see Fig.~\ref{fig:vel_sep}).

Similar to the impact parameter and Ly$\alpha$ luminosity results, we see no clear correlation between the \ion{H}{I} column density of the absorption systems and the velocity separation either when considering the full galaxy sample (red) or when restricting only to the highest confidence LAEs (black). When using either brightest LAEs or the closest LAEs (top right panels of Fig.~\ref{fig:cor_bright} and Fig.~\ref{fig:cor_close} respectively), we see absorbers across the full range of column density from LLSs to DLAs at both small and large velocity separations.

When we investigate the relationship between the velocity separation and the metallicity of the absorber, we do not observe any correlation, neither for the case of the brightest galaxy nor for the case of the nearest one (middle right panel of Fig.~\ref{fig:cor_bright} and Fig.~\ref{fig:cor_close} respectively). 
Hence, while we did see some suggestion of a possible anti-correlation between the metallicity and the impact parameter, this trend is not confirmed by the line of sight velocity separation. 
Although the line-of-sight velocity is most likely dominated by peculiar motions and radiative transfer effects, we have also completed the exercise of combining the projected impact parameter with the equivalent distance along the line of sight (assuming Hubble flow velocities). Even when considering this three-dimensional distance, no obvious correlations emerge. This is not surprising given the results discussed above and the fact that the three-dimensional distance is, in this case, dominated by the LOS velocity given the small impact parameters we are probing with MUSE.

The last correlation we consider is that between the line width of the absorber ($\Delta$v$_{90}$) with respect to the relative velocity separation along the line of sight (bottom right panel of Fig.~\ref{fig:cor_bright} and Fig.~\ref{fig:cor_close}). We again do not observe any correlation either for the brightest LAE or for the closest LAE. Indeed, the Kendall's $\tau$ test gives probabilities that the two parameters are uncorrelated of $\approx 0.71$ when choosing the brightest LAE and $\approx 0.85$ for the closest LAE.

\section{Covering Fraction of optically-thick gas around LAE\texorpdfstring{\MakeLowercase{s}}{s}}

\begin{figure}
    \centering
    \includegraphics[width=0.49\textwidth]{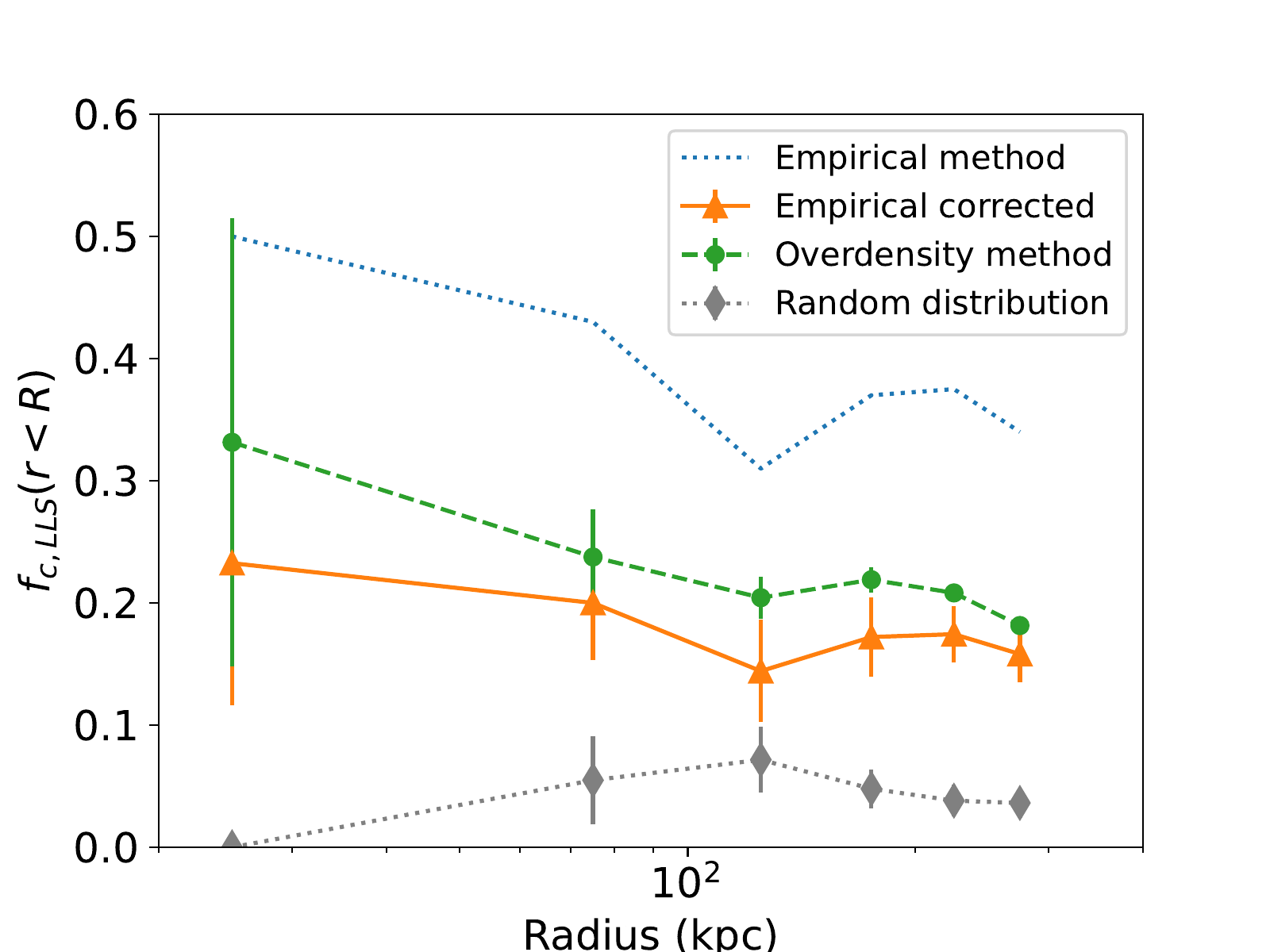}
    \caption{The cumulative covering fraction of \HI\ absorption with $\rm N_{\rm HI} > 10^{17.2}~\rm cm^{-2}$ around LAEs detected at $S/N>7$ as a function of impact parameter from the quasar line of sight.
    Empirical values from raw data are shown by the blue dotted line, while orange triangles mark the values corrected for the biased selection of LLSs in MAGG. The green circles show instead the covering fraction computed from the overdensity of LAEs, as described in the text. Both estimates are in good agreement. The results obtained for a random distribution of LAEs is also shown (grey diamonds).}
    \label{fig:cov_frac}
\end{figure}

Having explored the link between absorbers and galaxies globally and in individual systems, we next move on to investigate, in a statistical sense, the distribution of gas around LAEs by measuring the covering fraction of neutral hydrogen.
Empirically, the covering fraction can be defined as:
\begin{equation}\label{eq:cfdef}
\tilde{f_c} (<R) = \frac{N_{\rm det} (S/N \geq 7; N_{\rm HI}>\bar{N}_{\rm HI};r<R)}{N_{\rm tot}  (S/N \geq 7; r<R)}
\end{equation}
where $\rm N_{\rm det}$ is the number of galaxies that show absorption above a given \ion{H}{I} column density in the quasar spectrum within $\pm 500$~\kms\ of their redshift (corrected by 245~\kms\ due to the scattering of Ly$\alpha$) within a radius R; $\rm N_{~\rm tot}$ is instead the total number of galaxies in which we could potentially detect gas in absorption in the quasar spectrum if it was present (i.e. those lying in the path-length searchable for ALSs). 
To calculate this fraction, we need to extend the identification of LAEs beyond the regions occupied by ALSs. To do so, four authors (EKL, MiF, MF, MG) have visually inspected all sources extracted by {\sc cubex} with a $\rm S/N\geq 7$ across the entire MAGG cubes. We refer the readers to the forthcoming paper by Galbiati et al. (in prep) for more details on the full sample of LAEs across the entire MAGG survey. 
We exclude any systems within $3000~$\kms\ of the quasar redshift from this analysis so that our results are not affected by the environment around the quasar itself. In this calculation, we further include only the highest redshift LLS of each sightline, as for optically-thick systems without damping wings we cannot quantify the presence of additional ALSs at lower redshift due to the presence of the Lyman edge. 

Given the selection of MAGG sightlines, which favours the presence of $z>3$ LLSs, the incidence of optically-thick absorbers does not reflect the cosmological value, $\ell (z) \equiv dn/dz$.  Indeed, considering the total pathlength to the first  LLS in MAGG and the number of ALSs detected, we find $\ell_{\rm MAGG} \approx  4.3$, which is a factor $\approx 2.15$ higher than the cosmological incidence measured at $z\approx 3.5$, equal to $\ell_{\rm cosm} \approx  2$ \citep{prochaska2010,fumagalli2020}.
Due to this, the average pathlength to the first LLS in MAGG is $\approx 2.15$ times shorter than what found in an unbiased survey. In turn, this shorter pathlength boosts the empirical estimate of the covering factor by the same amount, as it decreases the number of LAEs uncorrelated to LLSs by the ratio of the true to measured pathlength. 
Therefore, Equation~\ref{eq:cfdef} needs to be corrected by a factor $\xi = \ell_{\rm MAGG}/\ell_{\rm cosm}$ to yield an accurate estimate of the true covering fraction, $f_c = \tilde{f_c}/\xi$.
In Figure~\ref{fig:cov_frac}, we show the cumulative covering fraction as a function of impact parameter for the $N_{\rm HI}  \ge 10^{17.2}~\rm  cm^{-2} $ systems, before (blue dotted line) and after (orange triangles) the correction factor $\xi$ is applied. The error bars show the 1$\sigma$ Wilson score confidence intervals. 
We note that a non-negligible uncertainty, up to a factor $\approx 2$, exists in the literature for $\ell_{\rm cosm}$ at $z\approx 3.5$ \citep[see a discussion in ][]{fumagalli2020}. Here, as noted, we assume a value of $\ell_{\rm cosm}\approx 2$ which brackets current estimates; however, values both below and above $\approx 2$ have been reported in the literature. As such, the covering fraction is subject to this uncertainty, and readers wishing to make a different assumption on the incidence of LLSs can simply rescale the values presented here by the factor $\xi$ defined above. 

Using MAGG data, we can derive a second estimate of the covering fraction based on the overdensity of LAEs near LLSs compared to the field, following the formalism outlined next. Starting from Equation~\ref{eq:cfdef}, we can write
\begin{equation}
    N_{\rm det}(<R) = \delta(<R) \rho_{\rm LAE} A_{\rm FOV} \Delta z N_{\rm LLS} \:,
\end{equation}
where $\delta(<R)$ is the overdensity of LAEs compared to the mean density $\rho_{\rm LAE}$ within an impact parameter $R$, $A_{\rm FOV}$ is the FOV area, $\Delta z$ is the search window for LAEs near LLSs ($\pm 500~$\kms), and $N_{\rm LLS}$ is the number of LLSs detected. 
Moreover,
\begin{equation}
    N_{\rm tot}(<R) = \rho_{\rm LAE} A_{\rm FOV} \Delta z_{\rm tot} \:,
\end{equation}
where $\Delta z_{\rm tot}$ is the total pathlength to the first LLS in MAGG. 
Combining, we obtain that 
\begin{equation}
     \tilde{f_c} (<R) = \frac{\delta(<R) \Delta z N_{\rm LLS}}{\Delta z_{\rm tot} } = \delta(<R) \Delta z \ell_{\rm MAGG}\:.
\end{equation}
Hence, the unbiased estimator becomes
\begin{equation}
     f_c (<R) = \delta(<R) \Delta z \ell_{\rm cosm}\:,
\end{equation}
where we have replaced the incidence of LLSs from MAGG with the cosmological value. 
Hence, the covering fraction can be evaluated simply by measuring $\delta(<R)$ from MAGG data. For this, we integrate the distribution of LAEs detected  within $\pm 500~$\kms\ of LLSs, dividing by the random expectation in bins of radius (see Fig.~\ref{fig:nEmitter_hist}). 
The resulting covering fraction based on the overdensity measure is shown by green circles in Fig.~\ref{fig:cov_frac}. The agreement with the empirical estimate is quite good, also considering the uncertainties.

The covering fraction profile for LLSs shows a mildly decreasing trend with impact parameter, reaching a maximum covering fraction of $f_{\rm c} \approx 0.3$ for $<50$~kpc, dropping down to just below $f_{\rm c} \approx 0.2$ at $\approx 300$~kpc.
The covering fraction also appears to have an inflection point near $\approx 100~$kpc, a feature that is visible both in the empirical estimate and in the one based on overdensity. As discussed below, we interpret this dip as a transition between the signal arising from the CGM of individual LAEs to the larger scale environment. 
For comparison, we include the results expected from a random distribution of LAEs (grey). This was obtained in the same manner as the N$_{\rm HI}$ > 10$^{17.2}$cm$^{-2}$ sample but the redshifts of the LAEs were randomly shuffled before performing the calculation. For this distribution we find lower covering fractions than for the actual sample. 

\begin{figure}
    \centering
    \includegraphics[width = 0.49\textwidth]{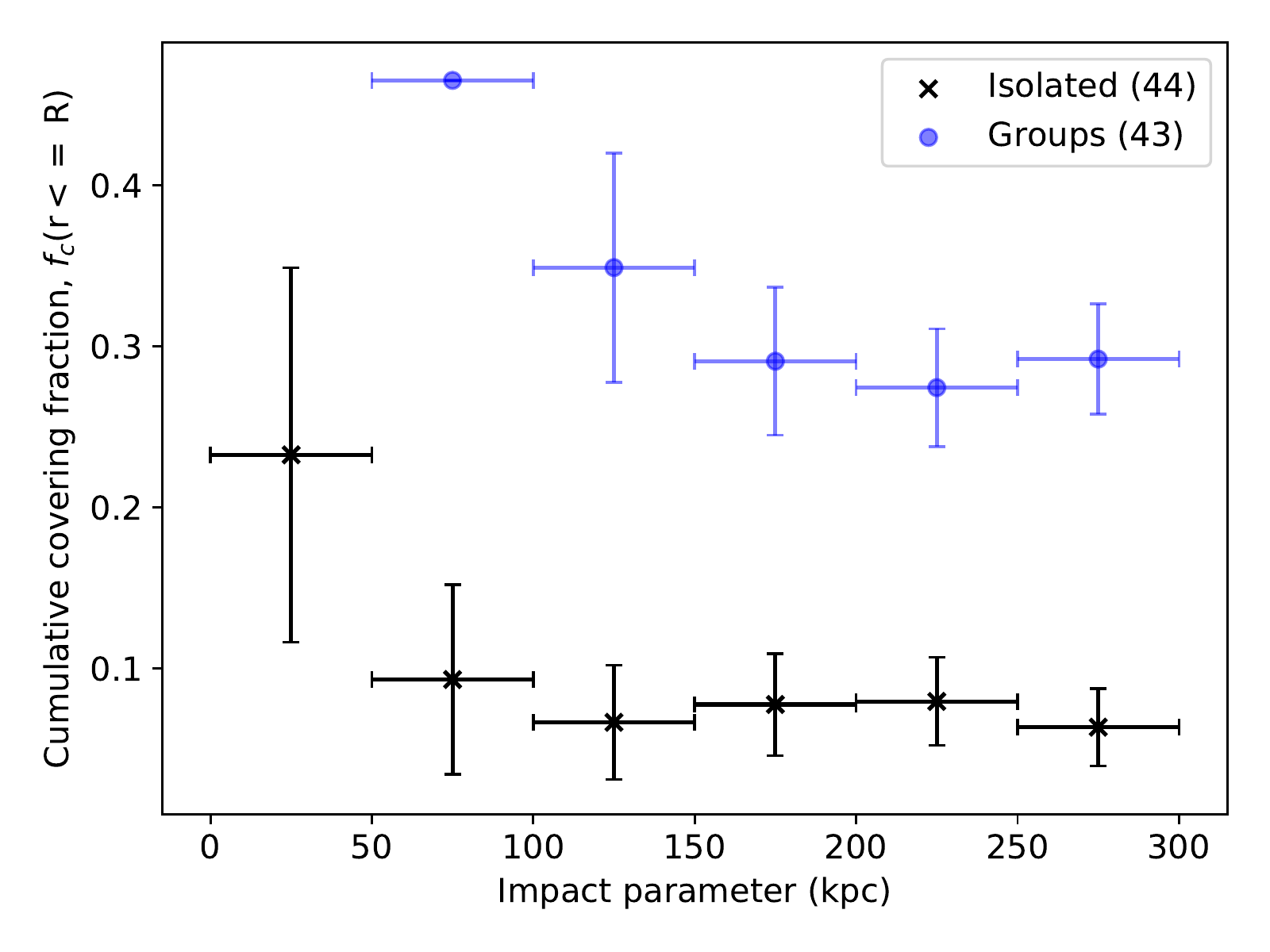}
    \caption{The cumulative corrected covering fraction as a function of impact parameter for $N_{\rm HI}  \geq 10^{17.2}~\rm  cm^{-2}$ absorbers around LAEs detected at $S/N>7$. The sample is split into isolated LAEs (black crosses) and LAEs in groups of 2 or more (blue circles). }
    \label{fig:cov_frac_isogroup}
\end{figure}

In Fig.~\ref{fig:cov_frac_isogroup}, we repeat the measure of the covering fraction for $N_{\rm HI} > 10^{17.2}~\rm cm^{-2}$ absorbers, but splitting the sample into `isolated' LAEs and LAEs in `groups'. LAEs within groups are defined as those that have at least one other LAE within $\pm 500$~\kms. Using this selection we see a clear difference between the covering fractions for isolated and group galaxies, with the latter having a higher covering fraction at all separations. For example, at separations $<300$~kpc the covering fraction for isolated galaxies is only $f_{\rm c} = 0.08 \pm 0.02$ while for groups it is $f_{\rm c} = 0.29 \pm 0.03$. At smaller impact parameters the covering fractions increase for both samples with $f_{\rm c}$ reaching $0.23 \pm 0.11$ for isolated galaxies at impact parameters $<50$~kpc. For groups of galaxies, the covering fraction approaches 50 per cent at $<100$~kpc.

With the covering fraction in hand, we can also revisit the argument by \citet{gallego2021}, who presented a method to constrain the amplitude of the ultraviolet background (UVB) background at $z>3$ starting from a measurement of the Ly$\alpha$ fluorescence in optically-thick clouds. In their original analysis, however, the LLS covering fraction was not known, hence they relied on results from numerical simulations. With our new measurement, we can independently verify their assumptions on the covering fractions. For this, we repeat the analysis of the covering fraction starting from a measurement of the overdensity using the HI column density cut of N$_{\rm HI}$ > 10$^{17.5}$cm$^{-2}$ and the same radial binning as in \citet{gallego2021}, focusing in particular on the interval $8-20~$arcsec ($60-150~$kpc), where we have the most constraining power from MAGG data. 
In this interval, we find a covering factor of $\approx 17$ per cent, again assuming $\ell_{\rm cosm} \approx 2$. 
In this bin, \citet{gallego2021} estimated with simulations $f_{\rm c} = 0.083 \pm 0.002$ for $z\approx 3$ and $f_{\rm c} = 0.106 \pm 0.004$ for $z\approx 4$, which is a factor $\lesssim 2$ smaller than estimated in MAGG. This difference is, however, not particularly significant at the current precision with which $\ell_{\rm cosm}$ is measured.
At face value, our determination of the covering fraction combined with the fluorescence measure in \citet{gallego2021} would yield an estimate of the UVB in good agreement with the \citet{haardt2001} UVB model, which is slightly lower than what is inferred by \citet{gallego2021} using simulations.

\section{Discussion}\label{Discussion}

Through a dedicated survey of galaxies around optically-thick absorbers, this analysis of the MAGG data provides the most complete determination of the UV-selected galaxy population surrounding partially neutral gas regions at $z\sim 3-4$, including UV-bright LBGs and, as a particularly novel aspect, LAEs with $L_{\rm Ly\alpha} \gtrsim 10^{41}~\rm erg~s^{-1}$. Our in-depth study of the environment of LLSs reveals that: i) a fraction of LLSs ranging from 57 to 82 per cent (depending on the confidence of detection) is near one or more galaxies, the near-totality of which is an LAE; ii) there is solid evidence of a physical association between the LAEs and the LLSs based on the clustering in velocity and space, and on the normalization of the luminosity function; iii) the bulk of these associations is found at somewhat large distances from the absorbing gas, with impact parameters ranging from 50 to 250 kpc, and along filaments connecting LAEs and LLSs; iv) there is no evident correlation between the LLS absorption and the LAE emission properties, except when considering the absorber line widths and the Ly$\alpha$ luminosity.
Building on these pieces of evidence, and in comparison with literature studies, we can now tackle a series of questions relating to the nature and origin of high-redshift LLSs and to what extent LAE-LLS associations provide information on the role of the baryon cycle at high redshift.

\subsection{A rising detection rate of galaxies around \texorpdfstring{\HI}{HI} absorbers}

Strong links between LLSs or DLAs and galaxies at $z\gtrsim 2$ have been documented in the literature for some time now. The pioneering searches of galaxies associated with DLAs using traditional spectrographs \citep{fynbo2010,krogager2017} have been some of the most successful efforts to build a statistical sample of galaxies associated to DLAs before the deployment of MUSE. This sample counted however only $\approx 10$ associations.
The advent of large-format IFUs and especially MUSE has accelerated the rate of discovery near both DLAs and LLSs, following in the footsteps of programmes with smaller IFUs such as SINFONI \citep{peroux2012}. Indeed, early works based on MUSE often detected galaxies within the local environment of absorbers \citep[e.g.][]{fumagalli2016, mackenzie2019, lofthouse2020}, an effort that culminated in the full sample of MAGG where we find a high detection rate of galaxies around absorbers of $\approx 82$ per cent. This is higher than previous surveys with SINFONI, e.g., by \citet{peroux2012} who quote detection rates ranging from $\approx 14$ to $\approx 37$ per cent.
The main factor behind this increase is the larger impact parameter probed by MUSE around absorbers compared with what is possible in SINFONI (hundreds of kpc versus tens of kpc). Indeed, the majority, $\approx 79$ per cent, of the LAEs in the MAGG sample are found at impact parameters above $100~\rm kpc$. 
Another factor is the increased sensitivity to faint emission in the optical with MUSE compared to near-IR detections with SINFONI which improves the detection efficiency.
Additionally, Ly$\alpha$ is intrinsically brighter than H$\alpha$ emission in systems with low-dust obscuration. When comparing to results from similar studies using MUSE, and hence similar search volumes albeit with smaller samples of absorbers, we find similar results such as \citet{mackenzie2019} who achieve a detection rate of $\sim$80 per cent from a study of six DLAs. 

Despite the fact that the detection of galaxies near absorbers is now becoming the norm, there are still examples both in MAGG and in the literature of systems with no detectable galaxies within hundreds of kpc from the absorbing gas \citep[e.g.][]{fumagalli2016}. An obvious reason is related to the intrinsic bias of MUSE towards less dusty galaxies and galaxies that are line emitters compared to UV bright galaxies with no emission lines. Both ALMA surveys and targeted spectroscopic searches sensitive to absorption lines indeed provide examples of these types of associations. Moreover, especially in LLSs, the bright glare of the background quasar still affects the detectability of faint sources near $\approx 10-15~\rm kpc$ where PSF residuals are non-negligible.   

Nevertheless, comparing the variation of the rate of non detections in confidence $>2$ versus confidence $\le 2$ systems ($\approx 18$ per cent in all LAEs and $\approx 40$ per cent for high-confidence ones; Fig.~\ref{fig:nEmitter_hist}), it is clear that the fraction of detections rapidly rises with increasing sensitivity, as expected given the shape of the luminosity function. Deeper surveys have therefore the potential to reach detection rates approaching 100 per cent. Moreover, as argued in the next section, there is evidence that a significant percentage if not the entirety of the optically-thick gas observed at $z\approx 3-4$ can be associated with LAEs of comparable luminosity to the ones observed by MAGG within a radius of $\approx 150~\rm kpc$. This implies that, despite the completeness issues noted above, LAEs have a sufficient number density to be very often if not always located in proximity to optically-thick gas resulting in associations. We therefore predict that, in virtue of the clustering of other types of galaxies (e.g. more massive systems) with LAEs, there should be a significant overlap in associations between, e.g. dusty or UV bright galaxies and LAEs. Indeed, already within MAGG and as already reported by other studies, we see instances of multiple detections near a single LLS, with up to eight LAEs detected around one absorber. A combination of multi-wavelength and deep surveys of fields hosting LLSs is therefore likely to reveal a progressively higher number of multiple associations composed of systems which are detected at different wavelengths. 

This rapid increase of detections made possible by MUSE and the recognition that optically-thick absorbers could trace either the CGM of galaxies clustered to other systems or even large-scale structures shared among multiple galaxies (see the next section) prompts us to move away from the traditional one--to--one associations between galaxies and absorbers, and instead consider these absorbers as residing in more complex galaxy environments. The direct consequence is that the interpretation of signatures of the baryon cycle (e.g. inflows or outflows) encoded in ALSs is made more complex, discouraging straightforward conclusions which could lead to incorrect or conflicting inferences from data. To fully exploit the constraining power of gas-galaxy associations for the baryon cycle it is therefore necessary to resort to more statistical analysis, requiring detailed numerical models or even fully cosmological hydrodynamic simulations (see \citealt{mackenzie2019}, for an example). Thankfully, the increasing number of galaxy-absorber associations makes this approach more realistic compared to what is possible in previous surveys.   

\subsection{The nature of optically-thick gas}

The nature of optically-thick gas giving rise to high-redshift LLSs has been long studied in the literature, particularly through the statistical analysis of LLSs and numerical simulations. Globally, the incidence \citep[e.g.][]{sargent1989,prochaska2010,fumagalli2020} and the detailed absorption properties of LLSs such as density and metallicity \citep[e.g.][]{steidel1990,prochaska1999,fumagalli2013}, place these optically-thick absorbers in overdensities comparable to what is found in the virial regions of galaxies. Such a conclusion finds general agreement with the results of numerical simulations, which place a non-negligible fraction of LLSs in close proximity to (or even within) the halo of high-redshift galaxies \citep[e.g.][]{faucherGiguere2011,fumagalli2011,vandeVoort2012}. However, the picture is complicated by the fact that, as the neutral fraction of the IGM increases with redshift, progressively more systems are expected to arise from intergalactic regions \citep{mcquinn2011,fumagalli2013}.
Leveraging the blind survey of galaxies made possible by MAGG, we can revisit the question of where LLSs reside with respect to galaxies.  

Being \HI-selected systems, we consider in turn three scenarios for the origin of LLSs: the ISM, the CGM, and the IGM. Starting from the ISM, we note that the majority of the associations lie at projected impact parameters $\gtrsim 50~\rm kpc$, larger than the expected extent of the gas ``disks'' at $z>3$ \citep{rafelski2016} also considering some of the most extreme cases known \citep{neeleman2020,cicone2015}.  
A question arises, however, on possible selection biases due to the presence of the quasar that hampers the visibility of the innermost $\approx 1~\rm arcsec$ (or $\approx 7-8~\rm kpc$). We exclude this possibility on three grounds. First, Fig.~\ref{fig:IPhist} does not suggest the presence of a truncated distribution with detections piling up close to the putative limit at which we would start missing sources. Second, when stacking spectra extracted from circular apertures of $1-2~$arcsec at the quasar locations in the rest-frame of the absorbers, we do not observe evidence of Ly$\alpha$ emission which would arise from a large population of bright LAEs on top of quasars. Third, we inspect individual narrow-band images extracted from the cube at the redshift of the absorbers (also accounting for the typical shift of Ly$\alpha$) as well as stepping through each wavelength slice of the cube around the wavelengths of the absorbers searching for emission near the quasar. Through this visual inspection we do not find any additional sources that were missed by our {\sc Cubex} search.  We also note that there is minimal (if any) contribution of the quasar residual in the absorption trough of DLAs.    
Such systems close to quasars have been previously identified in connection with extremely strong \HI\ absorption systems \citep{ranjan2020}, which are however absent in MAGG due to the intrinsic rarity of systems with $\log (N_{\rm HI}/\rm cm^{-2}) > 21.6$.

Disentangling the contribution of the CGM and IGM to the origin of LLSs near LAEs is less trivial, due to the fact there is no well defined transition marking the boundary between these two gas components. Following a similar argument to the one proposed for the ISM, however, we can rule out the inner CGM as the main contributor. Indeed, the clustering analysis of spectroscopic samples with comparable luminosity suggests a halo mass for LAEs that approaches $\approx 10^{11}~\rm M_\odot$ \citep{herrero2021}. The virial radius of such a halo at $z\approx 3$ is $\approx 35~\rm kpc$, implying that the bulk of the associations arise from beyond twice the virial radius. This seems to disfavour the inner CGM of LAEs as a primary source of optically-thick gas, but still leaves circumgalactic regions extending at $3-4$ times the virial radius as plausible sites where LLSs can be found. At these distances, the divide between the CGM and IGM blurs. As argued in Section~\ref{subsection:align}, there is also evidence of filaments connecting multiple LAEs and the LLSs. 
\cite{moller2001} provided one of the earliest detections of such structures by mapping out a string of galaxies which covered almost 5 Mpc \citep[see also][]{fumagalli2016b,mackenzie2019}. 
In more recent years, we are now also able to see these filaments directly through Ly$\alpha$ emission. While these observations are difficult due to the intrinsic faintness of such emission, progress was made by observing regions where there are sources of ionizing photons which locally enhance the Ly$\alpha$ line. For example, \citet{cantalupo2014} observed a 460~kpc cosmic web filament in Ly$\alpha$ which has been illuminated by emission from a quasar at $z\sim2.3$.
In \citet{umehata2019}, the authors observed SSA22, a protocluster at $z=3.1$, and reported the detection of a 1.3~Mpc filament with several star-forming galaxies embedded in it. \cite{bacon2021} used the MUSE Extremely deep field, a 140h MUSE program in the Hubble Ultra Deep Field, to detect extended Ly$\alpha$ emission from the cosmic web in proximity to AGNs beyond the extreme environments probed in  previous studies.
Therefore, combining our analysis and mounting evidence in the literature of filaments connecting galaxies, we propose that a significant fraction of optically-thick gas could indeed arise from gas inside these IGM structures. Finally, the fraction of LLSs without associated LAEs within $\approx 250~\rm kpc$ is sufficiently low to exclude an abundant population of LLSs arising in IGM filaments far from galaxies. Furthermore, no dependence on redshift is found for the fraction of LLSs associated with galaxies. Therefore we exclude the idea of an increased contribution from IGM regions as redshift increases, although with large uncertainties due to the limited sample.

We must also acknowledge that -- despite the very complete survey enabled by MUSE -- our selection is still biased in favour of line emitters, so galaxies lacking prominent Ly$\alpha$ emission are underrepresented in our study. Hence, we cannot derive an absolute statement on the origin of LLSs with these data alone. However, while the fraction of star-forming galaxies emitting LAEs is not particularly high ($\lesssim 20$ per cent for comparable samples; see \citealt{kerutt2022}), there should be no preference in selecting LAEs at large separations with respect to the position of LLSs. Therefore, we expect that the distribution of impact parameters is representative of the underlying LAE population. Therefore, relative to these sources, we can affirm that LLSs originate primarily either from the outer CGM (defined here as $3-4~R_{\rm vir}$) or in the IGM but in close proximity to galaxies.

\begin{figure}
    \centering
    \includegraphics[scale=0.48]{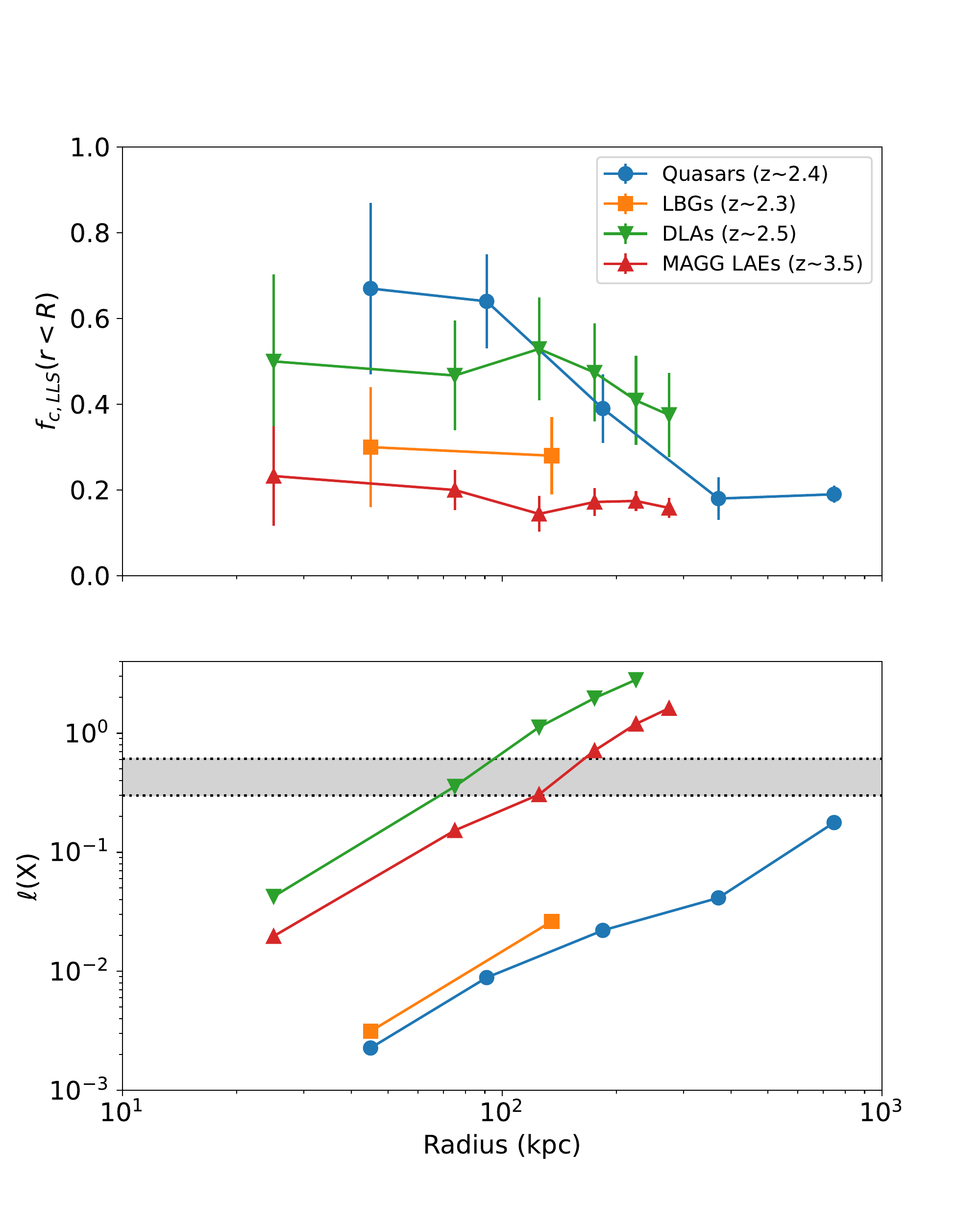}
    \caption{Top: Comparison between the covering fraction of optically-thick hydrogen around LAEs (red upward triangles) from this work, quasars (blue circles) from \citet{prochaska2013}, LBGs (orange squares) from \citet{rudie2012}, and DLAs (green downward triangles) from \citet{rubin2015}. Bottom: The contribution to the cosmic incidence of $\tau \gtrsim 2$ LLSs from the optically-thick gas detected around different populations (colour-coded as above) as a function of radius. The LLS incidence measured in quasar absorption line surveys is shown by the grey band, where the lower limit is the $z\approx 2$ value from \citet{omeara2013} and the upper limit is the $z\approx 4$ value from \citet{prochaska2010}. Due to their high number density, LAEs account for the near totality of LLSs by $\approx 150~\rm kpc$, and are thus overlapping with the DLA population.}
    \label{fig:cfandlx}
\end{figure}

The search for galaxies near LLSs in MAGG provides a direct way to address independently the question on the incidence of optically-thick gas within extended Ly$\alpha$ halos surrounding galaxies. Deep long-slit and IFU observations have indeed frequently uncovered extended halos near Ly$\alpha$ emitters with luminosity comparable to the one selected by our survey \citep{rauch2008,wisotzki2018}. Thanks to these studies, it was possible to compare the redshift incidence of Ly$\alpha$ emitting regions with the incidence of optically-thick gas derived from absorption spectroscopy. Based on the number of LAEs and the extent of their Ly$\alpha$ halos, these authors found that LLSs have a comparable incidence to halos above a surface brightness of $10^{37}~\rm erg~s^{-1}~kpc^{-2}$ out to $\approx 50-70~\rm kpc$ ($10^{38}~\rm erg~s^{-1}~kpc^{-2}$ out to $\approx 20-30 ~\rm kpc$ for DLAs). At face value, this correspondence would imply that a large fraction of LLSs (DLAs) arise from within $\approx 2$ ($\approx 1$) times the virial radius of LAEs with $L_{\rm Ly\alpha} \gtrsim 10^{41}~\rm erg~s^{-1}$. By performing the opposite experiment (i.e., by searching for galaxies near LLSs), our MAGG survey places stringent constraints on this inference and, based on the findings discussed above, rules out a unique correspondence between the innermost part of halos and the distribution of optically-thick gas. Indeed, we have shown that LLSs are rarely found at such close proximity from LAEs. We conclude that comparisons between emission and absorption properties need to account for a more complicated distribution of optically-thick gas that extends to larger radii with covering factors much lower than unity.

We can further extend this investigation into the nature of optically-thick gas by combining additional tracers known from the literature, such as LBGs  \citep[e.g.][]{rudie2012}, DLAs \citep[e.g.][]{rubin2015}, and quasars \citep[][]{prochaska2013}. This comparison in shown in Fig.~\ref{fig:cfandlx}. For covering fractions that are presented in the literature as differential values (i.e., in annuli of impact parameter where $R_1<r<R_2$), we have converted the values to cumulative covering fractions (i.e., $r<R$) in order to directly compare them with the cumulative values used in MAGG.
Considering the covering fraction, quasars are by far the population with the most extended optically-thick gas in their surroundings, with elevated values up to $\approx 100$~kpc which drop to lower but still considerable values up to 1~Mpc. \citet{prochaska2013} attribute this rapid decline to a boundary in the quasar CGM and also note that, due the number of ambiguous cases in which they could not disentangle between optically-thick and optically-thin hydrogen based on partial coverage of the Lyman series, these values are believed to be lower limits. 
LAEs from our study exhibit a similar behaviour, but with a clear shift to smaller radii and with intrinsically lower values. Similarly to the interpretation of the covering fraction for quasars, we attribute the enhanced values at $\lesssim 60-70~$kpc to the contribution of the inner CGM of LAEs (within twice their virial radius), with a further contribution at larger distances arsing from the outer CGM and the IGM connecting LAEs, as already noted above. 
LBGs appear to show, in comparison to LAEs, a similar covering fraction but with a flatter radial dependence. It should be noted however that the survey by \citet{rudie2012} has only a very limited number of LLSs, and hence this determination is the one that is subject to the largest uncertainties. 

DLAs show instead an elevated covering fraction, but a peculiar radial dependence compared to the other populations. However, similar to the quasar case, this measurement suffers from ambiguous determinations of optically-thick versus optically-thin systems and hence these covering fractions should be, too, considered upper limits. Moreover, the measurement of the covering fraction near DLAs is conceptually different from the other three cases presented in this figure. DLAs arise from gas that is believed to be distributed around galaxies, and therefore the radial distance from a DLA is  different from the one measured from galaxies. More explicitly, assuming for the sake of argument that DLAs are identically associated to LAEs, one should not expect the same measurement of covering fraction. Indeed, while the LAEs are expected always at the centre of the halo, DLAs are themselves distributed inside the halos and hence the covering fraction of DLAs is a convolution of ``off-centre'' measurements of the hydrogen distribution \citep[see][for an explicit example]{fumagalli2014}.  

Besides the direct comparisons of the covering fraction, insight into the nature of optically-thick gas and its relation to galaxies can be obtained by considering the contribution of each population to the total incidence of LLSs measured in quasar spectroscopic surveys (i.e., the number of LLSs per unit redshift $\ell(z)$). As commonly assumed in the literature \citep[see e.g.][]{steidel2010,prochaska2013,fumagalli2013}, knowing the covering fraction of optically-thick gas as a function of radius in a population of known number density $n_{\rm com}$, one can estimate the incidence of LLSs arising at a given radius from that population by computing the product of the effective cross section of optically-thick gas and the population number density 
\begin{equation}\label{eq:lxcontr}
    \ell(X,r<R) = \frac{c}{H_0}f_c(r<R)\pi R^2 n_{\rm com}\:.
\end{equation}
Here, the cross section is computed in physical units while the number density is in comoving units, resulting in a measure of the incidence in terms of the absorption distance
\begin{equation}
    dX = dz \frac{(1+z)^2}{\sqrt{\Omega_{\rm m}(1+z)^3+\Omega_{\rm \Lambda}}}\:.
\end{equation}
The results are shown in the bottom panel of Fig.\ref{fig:cfandlx}, and compared with the incidence of LLSs from absorption spectroscopy at $z\approx 2$ \citep{omeara2013} and at $z\approx 4$ \citep{prochaska2010}. 

Despite their elevated covering fraction, due to their moderate number density ($n\approx 1.2\times 10^{-4}~\rm Mpc^{-3}$ for halos with $M\gtrsim 10^{12.5}~\rm M_\odot$), quasars contribute to a minimal fraction of LLSs to within a few hundred kpc. However, when measured at large distances, their covering fraction rapidly rises to a substantial contribution of the LLS population, reaching $\approx 70$ per cent at $\approx 1~\rm Mpc$. Similarly, the combination of somewhat modest covering fraction and number density ($n\approx 3.7\times 10^{-4}~\rm Mpc^{-3}$ for halos with $M\gtrsim 10^{12}~\rm M_\odot$) makes LBGs a comparably minor contribution of LLSs. 
The opposite behaviour is seen instead for both LAEs and DLAs: despite their covering fraction being almost a factor of $\approx 2-3.5$ lower than quasars at fixed radius, their number density is $n\approx 10^{-2}~\rm Mpc^{-3}$, which is obtained for LAEs by integrating the luminosity function calculated in the MAGG survey (Galbiati et al., in prep) for $L_{\rm Ly\alpha} > 10^{41.5}~\rm erg~s^{-1}$ and for DLAs integrating the halo mass function above $10^{11}~\rm M_\odot$. In both cases, the high number density is sufficient to account for the entire LLS population already between $\approx 80~\rm kpc$ and $\approx 150~\rm kpc$.  The comparable number density and the fact that both DLAs and LAEs produce a similar contribution to LLSs strengthen the argument presented already by, e.g., \citet{rauch2008} and \citet{wisotzki2018} that LAEs account for the main DLA population\footnote{In Fig.~\ref{fig:cfandlx}, we used the empirical estimate of covering fraction from MAGG. Assuming the one based on overdensity would lead to an even greater contribution to LLSs at all radii.}.

Two further considerations are warranted at this point.  
First, it may appear surprising that either DLAs and LAEs overpredict the observed incidence of LLSs beyond $\approx 100-170~\rm kpc$. In fact, there is reason to believe that these estimates are uncertain to at least a factor of two for several reasons: i) the minimum luminosity or halo mass function assumed for the calculation of the number density is subject to significant uncertainty especially for DLAs, but the resulting number density, $n$, is a steep function of this cutoff; ii) the determination of the abundance of LLSs is in itself uncertain \citep[see a discussion in ][]{fumagalli2020}, with repercussions on the the precision with which we can estimate the covering fraction in MAGG; iii) the calculation in Equation~\ref{eq:lxcontr} is rather crude as it attributes average binned values of covering fraction to all halos regardless of any underlying mass dependence. Furthermore and most importantly, this calculation breaks down at large radii due to the clustering of multiple LAEs near LLSs. If optically-thick gas arises from an astrophysical structure engulfing multiple LAEs, the number density of LAEs should be replaced by the number density of underlying structures giving rise to the absorption. Hence, the use of the number density of LAEs leads to an overestimate of the incidence. If DLAs are linked to LAEs as argued above, a similar consideration applies to the case of DLAs. The second consideration stems from the fact that LAEs (hence DLAs) account for all LLSs by $\approx 80-150~\rm kpc$ but that quasars themselves contribute with equal importance to the incidence out to $\approx 1~\rm Mpc$. This apparent contradiction can be reconciled by the fact that, at such large distances from quasars, one is likely to be probing the covering fraction of populations clustered to quasars themselves. For example, the analysis of MAGG data in proximity of quasars reveals a strong clustering of LAEs up to $\approx 0.5~\rm Mpc$ \citep[][see also Fig.~\ref{fig:LF_xcor}]{fossati2021}. Hence, it is reasonable to believe that LAEs and quasars (at large distances) are not independent populations, and the two contributions shown in Fig.~\ref{fig:cfandlx} should not be added together but rather regarded as arising from the same astrophysical structures. 

In summary, based on the following three pieces of evidence revealed by the MAGG survey we can add new constraints to the nature of optically-thick gas in the Universe: i) LAEs strongly cluster near LLSs but with associations extending to beyond twice the typical virial radius; ii) there is some degree of evidence of an alignment of multiple LAEs in filaments where LLSs are found; iii) LAEs easily account for the near totality of LLSs. In light of this, MAGG confirms previous claims that LAEs are indeed at the origin of the LLSs seen in absorption spectra, but extends this picture put forward by previous studies, shifting the importance from the inner CGM of individual LAEs to the filamentary cosmic structures that connect multiple LAEs and host a substantial fraction of these LLSs.

\begin{figure*}
    \centering
    \begin{tabular}{c|c}
    \includegraphics[width=0.49\textwidth, trim= 0.5cm 0.5cm 0cm 0cm, clip]{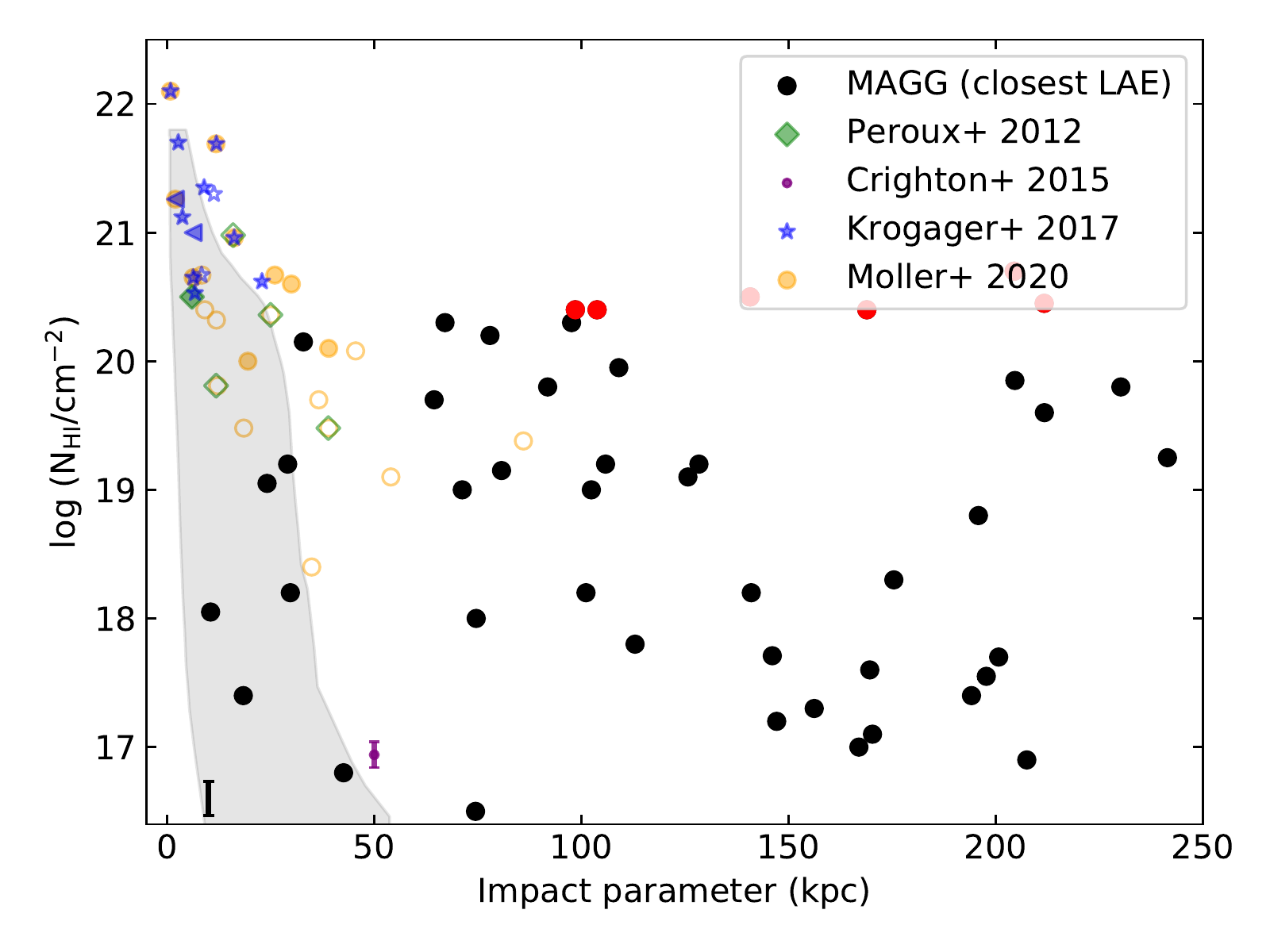}
  & \includegraphics[width=0.49\textwidth, trim= 0.5cm 0.5cm 0cm 0cm, clip]{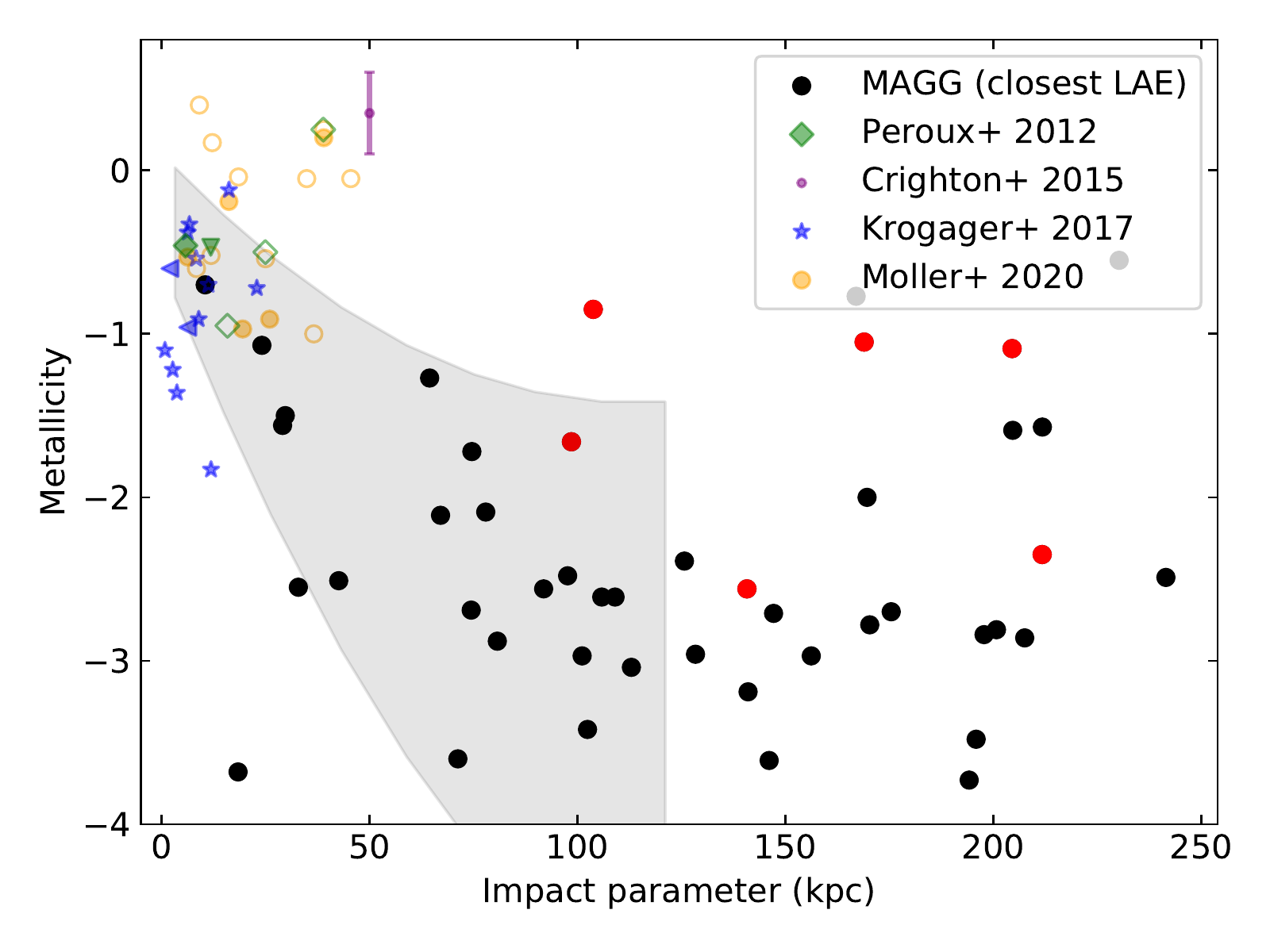}
 \\
    \end{tabular}
        \caption{Comparisons between the MAGG sample and literature values. \textit{Left:} \HI\ column density of absorbers versus the impact parameter to the closest galaxy. We include both observational and theoretical results from the literature for comparison. The typical error on the MAGG results is shown in the lower left corner. 
        \textit{Right:} Comparison of the absorber metallicity as a function of impact parameter. For both plots, we show the MAGG results as dots (red for DLAs, black for LLSs). Other literature results are shown with the green diamonds (DLA sample from \citet{peroux2012}), blue stars \citep{krogager2017}, purple dot \citep{crighton2015} and orange circles \citep{moller2020}. Absorbers at $z<2$ are shown with open markers while those at $z\gtrsim2$ are shown with filled markers. The grey shaded region shows the extent of \HI\ systems drawn from \citet{rahmati2014} at $z\sim 3$ (left panel) and from the FOGGIE simulation \citep{lehner2021} at $2.2<z<3.6$ (right panel).     }

    \label{fig:NHImetscatter}
\end{figure*}

\subsection{Environmental effects on the gas distribution}

Following pioneering work based on multi-object spectroscopy \citep[e.g.,][]{bordoloi2011,nielsen2018}, surveys at high density enabled by MUSE have opened up a new window into exploring links between the density of galaxies and the properties of the surrounding gas. Several authors have reported at various redshifts a marked difference in the gas distribution near galaxies in pairs or groups versus those in isolation. The work by \citet{fossati2019} is among the first studies with MUSE that reports an increase in the strength and extent of \ion{Mg}{II} in $z\approx 1$ group galaxies, a trend later confirmed with MAGG by \citet{dutta2020} and, more systematically and with a larger sample, by \citet{dutta2021}. At these redshifts, the authors suggest that environmental effects act as a perturber of the galaxy CGM within groups, displacing cool and enriched material at larger distances compared to more isolated galaxies.  At higher redshift, excess in the optical depth of both hydrogen and \ion{C}{IV} near LAEs in groups compared to more isolated ones at $z\approx 3$ has been reported by \citet{muzahid2020}. Similarly, Galbiati et al. (in prep.) have found using MAGG data an excess in the covering factor of \ion{C}{IV} of $z\approx 3-4$ LAEs in groups compared to field galaxies. 

Combining the information gathered by our analysis of the MAGG fields, we can further add to the study of \citet{muzahid2020} and Galbiati et al. through a detailed exploration of the correlations between LAEs and LLSs. Our analysis reveals that $\approx 30-40\%$ of LLSs are associated with multiple galaxies (Fig.~\ref{fig:nEmitter_hist}, depending on the confidence level considered), suggesting that a significant fraction of optically-thick clouds live near multiple LAEs, especially considering that our survey is incomplete at low luminosity and likely biased against obscured and non line-emitting systems. When examining closely whether the detailed absorption properties (column density, metallicity, kinematics; Fig.~\ref{fig:DR_NHI_met}) vary systematically with galaxy environment, however, there does not appear to be any noticeable variation in the absorption properties as a function of the galaxy environment. Conversely, in line with what is observed by \citet{muzahid2020} and Galbiati et al.,  we see a much elevated covering factor of optically-thick gas at all the impact parameters probed once considering galaxies that are in pairs or multiples compared to isolated ones (Fig.~\ref{fig:cov_frac_isogroup}). 

We therefore conclude that the galaxy environment plays an active role in increasing the distribution of optically-thick (and metal enriched) gas near LAEs, although there is no individual connection between the line-of-sight properties measured in absorption and the environment. These pieces of observational evidence can find an explanation in light of the above discussion on the origin of optically-thick gas near LAEs. In the previous section, we argued that the majority of LLSs arises from the outer CGM and IGM along filamentary structures connecting LAEs. In this scenario, it is reasonable to expect an enhanced cross-section of optically-thick gas in the presence of multiple LAEs which trace regions at the intersection of multiple filaments and more massive nodes. This scenario would also explain the more enhanced \ion{C}{IV} distribution, provided that the gas within these structures is enriched, e.g., by outflows from galaxies as argued in some simulations \citep[e.g.,][]{shen2012}. 
Due to the local nature of quasar absorption spectroscopy and the fact that the hydrogen and metal distribution are expected to be non uniform (see below for additional discussion), there is no reason to expect a tight correlation between column density, metallicity, kinematics and the number of LAEs detected.

This hypothesis for the origin of an enhanced gas distribution of LAEs relies primarily on the large-scale structures within which galaxies resides, and thus differs from the hypothesis put forward at lower redshift to explain the cool gas distribution in groups, where environmental effects such gravitational and hydrodynamic interactions are deemed responsible for pushing cool gas at larger distances. At present, there is only limited possibility to accurately test this hypothesis with data, and therefore future efforts should focus on assessing the plausibility of this interpretation within cosmological simulations \citep[e.g.,][]{nelson2021} and with new deep observations that can reveal, e.g., perturbations in the gas and stellar content.   

\subsection{A patchy gas distribution near LAEs}

The study of scaling relations involving, on the one side, absorption properties derived from ALSs and, on the other side, emission properties from associated galaxies offers the potential to probe the baryon processes relevant for galaxy evolution on the gas surrounding galaxies inside the CGM. To this aim, particular attention has been given to the variation of hydrogen column density and metallicity as a function of distance from galaxies. Besides being quantities that can be easily computed also at high redshift, their relevance stems from the fact that both the hydrogen and metallicity distributions contain the signatures induced by accretion and feedback processes, as often suggested by simulations \citep[e.g.][]{hummels2013}. 

Two complementary approaches have been followed in the literature to pursue this investigation. One is the so-called `galactocentric' approach, in which samples of galaxies are selected along the line of sight of quasars and then their CGM is studied in absorption \citep[e.g.][]{steidel2010,rudie2012}. The other approach, which we follow primarily here, starts instead with samples selected based on a set of absorption properties and then proceeds to identify associated galaxies. With respect to \ion{H}{I} selected ALSs, at $z\gtrsim 2$, the latter approach has been primarily exploited for DLAs in particular using X-Shooter observations at multiple orientations \citep{fynbo2010} or the SINFONI IFU \citep{peroux2012}. Before our study however, only very small samples of lower column density systems have been examined. Thus, MAGG provides the largest sample for this kind of analysis to date. 

In Fig.~\ref{fig:NHImetscatter}, we focus on the two main scaling relations explored at high redshift: the dependence of \ion{H}{I} column density and metallicity as a function of projected distance from galaxies. We consider both the main DLA samples available in the literature and the LLS sample from MAGG. 
Two main features stand out. First, literature samples almost exclusively occupy the innermost 50~kpc, while MAGG data span a larger range of impact parameters. As already noted, this effects arises from technical reasons with X-Shooter and SINFONI probing most effectively or exclusively the inner $\lesssim 10~$arcsec near the quasars. The second feature is that, combined, the various samples suggest a generally decreasing trend of both hydrogen column density and metallicity as a function of distance from galaxies up to $\approx 30-60$~kpc, at which point the relations flatten and are dominated by a very large scatter, with column densities spanning a range of over three decades and metallicities covering over two orders of magnitude at fixed impact parameter. A large scatter is also evident at small impact parameters, where \ion{H}{I} column densities and metallicities are on average more elevated but with data points covering almost 2 dex in hydrogen column density and a decade in metallicity also for impact parameters of $\lesssim 50~$kpc. 

The existence of such a large scatter is not surprising in light of our current appreciation of the structure of the CGM and IGM. Observations against multiple lines of sight reveal at all redshifts large variations in the hydrogen and metal content on distances of few kiloparsecs, or even tens of parsecs \citep[e.g.][]{rauch2001,lopez2018,decia2021,bordoloi2022}.
Moreover, according to the picture painted above, LAEs in groups or along filaments extending for hundreds of kiloparsecs are likely to experience even larger scatter not only due to the patchy distribution of the CGM/IGM, but also to the superposition of different structures along the line of sight \citep[e.g.][]{bordoloi2011,dutta2020}. Nevertheless, a general systematic trend is still expected provided that sufficient samples are available to unveil robust mean values over distributions of large intrinsic scatter. This is in fact seen in numerical simulations, where mean trends emerge both in the neutral hydrogen and metallicity profiles albeit with dispersions of several orders of magnitude. While systematic comparisons with simulations are not our main aim, for illustration purposes we offer two examples in Fig.~\ref{fig:NHImetscatter}, where we show (to the left) the expected \ion{H}{I} radial profile reconstructed by \citet{rahmati2014} using the EAGLE simulation \citep{schaye2015} and the metallicity distribution (to the right) predicted by the FOGGIE simulations \citep{peeples2019} as shown by \citet{lehner2021}. In both cases, a decreasing trend with radius is predicted with a scatter in the distributions that spans several orders of magnitudes and that increases with distance from the galaxies.

An important caveat on Fig.\ref{fig:NHImetscatter} applies to this discussion when combining DLAs samples from the literature with MAGG observations. The MAGG sample has been purely selected based on \ion{H}{I} column density, with no information on metallicity and with the aim to cover a broad range of column densities. DLA samples, especially from \citet{krogager2017} and collaborators, not only target systematically $\log (N_{\rm HI}/~\rm cm^{-2}) \ge 10^{20.3}$ but focus primarily on metal rich absorbers for which modelling suggests a higher probably of identifying galaxy counterparts. Hence, the suggestion of a marked trend both in column density and metallicity in the data with radius is, at this point, a possible artefact of the different selection effects with literature surveys probing galaxies only at projected distances $\lesssim 50~\rm kpc$ around higher column density and higher metallicity absorbers. 
A second point to note is that, differently from NIR surveys, MAGG targets only Ly$\alpha$ at $z>3$. While this means that we are potentially more sensitive to lower mass galaxies and in general reach lower flux limits, dust attenuation may become a non-negligible bias for the most metal rich absorbers in our sample and galaxies at closer separations may exist but go undetected in our survey.
 
In summary, while MAGG has drastically increased the number of galaxy detections near optically-thick absorbers, it has uncovered very large scatters in the distribution of hydrogen and metals due to the patchy nature of the CGM and the presence of large scale structures within which multiple LAEs are embedded. This intrinsic property of the data thus prevents us from deriving firm constraints on the impact of accretion and feedback on the CGM. Going forward, future efforts should increase the sample size to identify reliable trends within distributions that are characterised by a large dispersion. Moreover, multi-wavelength observations of at least UV and optical lines should identify and minimise possible systematic effects linked to dust near metal rich absorbers. Finally, although we admit this is not a trivial effort, larger scales reaching at least $\approx 1-2~\rm Mpc$ should be explored to investigate more the clustered nature of LAEs in large scale structures.

\section{Summary and conclusions}\label{sec:summary}

In this paper, we have presented the study of the link between $z\sim 3-4$ optically-thick ALSs and galaxies from the full MUSE Analysis of Gas around Galaxies (MAGG) sample. The MAGG sample covers 28 quasar fields with medium-depth IFU observations using VLT MUSE for at least 4h on-source. These quasar sightlines were selected to include at least one \ion{H}{I} absorption system with $N_{\rm HI} \gtrsim 10^{16.5}~\rm cm^{-2}$ from archival high-resolution quasar spectroscopy. In total, the full sample includes 61 \ion{H}{I} optically-thick absorbers. With MUSE, we identified galaxies via Ly$\alpha$ emission by searching a velocity window of 1000~\kms\ around these absorbers in the $(500~\rm kpc)^2$ FOV  of MUSE, and we studied the connection between gas and galaxies across a wide range of absorption properties with no pre-selection on parameters such as the metallicity. 

The key results emerging from our analysis are:
\begin{enumerate}

    \item We identified \nem~galaxies around 61 high column density \ion{H}{I} absorbers at $2.9 < z < 4.2$. For 10 absorbers we detected no galaxies within the volume searched (impact parameters $\lesssim 300~\rm kpc$ and velocity window of $\pm 1000~$\kms). This resulted in a detection rate of 82 per cent ($50/61$). The highest number of detected galaxies around a single absorber was 8 while most of the absorption systems had a single galaxy detection.
    
    \item The \ion{H}{I} absorbers were found to be clearly correlated with galaxy overdensities. Specifically, galaxies were found to cluster in velocity space with the majority of them lying within $\rm 500 km~s^{-1}$ of the absorbers' redshift, once accounting for a typical velocity offset of $\approx 250~\rm km~s^{-1}$ due to the resonant Ly$\alpha$ emission. The luminosity function for galaxies around absorbers has an $\approx 5\times$ higher normalisation than field galaxies, further supporting the physical connection between absorbers and LAEs. This enhancement is, however, less strong than the one observed for galaxies around the quasars themselves \citep{fossati2021}.
    We also showed that the transverse cross-correlation between galaxies and absorbers has a positive correlation, indicating the presence of clustering on the scales probed within the MUSE FOV. We reported a clustering length of $R_0 = 1.43^{+0.26}_{-0.29} h^{-1}$~cMpc.
        
    \item We observed no differences in the distributions of either the \ion{H}{I} column density or metallicity for samples of absorbers for which there are no, 1 or $\ge 2$ associated galaxies, indicating that the galaxy environment does not alter significantly the absorption properties. 
    
    \item When multiple LAEs are detected near an LLS, we observed a preferential alignment along the galaxy-absorber axis with an excess of galaxies along this direction relative to random distribution, suggestive of LAEs and optically-thick absorbers lying in filamentary structures.
    
    \item No strong correlations were observed in individual systems between the ALS properties (\ion{H}{I} column density, metallicity, metal ion velocity width) and the galaxy emission properties (Ly$\alpha$ emission, projected distance between galaxy and absorber, and line of sight velocity separation). Only a mild correlation between the ALS velocity width ($\Delta v_{90}$) and Ly$\alpha$ emission was found, suggesting that the brightest LAEs live in regions of higher velocity dispersion, possibly due to effects induced by their halo mass or outflows. These results hold when considering either the brightest galaxy detected around each absorber or when selecting the closest in terms of impact parameter. 
    
    \item Once considering the full sample of $>1000$ LAEs identified in the MUSE cubes, we calculated the covering fraction of optically-thick gas around galaxies.
    We found an elevated covering factor on scales of $\lesssim 50~\rm kpc$, typical of the inner CGM, within $1-2~R_{\rm vir}$. We also observed an inflection in the radial dependence of the covering factor around $\approx 100~\rm kpc$, which we interpret as a transition between the contribution of the CGM of individual LAEs and the contributions from the IGM/CGM of other galaxies embedded in the same structures. When considering galaxies that reside in pairs or groups, we found an elevated covering factor (up to $3\times$) compared to what is seen around isolated galaxies. 
\end{enumerate}

The empirical findings described above led us to the following conclusions on the nature of optically-thick gas at $z\gtrsim 3$ and the use of galaxy-absorber pairs to constrain the baryon cycle:

\begin{enumerate}
 \item The sensitivity of MUSE and the ability to uniformly probe impact parameters up to $\approx 250$~kpc from the quasars now routinely leads to very high detection rates of galaxies near optically-thick absorbers, with $>80$ per cent detection rate at the moderate depth of our survey regardless of the absorption properties. IFU surveys thus have the potential to reach 100 per cent detection rates in deeper searches. The majority of the identified galaxies ($\approx 80$ per cent) lie at large impact parameters ($\gtrsim 100~\rm kpc$) with respect to the absorbing gas, with $\approx 30-40$ per cent of LLSs being associated with more than one galaxy. Our survey thus reveals the full impact of the clustered nature of galaxies within cosmic structures containing hydrogen, forcing us to move from the study of individual galaxy counterparts of absorbers to a more statistical analysis of the galaxy environment near ALSs.    

    \item Based on the recovered statistics of the distribution of galaxies near optically-thick gas, we exclude the ISM and inner CGM ($\lesssim 1-2~R_{\rm vir}$) of the identified galaxies as the origin of the observed LLSs. We favour instead the outer CGM (defined here as $3-4~R_{\rm vir}$) and filaments of the IGM within which galaxies are embedded as the sites where LLSs reside. Furthermore, in line with conclusions from studies of Ly$\alpha$ in emission, we identify LAEs as the most likely contributors to the cosmic number density of optically-thick gas. Indeed, given the elevated number density and covering factor of optically-thick gas around LAEs, virtually all LLSs can be identified within 150~kpc of an LAE. Moreover, the clustered nature of LAEs with respect to themselves and with other galaxy populations (e.g. LBGs or quasars) imply that the covering factors measured at large distances in these other galaxy populations are not independent and are already accounted for in the budget attributed to LAEs. In other words, by virtue of their larger number density, LAEs contribute either directly with their CGM to the cosmic budget of LLSs or act as tracers of the larger scale structures which host both LLSs and other galaxies.  
    \item The emerging picture of LAEs embedded within filaments hosting optically-thick gas, combined with the literature evidence that the CGM/IGM is patchy and locally inhomogeneous, explains the lack of significant correlations between absorption and emission properties, and the very large scatter ($\gtrsim 2-3$ dex) observed in the hydrogen and metal distribution around LAEs on scales of $\approx 50-200~\rm kpc$. In turn, this implies that the signature of baryon processes (i.e., inflows and outflows) around galaxies are hard to discern in moderate-size samples. Larger samples are thus required to unveil systematic trends inside distributions with intrinsically high dispersion. 
    Furthermore, it is not particularly surprising that the galaxy environment does not have a clear influence on the detailed properties of ALSs, but the elevated covering factor in groups can be explained as enhanced hydrogen cross-sections in massive nodes where multiple filaments intersect or overlap.  
\end{enumerate}

In conclusion, the MAGG survey has provided the most detailed view of the connection between optically-thick gas and galaxies at $z\gtrsim 3$, enabling for the first time a systematic study of the clustering of line emitting galaxies with respect to LLSs, and of their individual absorption and emission properties. This analysis has led to a clearer understanding of how hydrogen is distributed around star-forming galaxies at high redshift, painting a view similar to the one proposed by cosmological simulations according to which galaxies are clustered inside filaments that contain substantial amount of neutral hydrogen. From a theoretical point of view, the next steps are to provide a more rigorous and detailed assessment of the level of agreement that exists between cosmological simulations and the MAGG observations. Empirically, extending observations to rest-frame optical and infrared emission will become a crucial element to complete this picture with dusty or non star-forming galaxies that are underrepresented in MAGG. Campaigns with the {\it James Webb Space Telescope} and ALMA should make this possible. Furthermore, studies like the one proposed here should be continued to further grow the samples and unveil the systematic trends between galaxy properties in emission and gas properties in absorption, which are critically needed for novel constraints on the baryon processes operating near and around galaxies.

\section{Data availability}
The VLT data used in this work are available from the European Southern Observatory archive (\url{https://archive.eso.org/}) either as raw data or phase 3 data products. Spectroscopy obtained at the Keck telescopes is available via the Keck Observatory Archive (KOA, \url{https://www2.keck.hawaii.edu/koa/public/koa.php}).
Cubextractor can be obtained upon request by contacting Sebastiano Cantalupo, and other codes used in this paper have been made available at http://www.michelefumagalli.com/codes.html.

\section{Online supporting information}
The fits of the metal absorption lines for all of the systems in this work are included as online material along with associated tables listing the properties for each ion including equivalent widths and column densities. We also provide the full figures from the photoionisation modelling used to determine the metallicity of each absorption systems.

\section*{Acknowledgements}

This project has received funding from the European Research Council (ERC) under the European Union's Horizon 2020 research and innovation programme (grant agreement No 757535) and by Fondazione Cariplo (grant No 2018-2329). RJC is supported by a Royal Society University Research Fellowship, and acknowledges support from STFC (ST/T000244/1). SC gratefully acknowledges support from the European Research Council (ERC) under the European Union's Horizon 2020 research and innovation programme grant agreement No 864361.
MTM thanks the Australian Research Council for \textsl{Discovery Project} grants DP130100568, DP170103470 and DP190100417, and for \textsl{Future Fellowship} grant FT180100194 which supported this work.
This work is based on observations collected at the European Organisation for Astronomical Research in the Southern Hemisphere under ESO programmes ID 
197.A-0384, 
065.O-0299,
067.A-0022,
068.A-0461,
068.A-0492,
068.A-0600,
068.B-0115,
069.A-0613,
071.A-0067,
071.A-0114,
073.A-0071,
073.A-0653,
073.B-0787,
074.A-0306,
075.A-0464,
077.A-0166,
080.A-0482,
083.A-0042,
091.A-0833,
092.A-0011,
093.A-0575,
094.A-0280,
094.A-0131,
094.A-0585,
095.A-0200,
096.A-0937,
097.A-0089,
099.A-0159,
166.A-0106,
189.A-0424.
This work used the DiRAC Data Centric system at Durham University, operated by the Institute for Computational Cosmology on behalf of the STFC DiRAC HPC Facility (www.dirac.ac.uk). This equipment was funded by BIS National E-infrastructure capital grant ST/K00042X/1, STFC capital grants ST/H008519/1 and ST/K00087X/1, STFC DiRAC Operations grant ST/K003267/1 and Durham University. DiRAC is part of the National E-Infrastructure. This research made use of Astropy, a community-developed core Python package for Astronomy \citep{astropy2013,astropy2018}. 
This research has made use of the NASA/IPAC Extragalactic Database (NED) which is operated by the Jet Propulsion Laboratory, California Institute of Technology, under contract with the National Aeronautics and Space Administration.
This research has made use of data from the Sloan Digital Sky Survey IV (\url{www.sdss.org}), funded by the Alfred P. Sloan
Foundation, the U.S. Department of Energy Office of Science, and the Participating Institutions.


\bibliographystyle{mnras}

\clearpage
\appendix

\bsp 
\label{lastpage}
\end{document}